\documentclass[11pt,a4paper]{article}

\usepackage{subcaption}
\usepackage{enumitem}
\usepackage{mathtools}
\usepackage[top=2.0cm, bottom=2.0cm, left=2.0cm, right=2.0cm]{geometry}
\usepackage{pgfplots}
\pgfplotsset{compat=newest}

\usepackage{ifthen}
\usepackage{colortbl}
\usepackage{array}
\usepackage{color}
\usepackage{soul}
\usepackage{amssymb}

\newboolean{IsWide}
\setboolean{IsWide}{true}

\newcommand{\FlaTwoByTwo}[4]{
\left(
\begin{array}{c I c}
#1 & #2 \\ \whline
#3 & #4
\end{array}
\right)
}

\newcommand{\FlaTwoByTwoSingleLine}[4]{
\left(
\begin{array}{c | c}
#1 & #2 \\ \hline
#3 & #4
\end{array}
\right)
}

\newcommand{\FlaTwoByOneSingleLine}[2]{
\left(
\begin{array}{c}
#1 \\ \hline
#2
\end{array}
\right)
}

\newcommand{\FlaOneByTwo}[2]{
\left(
\begin{array}{c I c}
#1 & #2
\end{array}
\right)
}

\newcommand{\FlaOneByTwoSingleLine}[2]{
\left(
\begin{array}{c | c}
#1 & #2
\end{array}
\right)
}

\newcommand{\FlaThreeByThreeTL}[9]{
\left(
\begin{array}{c | c I c}
#1 & #2 & #3 \\ \hline
#4 & #5 & #6 \\ \whline
#7 & #8 & #9
\end{array}
\right)
}

\newcommand{\FlaThreeByThreeBR}[9]{
\left(
\begin{array}{c I c | c}
#1 & #2 & #3 \\ \whline
#4 & #5 & #6 \\ \hline
#7 & #8 & #9
\end{array}
\right)
}

\newcommand{\FlaThreeByThreeSingleLine}[9]{
\left(
\begin{array}{c | c | c}
#1 & #2 & #3 \\ \hline
#4 & #5 & #6 \\ \hline
#7 & #8 & #9
\end{array}
\right)
}

\newcommand{\FlaOneByThreeR}[3]{
\left(
\begin{array}{c I c | c}
#1 & #2 & #3
\end{array}
\right)
}

\newcommand{\FlaOneByThreeL}[3]{
\left(
\begin{array}{c | c I c}
#1 & #2 & #3
\end{array}
\right)
}

\newcommand{\FlaPartition}[2]{
\ifthenelse{\boolean{IsWide}}{{\bf partition } \hspace{-1em} #1 \hspace{-1em} #2}
{{\bf partition } \+ \\ #1 \+ \\ #2 \- \-}
}

\newcommand{\FlaRepartition}[2]{
\ifthenelse{\boolean{IsWide}}{{\bf repartition } \hspace{-1em} #1 \hspace{-1em} #2}
{{\bf repartition } \+ \\ #1 \+ \\ #2 \- \-}
}

\newcommand{\FlaStartCompute}{%
\setlength{\unitlength}{1.6in}%
\begin{picture}(3,0.01)
\put(0,0){\line(1,0){3}}
\put(0,0.01){\line(1,0){3}}
\end{picture}%
}

\newcommand{\FlaEndCompute}{%
\noindent%
\setlength{\unitlength}{1.6in}%
\begin{picture}(3,0.01)
\put(0,0){\line(1,0){3}}
\put(0,0.01){\line(1,0){3}}
\end{picture}%
}

\newcommand{\operation}{ [ D, E, F, \ldots ] \becomes {\rm op}( A, B, C, D, \ldots ) }
\newcommand{\routinename}{ [ D, E, F, \ldots ] \becomes {\rm op}( A, B, C, D, \ldots ) }
\newcommand{\routinecost}{ X }
\newcommand{\precondition}{ Q }
\newcommand{\postcondition}{ R }
\newcommand{\invariant}{ P }
\newcommand{\costinv}{ \  }

\newcommand{\guard}{ R }
\newcommand{\partitionings}{
\begin{minipage}{3in}
$ S_I $
\end{minipage}
}
\newcommand{\initialize}{}

\newcommand{\partitionsizes}{ \hspace{ 3.25in} }

\newcommand{\blocksize}{1}

\newcommand{\repartitionings}{
\begin{minipage}[t]{3in}
\ \\
\ \\
\ \\
\end{minipage}
}

\newcommand{\repartitionsizes}{ \hspace{ 3.25in} }
\newcommand{\moveboundaries}{
\begin{minipage}[t]{3in}
\ \\
\ \\
\ \\
\end{minipage}
}
\newcommand{\beforeupdate}{
$ \QBefore $
}
\newcommand{\afterupdate}{
$ \QAfter $
}
\newcommand{\update}{%
\begin{minipage}[t]{4in}
$ S_U $
\end{minipage}
}

\newcommand{\resetsteps}{
\renewcommand{\blocksize}{1}
\renewcommand{\operation}{ [ D, E, F, \ldots ] \becomes {\rm op}( A, B, C, D, \ldots ) }
\renewcommand{\routinename}{ [ D, E, F, \ldots ] \becomes {\rm op}( A, B, C, D, \ldots ) }
\renewcommand{\routinecost}{ 0 }
\renewcommand{\precondition}{ \PPre }
\renewcommand{\postcondition}{ \PPost }
\renewcommand{\invariant}{ \PInv }
\renewcommand{\costinv}{ \  }
\renewcommand{\guard}{ G }
\renewcommand{\partitionings}{ %
\begin{minipage}[t]{3in}
\ \\
\end{minipage}
}
\renewcommand{\partitionsizes}{ \hspace{ 3.25in} }
\renewcommand{\repartitionings}{%
\begin{minipage}[t]{3in}
\ \\
\end{minipage}
}
\renewcommand{\repartitionsizes}{ \hspace{ 3.25in} }
\renewcommand{\moveboundaries}{%
\begin{minipage}[t]{3in}
\ \\
\end{minipage}
}
\renewcommand{\beforeupdate}{
\QBefore
}
\renewcommand{\afterupdate}{
\QAfter
}
\renewcommand{\update}{
$ S_U $
}
}

\newcommand{\WSguard}{
$ \guard $
}

\newcommand{\WSpartition}{%
\begin{minipage}[t]{3.0in}%
\begin{tabbing}
ind \= ind \= \kill
{\bf \color{blue} Partition}
\partitionings \+ \\
{\bf \color{blue} where } \hspace*{-2ex} \partitionsizes
\end{tabbing}
\end{minipage}
}

\newcommand{\WSinitialize}{
\initialize
}

\newcommand{\WSrepartition}{
\begin{minipage}[t]{3in}
\ifthenelse{ \equal{\blocksize}{1} }{}
{%
\ifthenelse{ \equal{\blocksize}{blank} }{~}
{{\bf \color{blue} Determine block size} $ \blocksize $} \\
}
{\bf \color{blue} Repartition}
\begin{tabbing}
in \= in \= \+ \kill
\repartitionings \+ \\
{\bf \color{blue} where } \hspace*{-2ex} \repartitionsizes
\end{tabbing}
\end{minipage}
}

\newcommand{\WSrepartitionNarrow}{
\begin{minipage}[t]{2.05in}
\ifthenelse{ \equal{\blocksize}{1} }{\phantom{Determine}~}
{%
\ifthenelse{ \equal{\blocksize}{blank} }{}
{{\bf \color{blue} Determine block size} $ \blocksize $} \\
}
{\bf \color{blue} Repartition}
\begin{tabbing}
i \= i \= \+ \kill
\repartitionings 
\+ \\
{\bf \color{blue} where }
\begin{minipage}[t]{1.5in}
\repartitionsizes
\end{minipage}
\end{tabbing}
\end{minipage}
}

\newcommand{\WSmoveboundary}{%
\begin{minipage}[t]{4in}%
{\bf \color{blue} Continue with}
\begin{tabbing}
ind \= \+ \kill
\moveboundaries
\end{tabbing}
\end{minipage}
}

\newcommand{\WSupdate}{
\update
}

\newcommand{\FlaAlgorithmWithInit}{
\begin{center}
\begin{tabular}{| p{0.98\textwidth} |} \hline
{\bf \color{blue} Algorithm:} $\routinename$
\\ \whline
\WSinitialize \\
{\WSpartition} \\[0.3in]
{\bf \color{blue} while} \WSguard { \bf \color{blue} do} \\
\ \hspace{0.15in} \WSrepartition \\
\color{red} {\hspace{0.0in} \FlaStartCompute} \\
{\hspace{0.0in} \WSupdate} \\
\color{red} {\hspace{0.0in} \FlaEndCompute} \\
{\ \hspace{0.15in} \WSmoveboundary} \\
{{\bf \color{blue} endwhile}} \\ \hline
\end{tabular}
\end{center}
}

\newcommand{ \PPre }{ P_{\it pre} }
\newcommand{ \PPost }{ P_{\it post} }
\newcommand{ \PInv }{ P_{\it inv} }

\newcommand{ \QBefore }{ P_{\it before} }
\newcommand{ \QAfter }{ P_{\it after} }

\newcommand{\Answer}[1]{%
\ifthenelse{\boolean{ShowAnswers}}{%
{\color{blue}%
\vspace{0.05in}%
\noindent%
{\bf Answer:} #1}}
{}%
}

\newcommand{\PartAnswer}[1]{%
\ifthenelse{\boolean{ShowAnswers}}{%
{\color{blue} #1}}%
{}
}

\newcommand{\QuestionAnswer}[2]{%
\ifthenelse{\boolean{ShowAnswers}}{%
{\color{blue} #2}}%
{\color{black} #1}
}

\newcommand{\ShowAnswer}[1]{
\ifthenelse{\boolean{ShowAnswers}}{
{\color{blue} #1}}
{\phantom{#1}}
}

{
\vspace{0.1in}
\end{minipage}
\\ \whline
\end{tabular}
}

\newenvironment{unboxedexercise}{
\addtocounter{homeworkcounter}{1}%
\noindent%
\begin{exercise}
\rm 
}
{
\end{exercise}%
}

\newcolumntype{I}{!{\vrule width 1.5pt}}
\newlength\savedwidth
\newcommand\whline{\noalign{\global\savedwidth\arrayrulewidth
                            \global\arrayrulewidth 1.5pt}%
           \hline
           \noalign{\global\arrayrulewidth\savedwidth}}

\newcommand{\NoShow}[1]{}

\newcommand{\becomes}{:=}

\definecolor{color0}{RGB}{117,195,255}
\definecolor{color1}{RGB}{0,0,255}
\definecolor{color2}{RGB}{255,255,255}
\definecolor{color3}{RGB}{255,0,0}
\definecolor{color4}{RGB}{255,0,174}
\definecolor{color5}{RGB}{179,0,0}
\definecolor{color6}{RGB}{0,255,0}
\definecolor{color7}{RGB}{255,255,0}
\definecolor{color8}{RGB}{235,0,0}
\definecolor{color9}{RGB}{0,162,0}
\definecolor{color0}{RGB}{255,0,255}
\definecolor{color11}{RGB}{100,100,177}
\definecolor{color12}{RGB}{172,174,41}
\definecolor{color13}{RGB}{255,144,26}
\definecolor{color14}{RGB}{2,255,177}
\definecolor{color15}{RGB}{192,224,0}
\definecolor{color16}{RGB}{66,66,66}
\definecolor{color17}{RGB}{255,0,96}
\definecolor{color18}{RGB}{169,169,169}
\definecolor{color19}{RGB}{169,0,0}
\definecolor{color20}{RGB}{0,109,255}
\definecolor{color21}{RGB}{200,61,68}
\definecolor{color22}{RGB}{200,66,0}
\definecolor{color23}{RGB}{0,41,0}
\definecolor{color24}{RGB}{139,121,177}
\definecolor{color25}{RGB}{116,116,116}
\definecolor{color26}{RGB}{200,50,89}
\definecolor{color27}{RGB}{255,171,98}
\definecolor{color28}{RGB}{0,68,189}
\definecolor{color29}{RGB}{52,43,0}
\definecolor{color30}{RGB}{255,46,0}
\definecolor{color31}{RGB}{100,216,32}

\newcommand{\defeq}{\vcentcolon=}
\newcommand{\si}{(i)} 
\newcommand{\sip}{(i+1)} 

\newlength{\topFigureVerticalSpace}
\newlength{\bottomFigureVerticalSpace}
\newcommand{\tfvspace}{\vspace*{\topFigureVerticalSpace}}
\newcommand{\bfvspace}{\vspace*{-\bottomFigureVerticalSpace}}

\newcommand{\vs}{\vspace*{2mm}}
\newcommand{\minusvs}{\vspace*{-2mm}}

\newcommand{\randUTV}{{\sc randUTV}}

\begin{document}


\title{
  Efficient algorithms for computing a rank-revealing UTV factorization
  on parallel computing architectures
}

\date{}

\author{%
N.~Heavner%
\footnote{Department of Applied Mathematics, 
          University of Colorado at Boulder, 
          526 UCB, Boulder, CO 80309-0526, USA.
          e-mail: \texttt{nathanheavner@hotmail.com}
          } \and
F.~D.~Igual%
\footnote{Depto.~de Arquitectura de Computadores y Autom\'atica,
          Universidad Complutense de Madrid,
          28040--Madrid, Spain.
          e-mail: \texttt{figual@ucm.es}
          } \and
G.~Quintana-Ort\'{i}%
\footnote{Depto.~de Ingenier\'{\i}a y Ciencia de Computadores,
          Universidad Jaume I,
          12.071--Castell\'on, Spain.
          e-mail: \texttt{gquintan@uji.es}
          } \and
P.G.~Martinsson%
\footnote{Department of Mathematics,
          University of Texas at Austin,
          Stop C1200,
          Austin, TX 78712-1202, USA.
          e-mail: \texttt{pgm@ices.utexas.edu}
          }
}

\maketitle


\begin{abstract}
The randomized singular value decomposition (RSVD) is by now a well
established technique for efficiently computing an approximate singular
value decomposition of a matrix.
Building on the ideas that underpin the RSVD, the recently proposed
algorithm ``randUTV'' computes a \textit{full} factorization of a given
matrix that provides low-rank approximations with near-optimal error.
Because the bulk of randUTV is cast in terms of communication-efficient
operations like matrix-matrix multiplication and unpivoted QR
factorizations, it is faster than competing rank-revealing factorization
methods like column pivoted QR in most high performance computational
settings.
In this article, optimized randUTV implementations are presented for both
shared memory and distributed memory computing environments.
For shared memory, randUTV is redesigned in terms of an {\em
algorithm-by-blocks} that, together with a runtime task scheduler,
eliminates bottlenecks from data synchronization points to achieve
acceleration over the standard {\em blocked algorithm}, based on a purely
fork-join approach.
The distributed memory implementation is based on the ScaLAPACK library.
The performances of our new codes compare favorably with competing
factorizations available on both shared memory and distributed memory
architectures.
\end{abstract}


\maketitle

\section{Introduction.}

\subsection{Overview.}
\label{sec:overview}

Computational linear algebra faces significant challenges as high performance computing moves further away from the serial into the parallel. Classical algorithms were designed to minimize the number of  floating point operations, and do not always lead to optimal performance on
modern communication-bound architectures. The obstacle is particularly apparent in the area of rank-revealing matrix factorizations. Traditional  techniques based on column pivoted QR factorizations or Krylov methods tend to be challenging to parallelize well, as they are most naturally viewed as a sequence of matrix-vector operations.

In this paper, we describe techniques for efficiently implementing a randomized algorithm for computing a so-called rank-revealing UTV decomposition \cite{martinsson2017randutv}. Given an input matrix $A$ of size $m \times n$, the objective is to compute a factorization
\begin{equation}
\label{eq:defUTVpre}
\begin{array}{ccccccccccc}
A &=& U & T & V^{*},\\
m\times n && m\times m & m\times n & n\times n
\end{array}
\end{equation}
where the middle factor $T$ is upper triangular (or upper trapezoidal in the case $m < n$) and the left and right factors $U, V$
are orthogonal. The factorization is rank-revealing in the sense that
\begin{equation}
\| A - U(:,1:k) T(1:k,:) V^* \| \approx \inf \{ \| A - B \| : B \text{ has rank } k \}.
\end{equation}
{In a factorization resulting from randUTV, the middle matrix $T$ often has elements above the diagonal that are very small in modulus, which means that the diagonal entries of $T$ become excellent approximations to the singular values of $A$.}
A factorization of this type is useful for solving tasks such as low-rank approximation, for determined basis to approximations to the fundamental subspaces of $A$, for solving ill-conditioned or over/under-determined linear systems in a least-square sense, and for estimating the singular values of $A$.

The randomized UTV algorithm \texttt{randUTV} that we implement has characteristics that in many environments make it preferable to classical rank-revealing factorizations like column pivoted QR (CPQR) and the singular value decomposition (SVD):
\begin{itemize}
\item It consistently produces matrix factors which yield low-rank approximations with accuracy comparable to the SVD. The particular use of randomization in the algorithm is essentially risk free. The reliability of the method is supported by theoretical analysis, as well as extensive numerical experiments.
\item It casts most of its operations in terms of matrix-matrix multiplications, which are highly efficient in parallel computing environments. It was demonstrated in \cite{martinsson2017randutv} that a straightforward blocked implementation of \texttt{randUTV} executes faster than even highly optimized implementations of CPQR in symmetric multiprocessing (SMP) systems. In this manuscript, we present an implementation that improves on the performances in \cite{martinsson2017randutv} for SMP and obtain similar findings for distributed memory architectures.

\item It processes the input matrix in sets of multiple columns, so it can be stopped part way through the factorization process if it is found that a requested tolerance has been met. If $k$ columns end up having been computed, only $O(mnk)$ flops will have been expended.
\end{itemize}

In this manuscript, we present two efficient implementations for computing the \texttt{randUTV} factorization: the first one for shared-memory machines, and the second one for distributed-memory machines.
Regarding shared-memory architectures, the implementation presented in our paper~\cite{martinsson2017randutv} proposed a blocked algorithm in which parallelism was extracted on a per-task basis, relying on parallel BLAS implementations, and hence following a fork-join parallel execution model. Here, we propose a novel algorithm-by-blocks~\cite{ab-toms-2009}, in which sequential tasks are dynamically added to a Directed Acyclic Graph (DAG) and executed by means of a runtime task scheduler ({\tt libflame}'s SuperMatrix~\cite{chan2007supermatrix}). This approach enhances performance by mitigating the effects of the inherent synchronization points in fork-join models, and has shown its potential in other linear algebra implementations~\cite{chan2008supermatrix}. In addition, given the recent improvements in terms of performance of modern SVD implementations (e.g. in Intel MKL), we show how runtime-based implementations of \texttt{randUTV} are still on par with them in terms of performance. Regarding our second proposal, it is the first time a distributed-memory version of \texttt{randUTV} is presented in the literature; performance results
reveal excellent scalability results compared with state-of-the-art distributed-memory implementations.

Specifically, the main contributions of the paper compared with the state-of-the-art are:

\begin{enumerate}

	\item We propose a novel algorithm-by-blocks for computing the {\tt randUTV} factorization that maximizes performance at no programmability cost.
	\item We have integrated our solution with an existing task-based software infrastructure ({\tt libflame} SuperMatrix), hence providing an out-of-the-box implementation based on tasks for {\tt randUTV}.
	\item On shared-memory architectures, we provide a detailed study of the optimal block sizes compared with a parallel-BLAS-based solution, and report qualitative and quantitative differences between them that can be of interest for the community. Similarly, we have carried out a detailed performance and scalability study on two highly-parallel shared-memory machines.
	\item Performance results reveal the benefits of the {\em algorithm-by-blocks} compared with the {\em blocked algorithm} on our target testbed, yielding performance improvements between $1.73\times$ and $2.54\times$ for the largest tested matrices. Accelerations compared with proprietary MKL SVD implementations also reveal substantial performance gains, with improvements up to $3.65\times$ for selected cases, and in general in all cases that involve relatively large matrices ($n>4000$).
	\item On distributed-memory architectures, the comparison in terms of execution time with ScaLAPACK SVD and CPQR and PLiC CPQR reveal consistent performance gains ranging from $2.8\times$ to $6.6\times$, and an excellent scalability on the tested platforms.
	\item On distributed-memory architectures, we provide a detailed performance study regarding block sizes, grid sizes, threads per process, etc. on several number of nodes.

\end{enumerate}

The paper is structured as follows: We first discuss the notation that will be used throughout the paper in Section \ref{sec:prelims}. In Section \ref{sec:utv}, we familiarize the reader with the \texttt{randUTV} algorithm that was recently described in \cite{martinsson2017randutv}. Sections \ref{sec:shared} and \ref{sec:distributed} describe the shared and distributed memory implementations that form the main contribution of this manuscript. In Section \ref{sec:performance_analysis}, we present numerical results that compare our implementations to highly optimized implementations of competing factorizations.  Section \ref{sec:conclusions} summarizes the key findings and outlines some possibilities for further improvements and extensions.

\section{Preliminaries.}
\label{sec:prelims}

We use the notation $A \in \mathbb{R}^{m \times n}$ to specify that $A$ is an $m \times n$ matrix with real entries. An \textit{orthogonal} matrix is a square matrix whose column vectors each have unit norm and are pairwise orthogonal. $\sigma_i(A)$ represents the $i$-th singular value of $A$, and $\displaystyle{\inf(A) = \min_i  \{ \sigma_i(A) \}}$. The default norm $\| \cdot \|$ is the spectral norm. We also use the standard matrix indexing notation $A(c:d,e:f)$ to denote the submatrix of $A$ consisting of the entries in the $c$-th through $d$-th rows of the $e$-th through $f$-th columns.

\subsection{The Singular Value Decomposition (SVD)}
\label{sec:svd}

Let $A \in \mathbb{R}^{m \times n}$ and $p = \min(m,n)$.
It is well known \cite{golub,trefethen1997numerical,1998_stewart_volume1} that any matrix $A$ admits a
singular value decomposition (SVD) of the form
\[
\begin{array}{ccccc}
A & = & U & \Sigma & V^*, \\
m \times n & & m \times m & m \times n & n \times n
\end{array}
\]
where $U$ and $V$ are orthogonal and $\Sigma$ is diagonal.  We may also speak of the \textit{economic} SVD of $A$, given by
\[
\begin{array}{ccccc}
A & = & U & \Sigma & V^*, \\
m \times n & & m \times p & p \times p & p \times n
\end{array}
\]
in which case $U$ and $V$ are not necessarily orthogonal (because they are not square), but their columns remain orthonormal. The diagonal elements of $\Sigma$ are the \textit{singular values} $\{\sigma_{i}\}_{i=1}^{p}$ of $A$. These are ordered so that $\sigma_1 \ge \sigma_2 \ge \ldots \ge  \sigma_{p-1} \ge \sigma_p \ge 0$. The columns $u_i$ and $v_i$ of $U$ and $V$ are called the \textit{left} and \textit{right singular vectors}, respectively, of $A$.

A key fact about the SVD is that it provides theoretically optimal rank-$k$ approximations to $A$. Specifically, the Eckart-Young-Mirsky Theorem \cite{eckart1936approximation,mirsky1960symmetric} states that given the SVD of a matrix $A$ as described above and a fixed $1 \le k \le p$, we have that
\[
\| A - U(:,1:k) \Sigma(1:k,1:k) (V(:,1:k))^* \| = \inf \{ \| A - B \| : B \text{ has rank } k \}.
\]
A corollary of this result is that the subspaces spanned by the leading $k$ left and right singular vectors of $A$ provide the optimal rank-$k$ approximations to the column and row spaces, respectively, of $A$.
For instance, if $P$ is the orthogonal projection onto the subspace spanned by the left singular vectors of $A$,
then $PA = U(:,1:k) \Sigma(1:k,1:k) (V(:,1:k))^*$, so
$\|A - PA\| = \inf \{ \| A - B \| : B \text{ has rank } k \}$.

\subsection{The QR decomposition}
\label{sec:cpqr}

Given a matrix $A \in \mathbb{R}^{m \times n}$, let $p = \min(m,n)$. A QR decomposition of $A$ is given by
\[
\begin{array}{cccc}
A & = & Q & R, \\
m \times n & & m \times m & m \times n
\end{array}
\]
where $Q$ is orthogonal and $R$ is upper triangular. If $m > n$, then any QR can be reduced to the ``economic'' QR
\[
\begin{array}{cccc}
A & = & Q & R. \\
m \times n & & m \times n & n \times n
\end{array}
\]

The standard algorithm for computing a QR factorization relies on Householder reflectors. We refer to this algorithm as \textsc{HQR} in this article. A full discussion of the HQR algorithm can be found in \cite{golub,trefethen1997numerical,1998_stewart_volume1}; for our purposes, it is only necessary to note that the outputs of \textsc{HQR} are the following:
an upper triangular matrix $R$ of the QR factorization, and
a unit lower triangular matrix $V \in \mathbb{R}^{m \times p}$
and a vector $v \in \mathbb{R}^p$ that can be used to build or to apply $Q$ (see Section \ref{sec:compactwy}).
\footnote{We should say that $V$ holds the ``Householder vectors''.}
In this article, we make critical use of the fact that for $m > n$, the leading $p$ columns of $Q$ form an orthonormal basis for the column space of $A$.

\subsection{Compact $WY$ representation of collections of Householder reflectors.}
\label{sec:compactwy}

Consider a matrix $A \in \mathbb{R}^{n \times n}$,
and let $H_i \in \mathbb{R}^{n \times n}, \; i=1,\ldots,b$ be
Householder transformations.
As a Householder transformation has the following structure:
$H_i = I - \tau_i v_i v_i^*$,
applying it to a matrix $A$ requires a matrix-vector product
and a rank-1 update.
If all $H_i$ are applied one after another,
the computation requires $\mathcal{O}(bn^2)$ flops in overall
because of the special structure of the Householder transformations.
Both operations are matrix-vector based, and
therefore they do not render high performances on modern architectures.

If several Householder transformations must be applied,
the product $H = H_b H_{b-1} \cdots H_2 H_1$ may be expressed in the form
\[
H = I - W T W^*,
\]
where $W \in \mathbb{R}^{n \times b}$ is lower trapezoidal and
$T \in \mathbb{R}^{b \times b}$ is upper triangular.
This formation of the product of Householder matrices is called
the compact $WY$ representation \cite{schreiber1989storage}.
If the Householder transformations used to form each $H_i$ are known,
matrices $W$ and $T$ of the compact $WY$ are inexpensive to compute.
The above expression can be used to build the product $HA$:
\[
HA = A - WTW^*A.
\]
In this case, the cost is about the same,
but only matrix-matrix operations are employed.
Since on modern architectures matrix-matrix operations
are usually much more efficient that matrix-vector operations,
this approach will render higher performances.
Recall that one flop (floating-point operation) in a matrix-matrix operation
can be much faster (several times) than a flop in a matrix-vector operation.

\section{The UTV factorization.}
\label{sec:utv}

In this section, we discuss the rank-revealing UTV matrix factorization, establishing its usefulness in computational linear algebra and reviewing efficient algorithms for its computation. In Section \ref{sec:urv}, we review the classical UTV matrix decomposition, summarizing its benefits over other standard decompositions like column-pivoted QR and the SVD. In Section \ref{sec:randutv}, we summarize recent work \cite{martinsson2017randutv} that proposes a randomized blocked algorithm for computing this factorization.

\subsection{The classical UTV factorization.}
\label{sec:urv}

Let $A \in \mathbb{R}^{m \times n}$ and set $p = \min(m,n)$. A UTV decomposition of $A$ is any factorization of the form
\begin{equation}
\label{eq:defUTV}
\begin{array}{ccccccccccc}
A &=& U & T & V^{*},\\
m\times n && m\times m & m\times n & n\times n
\end{array}
\end{equation}
where $T$ is triangular and $U$ and $V$ are both orthogonal. In this paper, we take $T$ to be \textit{upper} triangular, which is typically the more convenient choice when $m \geq n$. It is often desirable to compute a \textit{rank-revealing} UTV (RRUTV) decomposition. For any $1 \le k \le p$, consider the partitioning of $T$
\begin{equation}
T \rightarrow
\FlaTwoByTwoSingleLine
  { T_{11} }{ T_{12} }
  { T_{21} }{ T_{22} },
  \label{eq:Tpart}
\end{equation}
where $T_{11}$ is $k \times k$. We say a UTV factorization is rank-revealing if
\begin{enumerate}
\item $ \inf(T_{11}) \approx \sigma_k(A)$,
\item $\| T_{12} \| \approx \sigma_{k+1}(A)$.
\end{enumerate}

The flexibility of the factors in a UTV decomposition renders certain advantages over other canonical forms like CPQR and SVD (note that each of these are specific examples of UTV factorizations). Since the right factor in CPQR is restricted to a permutation matrix, UTV has more freedom to provide better low-rank and subspace approximations. Also, since UTV does not have the SVD's restriction of diagonality on the middle factor, the UTV  is less expensive to compute and has more efficient methods for updating and downdating (see, e.g.~\cite{1992_stewart_subspace_tracking,stewart1993updating,barlow2002modification,erbay2002modified,park1995downdating}).

\subsection{The \texttt{randUTV} algorithm.}
\label{sec:randutv}

In \cite{martinsson2017randutv}, a new algorithm called \texttt{randUTV} was proposed for computing an RRUTV factorization. \texttt{randUTV} is designed to parallelize well, enhancing the RRUTVs viability as a competitor to both CPQR and the SVD for a wide class of problem types. It yields low-rank approximations comparable to the SVD at computational speeds that match, and in many cases outperform both CPQR and SVD. Unlike classical methods for building SVDs and RRUTVs, \texttt{randUTV} processes the input matrix by blocks of $b$ contiguous columns. \texttt{randUTV} shares the advantage of CPQR that the factorization is computed incrementally, and may be stopped early to incur an overall cost of $\mathcal{O}(mnk)$, where $k$ is the rank of the computed factorization.

\begin{figure}
\begin{center}
\begin{tabular}{ccc}
\includegraphics[width=.2\textwidth]{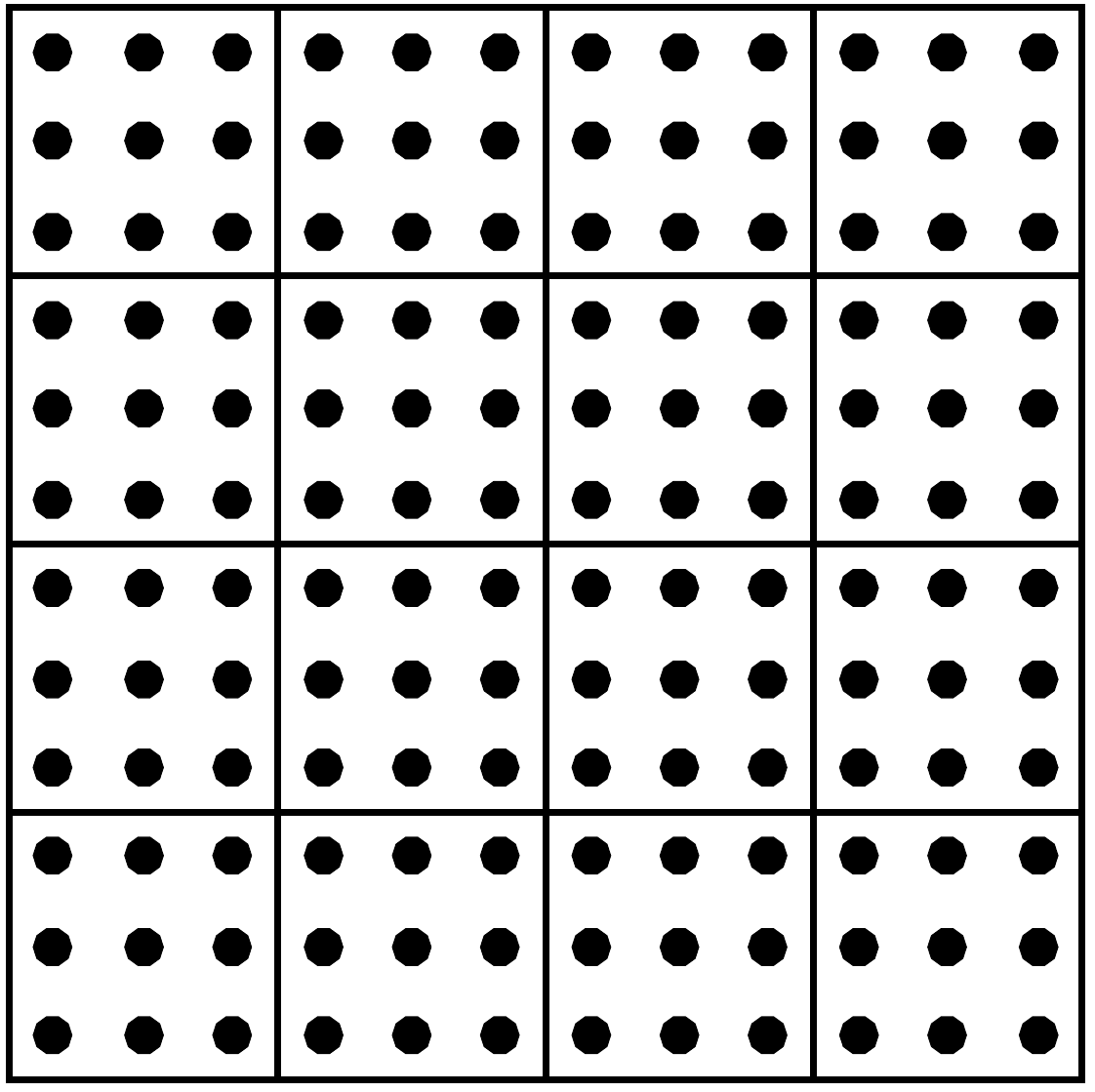}
&
\includegraphics[height=.2\textwidth]{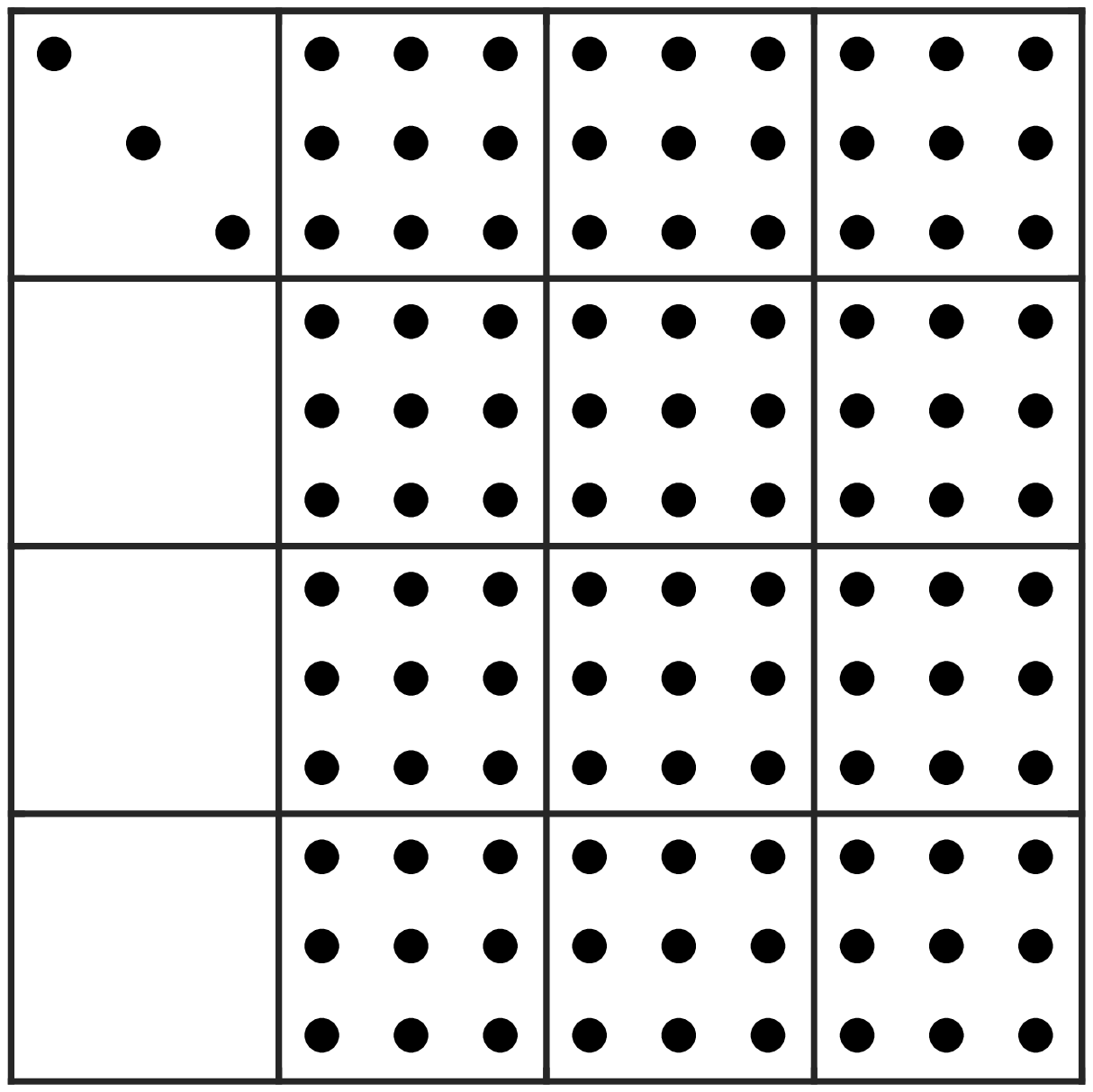}
&
\includegraphics[height=.2\textwidth]{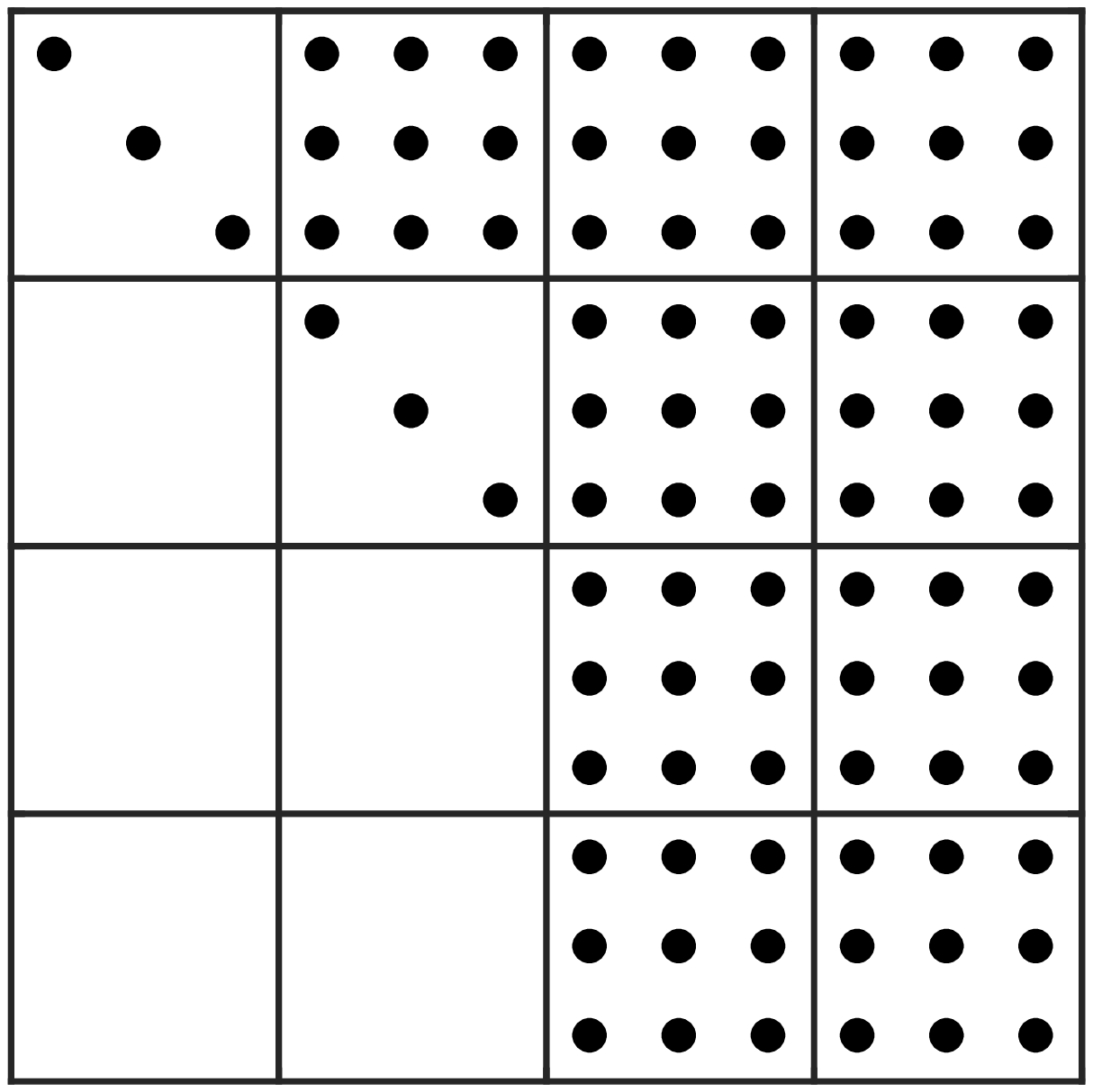} \\
after 0 steps:  & after 1 step:  & after 2 steps: \\
$T^{(0)} \defeq A$ & $T^{(1)} \defeq (U^{(0)})^* T^{(0)} V^{(0)}$ & $ T^{(2)} \defeq (U^{(1)})^* T^{(1)} V^{(1)}$
\end{tabular}
\end{center}
\caption{An illustration of the sparsity pattern followed by the first three $T^{\si}$ for \texttt{randUTV} if $n=12, b=3$. \texttt{randUTV} continues until the entirety of $T^{\si}$ is upper triangular.}
\label{fig:randutv-pattern}
\end{figure}

The driving idea behind the structure of the \texttt{randUTV} algorithm is
to build the middle factor $T$ with a right-looking approach,
that is,
in each iteration multiple columns of $T$ (a column block) are obtained
simultaneously and only the right part of $T$ is accessed.
To illustrate, consider an input matrix $A \in \mathbb{R}^{m \times n}$, and let $p = \min(m,n).$ A block size parameter $b$ with $1 \le b \le p$ must be chosen before \texttt{randUTV} begins. For simplicity, assume $b$ divides $p$ evenly. The algorithm begins by initializing $T^{(0)} \defeq A$. Then, the bulk of the work is done in a loop requiring $\frac{p}{b}$ steps. In the $i$-th step, a new matrix $T^{\sip}$ is computed with
\[
T^{\sip} \defeq (U^{\si})^* T^{\si} V^{\si},
\]
for some orthogonal matrices $U^{\si}$ and $V^{\si}$. $U^{\si}$ and $V^{\si}$ are chosen such that:
\begin{itemize}
\item the leading $ib$ columns of $T^{\sip}$ are upper triangular with $b \times b$ diagonal blocks on the main diagonal.
\item using the partitioning in Equation \ref{eq:Tpart} to define $T^{\si}_{11}$ and $T^{\si}_{22}$, we have $\inf(T^{\si}_{11}) \approx \sigma_k(A)$ and $\| T^{\si}_{22}\| \approx \sigma_{k+1}(A)$ for $1 \le k \le ib$.
\item $T^{\si}_{11}(k,k) \approx \sigma_k(A)$ for $1 \le k \le ib$.
\end{itemize}
An example of the sparsity patterns for each $T^{\si}$ is shown in Figure \ref{fig:randutv-pattern}.

Once $T^{\si}$ is upper triangular, $U$ and $V$ can then be built with
\[
V \defeq V^{(0)} V^{(1)} \cdots V^{(p/b-1)}
\]
and
\[
U \defeq U^{(0)} U^{(1)} \cdots U^{(p/b-1)}.
\]

In practice, $V^{\si}$ and $U^{\si}$ are constructed in two separate stages and applied to $T^{\si}$ at different points in the algorithm. We will henceforth refer to these matrices as $V^{\si}_{\alpha}, U^{\si}_{\alpha}$ and $V^{\si}_{\beta}, U^{\si}_{\beta}$ for the first and second stages, respectively. Also, just one $T$ matrix is stored, whose contents are overwritten with the new $T^{\sip}$ at each step.
Similarly, in case the matrices $U$ and $V$ are required to be formed,
only one matrix $U$ and one matrix $V$ would be stored.
The outline for \texttt{randUTV} is therefore the following:
\begin{enumerate}
\item Initialize $T \defeq A, \, V \defeq I, \, U \defeq I$.
\item for $i=0,1,\ldots,b/p-1$:
	\begin{enumerate}[label=\roman*.]
	\item Build $V^{\si}_\alpha$.
	\item Update $T$ and $V$: $T \leftarrow T V^{\si}_\alpha, \, V \leftarrow V V^{\si}_\alpha$.
	\item Build $U^{\si}_\alpha$.
	\item Update $T$ and $U$: $T \leftarrow (U^{\si}_\alpha)^* T, \, U \leftarrow U U^{\si}_\alpha$.
	\item Build $V^{\si}_\beta$ and $U^{\si}_\beta$ simultaneously.
	\item Update $T$, $V$, and $U$: $T \leftarrow (U^{\si}_\beta)^* T V^{\si}_\beta$, $V \leftarrow V V^{\si}_\beta$, $U \leftarrow U U^{\si}_\beta$.
	\end{enumerate}
\end{enumerate}

A matlab code for an easily readable (but inefficient) implementation of \texttt{randUTV} is given in Figure \ref{fig:randUTVmatlab}.

\subsubsection{Building $V^{\si}_{\alpha}$.}
\label{sec:buildingv}

$V^{\si}_\alpha$ is constructed to maximize the rank-revealing properties of the final factorization. Specifically, consider the partitioning at step $i$ of matrices $T$ and $V^{\si}_\alpha$
\[
T \rightarrow \FlaTwoByTwoSingleLine
  { T_{11} }{ T_{12} }
  { T_{21} }{ T_{22} }, \,
V^{\si}_\alpha \rightarrow\FlaTwoByTwoSingleLine
  { I }{ 0 }
  { 0 }{ (V^{\si}_\alpha)_{22} },
\]
where the top left block of each partition is $ib \times ib$. Then $V^{\si}_\alpha$ is constructed such that the leading $b$ columns of $(V^{\si}_\alpha)_{22}$ form an  orthonormal approximate basis for the leading $b$ right singular vectors of $T_{22}$. An efficient method for such a construction has been developed recently (see, \textit{e.g.}~\cite{rokhlin2009randomized,2011_martinsson_randomsurvey,martinsson2020randomized,2006_martinsson_random1_orig,2011_martinsson_random1}) using ideas in random matrix theory. $V^{\si}_\alpha$ is built as follows:
\begin{enumerate}
\item Draw a thin Gaussian random matrix $G^{\si} \in \mathbb{R}^{(m-ib) \times b}$.
\item Compute $Y^{\si} \defeq \left ( T_{22}^* T_{22} \right ) ^q (T_{22})^* G^{\si}$ for some small integer $q$.
\item Perform an unpivoted QR factorization on $Y^{\si}$ to obtain an orthogonal $Q^{\si}$ and upper triangular $R^{\si}$ such that $Y^{\si} = Q^{\si}R^{\si}$.
\item Set $(V^{\si}_\alpha)_{22} \defeq Q^{\si}$.
\end{enumerate}
The parameter $q$, often called the ``power iteration'' parameter, determines the accuracy of the approximate basis found in $(V^{\si}_\alpha)_{22}$. Thus, raising $q$ improves the rank-revealing properties of the resulting factorization but also increases the computational cost. For more details, see, \textit{e.g.}~\cite{2011_martinsson_randomsurvey}.

\subsubsection{Building $U^{\si}_{\alpha}$.}
\label{sec:buildingu}

$U^{\si}_\alpha$ is constructed to satisfy both the rank-revealing and upper triangular requirements of the RRUTV. First, we partition $U^{\si}_\alpha$
\[
U^{\si}_\alpha \rightarrow
\FlaTwoByTwoSingleLine
  { I }{ 0 }
  { 0 }{ (U^{\si}_\alpha)_{22} }.
\]
To obtain $(U^{\si}_\alpha)_{22}$ such that $((U^{\si}_\alpha)_{22})^* T_{22}(:,1:b)$ is upper triangular, we may compute the unpivoted QR factorization of $T_{22}(:,1:b)$ to obtain $W^{\si}, S^{\si}$ such that $T_{22}(:,1:b) = W^{\si} S^{\si}$. Next, observe that when the building of $U^{\si}_\alpha$ occurs, the range of the leading $b$ columns of $T_{22}$ is approximately the same as that of the leading $b$ left singular vectors of $T_{22}$. Therefore, the $W^{\si}$ from the unpivoted QR factorization also forms an orthonormal approximate basis for the leading $b$ left singular vectors of $T_{22}$, so $W^{\si}$ is an approximately optimal choice of matrix from a rank-revealing perspective. Thus we let $(U^{\si}_\alpha)_{22} \defeq W^{\si}$.

\subsubsection{Building $V^{\si}_{\beta}$ and $U^{\si}_{\beta}$.}
\label{sec:buildinguv2}

$V^{\si}_\beta$ and $U^{\si}_\beta$ introduce more sparsity into $T$ at low computational cost, pushing it closer to diagonality and thus decreasing $| T(k,k) - \sigma_k(A) |$ for $k=1,\ldots,(i+1)b$. They are computed simultaneously by calculating the SVD of $T_{22}(1:b,1:b)$ to obtain $U^{\si}_{SVD}, V^{\si}_{SVD}, D^{\si}_{SVD}$ such that $T_{22}(1:b,1:b) = U^{\si}_{SVD} D^{\si}_{SVD} (V^{\si}_{SVD})^*$. Then we set
\[
V^{\si}_\beta \defeq
\FlaThreeByThreeSingleLine
{I}{0}{0}
{0}{V^{\si}_{SVD}}{0}
{0}{0}{I}, \,
U^{\si}_\beta \defeq
\FlaThreeByThreeSingleLine
{I}{0}{0}
{0}{U^{\si}_{SVD}}{0}
{0}{0}{I}.
\]
Following the update step $T \rightarrow (U^{\si}_\beta)^* T V^{\si}_\beta,$ $T_{22}(1:b,1:b)$ is diagonal.

\begin{figure}
\begin{center}

\begin{tabular}{|c|c|}\hline
\begin{minipage}{0.50\textwidth}
\footnotesize
\begin{verbatim}

function [U,T,V] = randUTV(A,b,q)
  T = A;
  U = eye(size(A,1));
  V = eye(size(A,2));
  for i = 1:ceil(size(A,2)/b)
    I1 = 1:(b*(i-1));
    I2 = (b*(i-1)+1):size(A,1);
    J2 = (b*(i-1)+1):size(A,2);
    if (length(J2) > b)
      [UU,TT,VV] = stepUTV(T(I2,J2),b,q);
    else
      [UU,TT,VV] = svd(T(I2,J2));
    end
    U(:,I2)  = U(:,I2)*UU;
    V(:,J2)  = V(:,J2)*VV;
    T(I2,J2) = TT;
    T(I1,J2) = T(I1,J2)*VV;
  end
return

\end{verbatim}
\end{minipage}
&
\begin{minipage}{0.45\textwidth}
\footnotesize

\mbox{}

\begin{verbatim}

function [U,T,V] = stepUTV(A,b,q)
  G = randn(size(A,1),b);
  Y = A'*G;
  for i = 1:q
    Y = A'*(A*Y);
  end
  [V,~]    = qr(Y);
  [U,D,W]  = svd(A*V(:,1:b));
  T        = [D,U'*A*...
                V(:,(b+1):end)];
  V(:,1:b) = V(:,1:b)*W;
return

\end{verbatim}
\end{minipage}
\\[2mm] \hline
\end{tabular}
\end{center}
\caption{Matlab code for the algorithm \texttt{randUTV} that given an $m\times n$ matrix $A$
computes its UTV factorization $A = UTV^{*}$.
The input parameters $b$ and $q$ reflect the block size and the number of steps of power iteration,
respectively. This code is simplistic in that products of Householder reflectors are stored simply
as dense matrices, making the overall complexity $O(n^{4})$. (Adapted from Figure 3 of
\cite{martinsson2017randutv}.)}
\label{fig:randUTVmatlab}
\end{figure}

\section{Efficient shared memory \texttt{randUTV} implementation.}
\label{sec:shared}

%
%
Since the shared memory multicore computing architecture is ubiquitous
in modern computing,
it is therefore a prime candidate for an efficiently
designed implementation of the \texttt{randUTV} algorithm presented in
Section \ref{sec:randutv}.
Martinsson \textit{et al.}~\cite{martinsson2017randutv}
provided an efficient blocked implementation
of the \randUTV{} factorization
that was faster than competing rank-revealing factorizations,
such as SVD and CPQR.

Blocked implementations for solving linear algebra problems
are usually efficient since
they are based on matrix-matrix operations.
The ratio of flops to memory accesses
in vector-vector operations and matrix-vector operations
is usually very low: $\mathcal{O}(1)$
($\mathcal{O}(n)$ flops to $\mathcal{O}(n)$ memory accesses, and
$\mathcal{O}(n^2)$ flops to $\mathcal{O}(n^2)$ memory accesses,
respectively).
Performances are low on this type of operations
since the memory becomes a significant bottleneck with a so low ratio.
In contrast,
the ratio of flops to memory accesses
in matrix-matrix operations
is much higher: $\mathcal{O}(n)$
($\mathcal{O}(n^3)$ flops to $\mathcal{O}(n^2)$ memory accesses).
This increased ratio provides much higher performances
on modern computers since they require many flops per each memory access.

As usual in many linear algebra codes,
this blocked implementation of \randUTV{} kept all the parallelism
inside the BLAS library.
However,
the performances of this type of implementations based on a parallel BLAS
are not so efficient
as the number of cores increases in modern computers~\cite{ab-toms-2009}.

In Section \ref{sec:abb-overview}, we discuss a scheme called
algorithms-by-blocks for designing highly efficient algorithms
on architectures with multiple/many cores.
Section \ref{sec:abb-randutv} explores the application of algorithms-by-blocks
to \texttt{randUTV}.
Finally, Sections \ref{sec:flame} and \ref{sec:scheduler}
familiarize the reader with software used to implement algorithms-by-blocks
and a runtime system to schedule the various matrix operations, respectively.

\subsection{Algorithms-by-blocks: an overview.}
\label{sec:abb-overview}

\texttt{randUTV} is efficient in parallel computing environments
mainly because it can be \textit{blocked} easily.
That is,
it drives multiple columns of the input matrix $A$ to upper triangular form
in each iteration of its main loop.
The design allows most of the operations to be cast in terms of
the Level 3 BLAS (matrix-matrix operations), and
more specifically in \texttt{xgemm} operations (matrix-matrix products).
As vendor-provided and open-source multithreaded implementations
of the Level 3 BLAS are highly efficient and close to the peak speed,
\texttt{randUTV} renders high performances.
Thus, a blocked implementation of \texttt{randUTV} relying largely on
standard calls to parallel LAPACK and parallel BLAS was found to be faster
than the highly optimized MKL CPQR implementation for a shared memory system,
\textit{despite \texttt{randUTV} having a much higher flop count than
the CPQR algorithm} \cite{martinsson2017randutv}.

However, the benefits of pushing all parallelism
into multithreaded implementations of the BLAS library are limited.
Most high-performance blocked algorithms for computing factorizations
(such as Cholesky, QR, LU, etc.)
involve at least one task in each iteration
that works on very few data,
and therefore its parallelization does not render high performances.
These tasks usually involve the processing of blocks with
at least one small dimension $b$,
where $b$ is typically chosen to be 32 or 64,
usually much smaller than the matrix dimensions.
For instance,
in the blocked Cholesky factorization this performance-limited task is
the computation of the Cholesky factorization of the diagonal block,
whereas
in the blocked QR and LU factorizations this performance-limited part is
the computation of the factorization of the current column block.
Thus, since these tasks form a synchronization point,
all but one core are left idle during these computations.
For only four or five total cores, time lost is minimal.
As the number of available cores increases, though,
a significant waste in efficiency builds up.
The \texttt{randUTV} factorization is also affected by this problem,
since each iteration contains three tasks of this type:
the QR factorization of matrix $Y$,
the QR factorization of the current column block of $T$,
and the SVD of the diagonal block of $T$.

We are therefore led to seek a technique other than blocking to obtain higher performances, although we will not abandon the strategy of casting most operations in terms of the Level 3 BLAS. The key lies in changing the method with which we aggregate multiple lower level BLAS flops into a single Level 3 BLAS operation. Blocked algorithms do this by raising the granularity of the algorithm's main loop. In \texttt{randUTV}, for instance, multiple columns of the input are typically processed in one iteration of the main loop. Processing one column at a time would require matrix-vector operations (Level 2 BLAS) in each iteration, but processing multiple columns at a time aggregates these into much more efficient matrix-matrix operations (Level 3 BLAS).

The alternative approach, called algorithms-by-blocks, is to instead raise the granularity of the \textit{data}. With this method, the algorithm may be designed as if only scalar elements of the input are dealt with at one time. Then, the algorithm is transformed into Level 3 BLAS by conceiving of each scalar as a submatrix or block of size $b \times b$. Each scalar operation turns into a matrix-matrix operation, and operations in the algorithm will, at the finest level of detail, operate on usually a few (between one and four, but usually two or three) $b \times b$ blocks. Each operation on a few blocks is called a task. This arrangement allows more flexibility than blocking in ordering the operations, eliminating the bottleneck caused by the synchronization points in the blocking method.
The performance benefits obtained by the algorithm-by-blocks approach
with respect to the approach based on blocked algorithms
for linear algebra problems on shared-memory architectures
are usually significant~\cite{ab-toms-2009,chan2007supermatrix}.

An algorithm-by-blocks for computing the \texttt{randUTV}
requires that the QR factorization performed inside it
works also on $b \times b$ blocks.
In order to design this internal QR factorization process
such that each unit of work requires only $b \times b$ submatrices,
the algoritm-by-blocks for computing the QR factorization
must employ an algorithm based on updating an existing QR factorization.
We shall refer to this algorithm as \texttt{QR\_AB}.
We consider only the part of \texttt{QR\_AB}
that makes the first column of blocks upper triangular,
since that is all that is required for \texttt{randUTV AB}.
This work can be conceptualized as occurring in an iteration
with a fixed number of steps or tasks.

Figure~\ref{fig:qr_ab} shows this process for a $9 \times 9$ matrix
with block size 3.
In this figure,
the continuous lines show the $3 \times 3$ blocks involved in the current task,
`$\bullet$' represents a non-modified element by the current task,
`$\star$' represents a modified element by the current task, and
`$\cdot$' represents a nullified element by the current task.
The nullified elements are shown because, as usual,
they store information about the Householder transformations
that will be later used to apply these transformations.
The first task, called \textit{Compute\_QR},
computes the QR factorization of the leading dense block $A_{00}$.
The second task, called \textit{Apply\_left\_Qt\_of\_dense\_QR},
applies the Householder transformations obtained in
the previous task (and stored in $A_{00}$) to block $A_{01}$.
The third task performs the same operation onto $A_{02}$.
The fourth task is the annhiliation of block $A_{10}$,
which is called \textit{Compute\_td\_QR}
(where `td` means triangular-dense).
The fifth task,
called \textit{Apply\_left\_Qt\_of\_td\_QR},
apply the transformations of the previous task to blocks $A_{01}$ and $A_{11}$.
The sixth task performs the same operation onto $A_{02}$ and $A_{12}$.
Analogously,
the seventh, eighth, and ninth tasks perform the
same as tasks fourth, fifth, and sixth to the first and third row of blocks.
%
%
By taking advantage of the zeros present in the factorizations for each
iteration, a well-implemented \texttt{QR\_AB}
cost essentially no more flops than the traditional blocked unpivoted QR.
The algorithm is described in greater detail in
\cite{ab-toms-2009,quintana2012runtime,chan2007supermatrix}.

%
%

\begingroup
\setlength{\tabcolsep}{2.0pt}       
\renewcommand{\arraystretch}{1.05}  

\newcommand{\nuevovs}{\vspace*{0.5cm}}
\newcommand{\nuevominusvs}{\vspace*{-0.1cm}}

\newcommand{\mybl}{$\bullet$}
\newcommand{\mybn}{\multicolumn{1}{c}{$\bullet$}} 
\newcommand{\mysl}{$\star$}
\newcommand{\mysn}{\multicolumn{1}{c}{$\star$}}
\newcommand{\mypl}{$\cdot$}
\newcommand{\mypn}{\multicolumn{1}{c}{$\cdot$}}

\renewcommand{\thesubfigure}{\arabic{subfigure}}

\newcolumntype{C}{>{\centering\arraybackslash}p{6.0pt}}

\begin{figure}
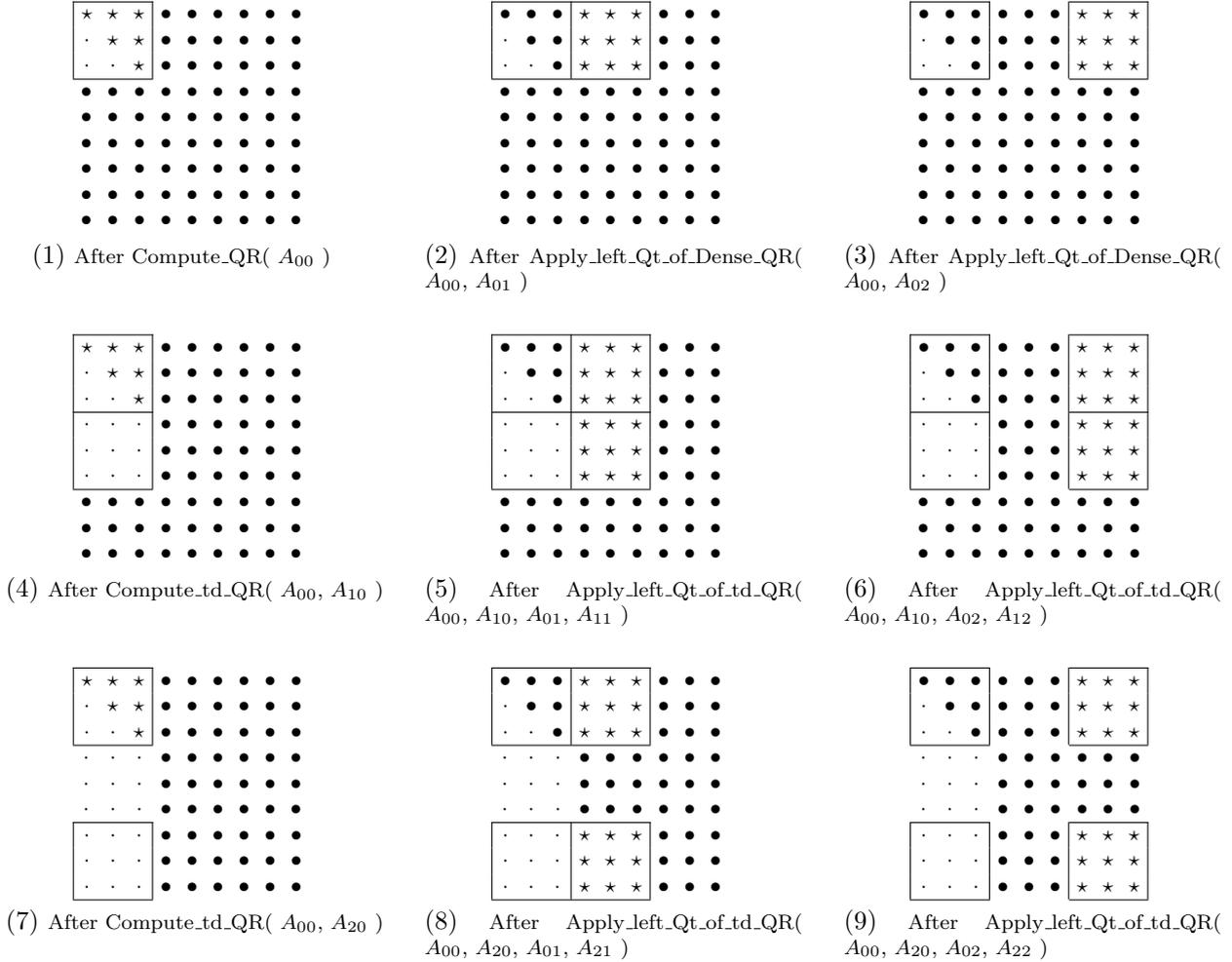


\begin{subfigure}{.30\linewidth}
\scriptsize
\centering
\begin{tabular}{|CCC|CCCCCC} 
\cline{1-3}
  \mysl & \mysl & \mysl & \mybl & \mybl & \mybl & \mybl & \mybl & \mybl \\
  \mypl & \mysl & \mysl & \mybl & \mybl & \mybl & \mybl & \mybl & \mybl \\
  \mypl & \mypl & \mysl & \mybl & \mybl & \mybl & \mybl & \mybl & \mybl \\ 
\cline{1-3}
  \mybn & \mybl & \mybn & \mybl & \mybl & \mybl & \mybl & \mybl & \mybl \\
  \mybn & \mybl & \mybn & \mybl & \mybl & \mybl & \mybl & \mybl & \mybl \\
  \mybn & \mybl & \mybn & \mybl & \mybl & \mybl & \mybl & \mybl & \mybl \\
  \mybn & \mybl & \mybn & \mybl & \mybl & \mybl & \mybl & \mybl & \mybl \\
  \mybn & \mybl & \mybn & \mybl & \mybl & \mybl & \mybl & \mybl & \mybl \\
  \mybn & \mybl & \mybn & \mybl & \mybl & \mybl & \mybl & \mybl & \mybl \\
\end{tabular}
\nuevominusvs
\caption{\scriptsize After Compute\_QR( $A_{00}$ ) \newline}
\nuevovs
\end{subfigure}
\hspace*{0.3cm}
\begin{subfigure}{.30\linewidth}
\scriptsize
\centering
\begin{tabular}{|CCC|CCC|CCC} 
\cline{1-6}
  \mybl & \mybl & \mybl & \mysl & \mysl & \mysl & \mybl & \mybl & \mybl \\
  \mypl & \mybl & \mybl & \mysl & \mysl & \mysl & \mybl & \mybl & \mybl \\
  \mypl & \mypl & \mybl & \mysl & \mysl & \mysl & \mybl & \mybl & \mybl \\ 
\cline{1-6}
  \mybn & \mybl & \mybn & \mybl & \mybl & \mybn & \mybl & \mybl & \mybl \\
  \mybn & \mybl & \mybn & \mybl & \mybl & \mybn & \mybl & \mybl & \mybl \\
  \mybn & \mybl & \mybn & \mybl & \mybl & \mybn & \mybl & \mybl & \mybl \\
  \mybn & \mybl & \mybn & \mybl & \mybl & \mybn & \mybl & \mybl & \mybl \\
  \mybn & \mybl & \mybn & \mybl & \mybl & \mybn & \mybl & \mybl & \mybl \\
  \mybn & \mybl & \mybn & \mybl & \mybl & \mybn & \mybl & \mybl & \mybl \\
\end{tabular}
\nuevominusvs
\caption{\scriptsize After Apply\_left\_Qt\_of\_Den\-se\_QR( $A_{00}$, $A_{01}$ )}
\nuevovs
\end{subfigure}
\hspace*{0.3cm}
\begin{subfigure}{.30\linewidth}
\scriptsize
\centering
\begin{tabular}{|CCC|CCC|CCC|} 
\cline{1-3} \cline{7-9}
  \mybl & \mybl & \mybl & \mybl & \mybl & \mybl & \mysl & \mysl & \mysl \\
  \mypl & \mybl & \mybl & \mybl & \mybl & \mybl & \mysl & \mysl & \mysl \\
  \mypl & \mypl & \mybl & \mybl & \mybl & \mybl & \mysl & \mysl & \mysl \\ 
\cline{1-3} \cline{7-9}
  \mybn & \mybl & \mybn & \mybl & \mybl & \mybn & \mybl & \mybl & \mybn \\
  \mybn & \mybl & \mybn & \mybl & \mybl & \mybn & \mybl & \mybl & \mybn \\
  \mybn & \mybl & \mybn & \mybl & \mybl & \mybn & \mybl & \mybl & \mybn \\
  \mybn & \mybl & \mybn & \mybl & \mybl & \mybn & \mybl & \mybl & \mybn \\
  \mybn & \mybl & \mybn & \mybl & \mybl & \mybn & \mybl & \mybl & \mybn \\
  \mybn & \mybl & \mybn & \mybl & \mybl & \mybn & \mybl & \mybl & \mybn \\
\end{tabular}
\nuevominusvs
\caption{\scriptsize After Apply\_left\_Qt\_of\_Den\-se\_QR( $A_{00}$, $A_{02}$ )}
\nuevovs
\end{subfigure}
\\
\begin{subfigure}{.30\linewidth}
\scriptsize
\centering
\begin{tabular}{|CCC|CCCCCC} 
\cline{1-3}
  \mysl & \mysl & \mysl & \mybl & \mybl & \mybl & \mybl & \mybl & \mybl \\
  \mypl & \mysl & \mysl & \mybl & \mybl & \mybl & \mybl & \mybl & \mybl \\
  \mypl & \mypl & \mysl & \mybl & \mybl & \mybl & \mybl & \mybl & \mybl \\ 
\cline{1-3}
  \mypl & \mypl & \mypl & \mybl & \mybl & \mybn & \mybl & \mybl & \mybn \\
  \mypl & \mypl & \mypl & \mybl & \mybl & \mybn & \mybl & \mybl & \mybn \\
  \mypl & \mypl & \mypl & \mybl & \mybl & \mybn & \mybl & \mybl & \mybn \\
\cline{1-3}
  \mybn & \mybl & \mybn & \mybl & \mybl & \mybn & \mybl & \mybl & \mybn \\
  \mybn & \mybl & \mybn & \mybl & \mybl & \mybn & \mybl & \mybl & \mybn \\
  \mybn & \mybl & \mybn & \mybl & \mybl & \mybn & \mybl & \mybl & \mybn \\
\end{tabular}
\nuevominusvs
\caption{\scriptsize After Compute\_td\_\-QR( $A_{00}$, $A_{10}$ ) \newline}
\nuevovs
\end{subfigure}
\hspace*{0.3cm}
\begin{subfigure}{.30\linewidth}
\scriptsize
\centering
\begin{tabular}{|CCC|CCC|CCC} 
\cline{1-6}
  \mybl & \mybl & \mybl & \mysl & \mysl & \mysl & \mybl & \mybl & \mybl \\
  \mypl & \mybl & \mybl & \mysl & \mysl & \mysl & \mybl & \mybl & \mybl \\
  \mypl & \mypl & \mybl & \mysl & \mysl & \mysl & \mybl & \mybl & \mybl \\ 
\cline{1-6}
  \mypl & \mypl & \mypl & \mysl & \mysl & \mysl & \mybl & \mybl & \mybn \\
  \mypl & \mypl & \mypl & \mysl & \mysl & \mysl & \mybl & \mybl & \mybn \\
  \mypl & \mypl & \mypl & \mysl & \mysl & \mysl & \mybl & \mybl & \mybn \\
\cline{1-6}
  \mybn & \mybl & \mybn & \mybl & \mybl & \mybn & \mybl & \mybl & \mybn \\
  \mybn & \mybl & \mybn & \mybl & \mybl & \mybn & \mybl & \mybl & \mybn \\
  \mybn & \mybl & \mybn & \mybl & \mybl & \mybn & \mybl & \mybl & \mybn \\
\end{tabular}
\nuevominusvs
\caption{\scriptsize After Apply\_left\_Qt\_of\_\-td\_QR( $A_{00}$, $A_{10}$, $A_{01}$, $A_{11}$ )}
\nuevovs
\end{subfigure}
\hspace*{0.3cm}
\begin{subfigure}{.30\linewidth}
\scriptsize
\centering
\begin{tabular}{|CCC|CCC|CCC|} 
\cline{1-3} \cline{7-9}
  \mybl & \mybl & \mybl & \mybl & \mybl & \mybl & \mysl & \mysl & \mysl \\
  \mypl & \mybl & \mybl & \mybl & \mybl & \mybl & \mysl & \mysl & \mysl \\
  \mypl & \mypl & \mybl & \mybl & \mybl & \mybl & \mysl & \mysl & \mysl \\ 
\cline{1-3} \cline{7-9}
  \mypl & \mypl & \mypl & \mybl & \mybl & \mybl & \mysl & \mysl & \mysl \\
  \mypl & \mypl & \mypl & \mybl & \mybl & \mybl & \mysl & \mysl & \mysl \\
  \mypl & \mypl & \mypl & \mybl & \mybl & \mybl & \mysl & \mysl & \mysl \\
\cline{1-3} \cline{7-9}
  \mybn & \mybl & \mybn & \mybl & \mybl & \mybn & \mybl & \mybl & \mybn \\
  \mybn & \mybl & \mybn & \mybl & \mybl & \mybn & \mybl & \mybl & \mybn \\
  \mybn & \mybl & \mybn & \mybl & \mybl & \mybn & \mybl & \mybl & \mybn \\
\end{tabular}
\nuevominusvs
\caption{\scriptsize After Apply\_left\_Qt\_of\_\-td\_QR( $A_{00}$, $A_{10}$, $A_{02}$, $A_{12}$ )}
\nuevovs
\end{subfigure}
\\
\begin{subfigure}{.30\linewidth}
\scriptsize
\centering
\begin{tabular}{|CCC|CCCCCC} 
\cline{1-3}
  \mysl & \mysl & \mysl & \mybl & \mybl & \mybl & \mybl & \mybl & \mybl \\
  \mypl & \mysl & \mysl & \mybl & \mybl & \mybl & \mybl & \mybl & \mybl \\
  \mypl & \mypl & \mysl & \mybl & \mybl & \mybl & \mybl & \mybl & \mybl \\ 
\cline{1-3}
  \mypn & \mypl & \mypn & \mybl & \mybl & \mybn & \mybl & \mybl & \mybn \\
  \mypn & \mypl & \mypn & \mybl & \mybl & \mybn & \mybl & \mybl & \mybn \\
  \mypn & \mypl & \mypn & \mybl & \mybl & \mybn & \mybl & \mybl & \mybn \\
\cline{1-3}
  \mypl & \mypl & \mypl & \mybl & \mybl & \mybn & \mybl & \mybl & \mybn \\
  \mypl & \mypl & \mypl & \mybl & \mybl & \mybn & \mybl & \mybl & \mybn \\
  \mypl & \mypl & \mypl & \mybl & \mybl & \mybn & \mybl & \mybl & \mybn \\
\cline{1-3}
\end{tabular}
\nuevominusvs
\caption{\scriptsize After Compute\_td\_\-QR( $A_{00}$, $A_{20}$ ) \newline}
\nuevovs
\end{subfigure}
\hspace*{0.3cm}
\begin{subfigure}{.30\linewidth}
\scriptsize
\centering
\begin{tabular}{|CCC|CCC|CCC} 
\cline{1-6}
  \mybl & \mybl & \mybl & \mysl & \mysl & \mysl & \mybl & \mybl & \mybl \\
  \mypl & \mybl & \mybl & \mysl & \mysl & \mysl & \mybl & \mybl & \mybl \\
  \mypl & \mypl & \mybl & \mysl & \mysl & \mysl & \mybl & \mybl & \mybl \\ 
\cline{1-6}
  \mypn & \mypl & \mypn & \mybl & \mybl & \mybn & \mybl & \mybl & \mybn \\
  \mypn & \mypl & \mypn & \mybl & \mybl & \mybn & \mybl & \mybl & \mybn \\
  \mypn & \mypl & \mypn & \mybl & \mybl & \mybn & \mybl & \mybl & \mybn \\
\cline{1-6}
  \mypl & \mypl & \mypl & \mysl & \mysl & \mysl & \mybl & \mybl & \mybn \\
  \mypl & \mypl & \mypl & \mysl & \mysl & \mysl & \mybl & \mybl & \mybn \\
  \mypl & \mypl & \mypl & \mysl & \mysl & \mysl & \mybl & \mybl & \mybn \\
\cline{1-6}
\end{tabular}
\nuevominusvs
\caption{\scriptsize After Apply\_left\_Qt\_of\_\-td\_QR( $A_{00}$, $A_{20}$, $A_{01}$, $A_{21}$ )}
\nuevovs
\end{subfigure}
\hspace*{0.3cm}
\begin{subfigure}{.30\linewidth}
\scriptsize
\centering
\begin{tabular}{|CCC|CCC|CCC|} 
\cline{1-3} \cline{7-9}
  \mybl & \mybl & \mybl & \mybl & \mybl & \mybl & \mysl & \mysl & \mysl \\
  \mypl & \mybl & \mybl & \mybl & \mybl & \mybl & \mysl & \mysl & \mysl \\
  \mypl & \mypl & \mybl & \mybl & \mybl & \mybl & \mysl & \mysl & \mysl \\ 
\cline{1-3} \cline{7-9}
  \mypn & \mypl & \mypn & \mybl & \mybl & \mybn & \mybl & \mybl & \mybn \\
  \mypn & \mypl & \mypn & \mybl & \mybl & \mybn & \mybl & \mybl & \mybn \\
  \mypn & \mypl & \mypn & \mybl & \mybl & \mybn & \mybl & \mybl & \mybn \\
\cline{1-3} \cline{7-9}
  \mypl & \mypl & \mypl & \mybl & \mybl & \mybl & \mysl & \mysl & \mysl \\
  \mypl & \mypl & \mypl & \mybl & \mybl & \mybl & \mysl & \mysl & \mysl \\
  \mypl & \mypl & \mypl & \mybl & \mybl & \mybl & \mysl & \mysl & \mysl \\
\cline{1-3} \cline{7-9}
\end{tabular}
\nuevominusvs
\caption{\scriptsize After Apply\_left\_Qt\_of\_\-td\_QR( $A_{00}$, $A_{20}$, $A_{02}$, $A_{22}$ )}
\nuevovs
\end{subfigure}

\caption{An illustration of the first tasks peformed by an algorithm-by-blocks for computing the QR factorization. 
The `$\bullet$' symbol represents a non-modified element by the current task,
`$\star$' represents a modified element by the current task, and
`$\cdot$' represents a nullified element by the current task
(they are shown because they store information 
about the Householder transformations that will be later used to apply them).
The continuous lines surround the blocks involved in the current task.}
\label{fig:qr_ab}
\end{figure}

\endgroup

\subsection{Algorithms-by-blocks for \texttt{randUTV}}
\label{sec:abb-randutv}

An algorithm-by-blocks for \texttt{randUTV}, which we will call \texttt{randUTV\_AB}, performs mostly the same operations as the original. The key difference is that the operations' new representations allow greater flexibility in the order of completion. We will discuss in some detail how this plays out in the first step of the algorithm. First,  choose a block size $b$ (in practice, $b=128$ or $256$ works well). For simplicity, assume $b$ divides both $m$ and $n$ evenly. Recall that at the beginning of \texttt{randUTV}, $T$ is initialized with $T \defeq A$. Consider a partitioning of the matrix $T$
\[
T \rightarrow
\left (
\begin{array}{c | c | c | c}
T_{11} & T_{12} & \cdots & T_{1N} \\ \hline
T_{21} & T_{22} & \cdots & T_{2N} \\ \hline
\vdots & \vdots & \ddots & \vdots \\ \hline
T_{M1} & T_{M2} & \cdots & T_{MN}
\end{array}
\right ) ,
\]
where each submatrix or block $T_{ij}$ is
$b \times b$, $N = n / b$, and $M = m / b$.
Note that the rest of matrices ($G$, $Y$, $U$, and $V$)
must also be accordingly partitioned.
The submatrices $T_{ij}$ (and those of the rest of matrices)
are treated as the fundamental unit of data in the algorithm,
so that each operation is expressed only in these terms.
For the first step of the algorithm, for instance:

\begin{enumerate}

\item
\textbf{Constructing $V^{(0)}$:}
The first step, $Y^{(0)} = (T^* T)^q T^* G^{(0)},$ is broken
into several tasks that each calculate the product of two blocks.
In the simplified case where $q=0$,
we have $M \times N$ products of two blocks.
The second step, the QR factorization of $Y^{(0)}$, uses an algorithm based on the idea of updating a QR factorization when more rows are added to the input matrix. Thus, the decomposition of each $Y_i^{(0)}$ is computed separately, and the resulting upper triangular factor $R^{(0)}$ is updated after each step. See, \textit{e.g.}~\cite{gunter2005parallel,quintana2012runtime,ab-toms-2009} for details on this approach to QR factorization.

\item
\textbf{Constructing $U^{(0)}$:}
This step requires an unpivoted QR factorization of the same size as $Y^{(0)}$, so same update-based algorithm used for $Y^{(0)}$ is used again here.

\item
\textbf{Computing SVD of $T_{11}$:}
This step is the same in \texttt{randUTV} and \texttt{randUTV\_AB}. In both cases, $T_{11}$ is interacted with as a single unit.

\item
\textbf{Updating $T$:}
The rest of \texttt{randUTV\_AB} involves the updating of $T$, \textit{i.e.}~the computations $T \leftarrow T V^{(0)}$ and $T \leftarrow (U^{(0)})^* T$.
The computations are broken down into separate stages such that the updating of each $T_{ij}$ is a different task.

\end{enumerate}

\subsection{The FLAME abstraction for implementing algorithm-by-blocks.}
\label{sec:flame}

A nontrivial obstacle to implementing an algorithm-by-blocks is the issue of programmability. Using the traditional approach of calls to a LAPACK implementation for the computational steps, keeping track of indexing quickly becomes complicated and error-prone.

The FLAME (Formal Linear Algebra Methods Environment) project \cite{gunnels2001flame,igual2012flame} is one solution to this issue. FLAME is a framework for designing linear algebraic algorithms that departs from the traditional index-based-loop methodology. Instead, the input matrix is interacted with as a collection of submatrices, basing its loops on re-partitionings of the input.

The FLAME API \cite{bientinesi2005representing} for the C language codifies these ideas, enabling a user of the API to code high performance implementations of linear algebra algorithms at a high level of abstraction. Furthermore, the methodology of the FLAME framework, and its implementation in terms of the {\tt libflame} library\cite{libflame_ref} makes it a natural fit for use with an algorithm-by-blocks. Thus, the actual code for the implementation of \texttt{randUTV\_AB} looks very similar to the written version of the algorithm given in Figure \ref{fig:FLArandutv}.

\begin{footnotesize}
\setlength{\arraycolsep}{2pt}
\resetsteps      


\renewcommand{\routinename}{ \left[ U,T,V \right] := \mbox{\sc randUTV\_AB}( A, q, n_b ) }


\renewcommand{\initialize}{
$
\begin{array}{@{}l}
 V \defeq \mbox{\sc eye}(n(A), n(A)) \\
 U \defeq \mbox{\sc eye}(m(A), m(A))
\end{array}
$
}


\renewcommand{\partitionings}{
  $
  A \rightarrow
  \FlaTwoByTwo{A_{TL}}{A_{TR}}
              {A_{BL}}{A_{BR}}
  $
,
  $
  V \rightarrow
  \FlaOneByTwo{V_{L}} {V_{R}}
  $
,
  $
  U \rightarrow
  \FlaOneByTwo{U_{L}}
              {U_{R}}
  $
}

\renewcommand{\partitionsizes}{
$ A_{TL} $ is $ 0 \times 0 $,
$ V_{L} $ has $ 0 $ columns,
$ U_{L} $ has $ 0 $ columns
}

\renewcommand{\guard}{
  m( A_{TL} ) < m( A )
}


\renewcommand{\blocksize}{b=\min(n_b,n(A_{BR}))}

\renewcommand{\repartitionings}{
$  \FlaTwoByTwo{A_{TL}}{A_{TR}}
              {A_{BL}}{A_{BR}}
  \rightarrow
  \FlaThreeByThreeBR{A_{00}}{A_{01}}{A_{02}}
                    {A_{10}}{A_{11}}{A_{12}}
                    {A_{20}}{A_{21}}{A_{22}}
$, 
$  \FlaOneByTwo{ V_L } { V_R }
\rightarrow
  \FlaOneByThreeR{V_0} {V_1} {V_2}
$
, 
$  \FlaOneByTwo{ U_L } { U_R }
\rightarrow
  \FlaOneByThreeR{U_0} {U_1} {U_2}
$
}

\renewcommand{\repartitionsizes}{
  $ A_{11} $ is $ b \times b $,
  $ V_1 $ has $ b $ rows,
  $ U_1 $ has $ b $ rows}


\renewcommand{\moveboundaries}{
$  \FlaTwoByTwo{A_{TL}}{A_{TR}}
              {A_{BL}}{A_{BR}}
  \leftarrow
  \FlaThreeByThreeTL{A_{00}}{A_{01}}{A_{02}}
                    {A_{10}}{A_{11}}{A_{12}}
                    {A_{20}}{A_{21}}{A_{22}}
$, 
$  \FlaOneByTwo{ V_L } { V_R }
\leftarrow
  \FlaOneByThreeL{V_0} {V_1} {V_2}
$
, 
$  \FlaOneByTwo{ U_L } { U_R }
\leftarrow
  \FlaOneByThreeL{U_0} {U_1} {U_2}
$
}


\renewcommand{\update}{
$
\hspace{1em} \begin{array}{lcl}
 \% \text{ Right transform } V \\
G & \defeq & \mbox{\sc generate\_iid\_stdnorm\_matrix}(m(A) - m(A_{00}), n_b) \\
Y & \defeq & \left ( \FlaTwoByTwoSingleLine {A_{11}}{A_{12}} {A_{21}}{A_{22}} ^*
       \FlaTwoByTwoSingleLine {A_{11}}{A_{12}} {A_{21}}{A_{22}} \right ) ^q
       \FlaTwoByTwoSingleLine {A_{11}}{A_{12}} {A_{21}}{A_{22}} ^*
       G \\
{[}Y,T_V] & \defeq & \mbox{\sc unpivoted\_QR}(Y) \\
\FlaTwoByTwoSingleLine {A_{11}}{A_{12}} {A_{21}}{A_{22}} & \defeq &
	\FlaTwoByTwoSingleLine {A_{11}}{A_{12}} {A_{21}}{A_{22}} - \FlaTwoByTwoSingleLine {A_{11}}{A_{12}} {A_{21}}{A_{22}} W_V T_V W_V^* \\
\FlaOneByTwoSingleLine {V_1}{V_2} & \defeq &
\FlaOneByTwoSingleLine {V_1}{V_2} - \FlaOneByTwoSingleLine {V_1}{V_2} W_V T_V W_V^* \\

\% \text{ Left transform } U \\
{[}\FlaTwoByOneSingleLine {A_{11}} {A_{21}},T_U{]} & \defeq & \mbox{\sc unpivoted\_QR}\left ( \FlaTwoByOneSingleLine {A_{11}} {A_{21}} \right ) \\
\FlaTwoByOneSingleLine {A_{12}} {A_{22}} & \defeq & \FlaTwoByOneSingleLine {A_{12}} {A_{22}} - W_U^*T_U^*W_U \FlaTwoByOneSingleLine {A_{12}} {A_{22}} \\
\FlaOneByTwoSingleLine {U_1}{U_2} & \defeq &
\FlaOneByTwoSingleLine {U_1}{U_2} - \FlaOneByTwoSingleLine {U_1}{U_2} W_U T_U W_U^* \\

\% \text{ small SVD} \\
{[}A_{11}, U_{SVD}, V_{SVD}] & \defeq & \mbox{\sc SVD}(A_{11})\\
A_{01} & \defeq & A_{01} V_{SVD} \\
A_{12} & \defeq & U_{SVD}^* A_{12} \\
V_1 & \defeq & V_1 V_{SVD} \\
U_1 & \defeq & U_1 U_{SVD}
\end{array}
$
}

\begin{figure}[tbp]
\FlaAlgorithmWithInit
\caption{The \texttt{randUTV} algorithm adapted for algorithms-by-blocks written with the FLAME methodology/notation.  In this algorithm, $
  W_V$ and $W_U$ are the unit lower trapezoidal matrices stored below the
  diagonal of $ Y $ and $\protect\FlaTwoByOneSingleLine {A_{11}} {A_{21}}$, respectively.}
\label{fig:FLArandutv}
\end{figure}

\end{footnotesize}

\subsection{Scheduling the operations for an algorithm-by-blocks.}
\label{sec:scheduler}

The runtime system called SuperMatrix \cite{chan2007supermatrix} is
an integral part of the {\tt libflame} distribution, and has been
leveraged to expose and exploit task-level parallelism in in \texttt{randUTV\_AB}.
To understand how SuperMatrix schedules and executes suboperations,
consider the problem of factorizing a matrix of $2 \times 2$ blocks
\[
A \leftarrow \FlaTwoByTwoSingleLine {A_{00}} {A_{01}} {A_{10}} {A_{11}},
\]
where each block is of size $b \times b$. We will consider the case where the power iteration parameter $q=0$ for simplicity.

Execution of the program proceeds in two phases:
the analysis stage and the execution stage.
In the first stage, instead of executing the code sequentially,
the runtime builds a list of tasks
recording the dependency information associated with each operation and
placing it in a queue.
An example of the queue built up by the runtime for \texttt{randUTV\_AB}
for the case that $A \in \mathbb{R}^{n \times n}$ and
the block size is $b=n/2$ is given in Figure \ref{fig:analyzer}.

\begin{figure}
\begin{center}
\includegraphics[width=0.35\textwidth,angle=90]{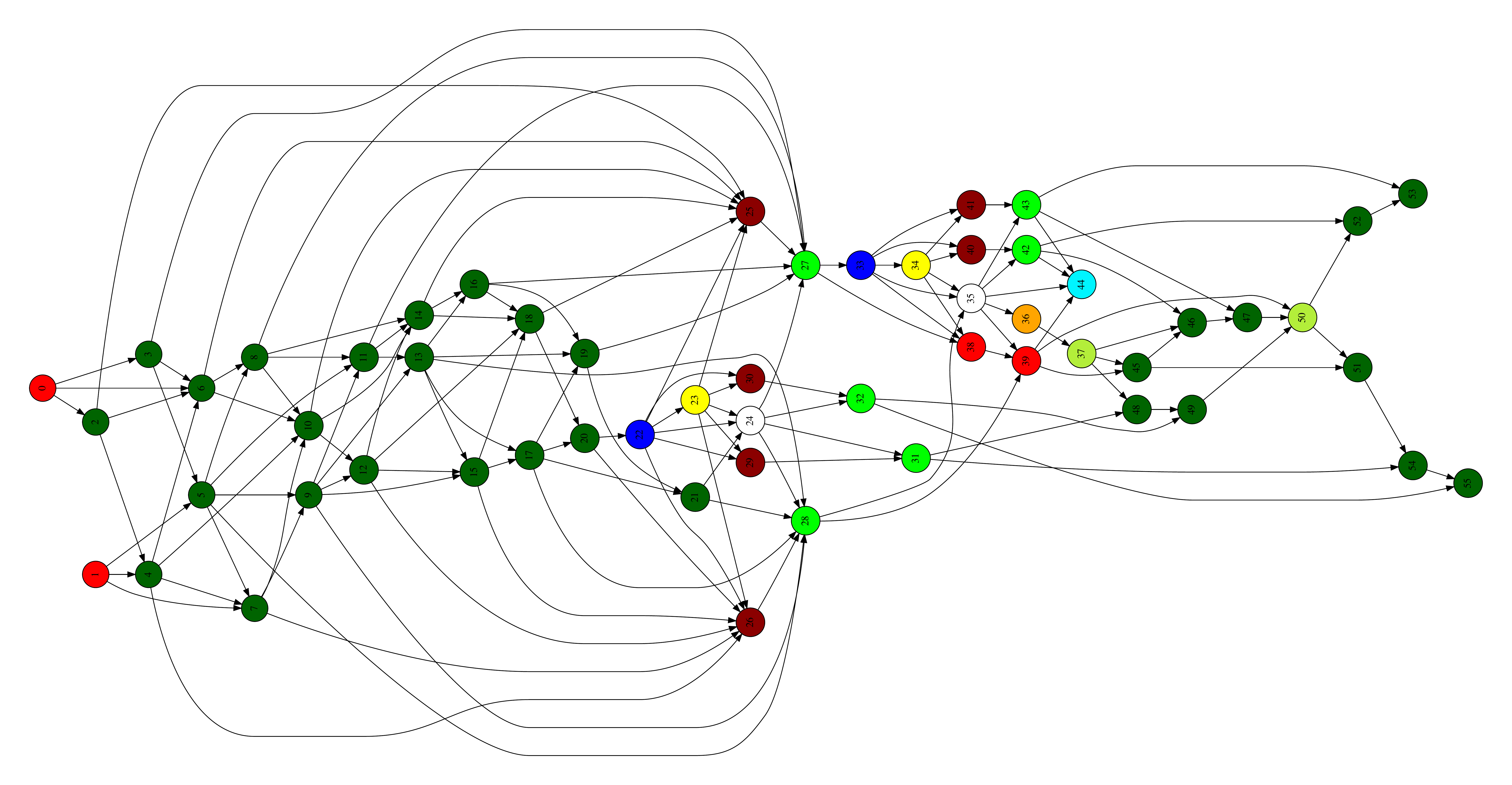}
\end{center}
\caption{Complete Directed Acyclic Graph 
exposed to the runtime task scheduler during the dispatching stage 
in the simplified case that the block size is $b = n/2$. \label{fig:DAG}}
\end{figure}

In the second stage, the scheduling/dispatching stage,
the tasks in the queue are dynamically scheduled and executed.
Each task is executed as soon as its input data becomes available and
a core is free to complete the work.
Figure \ref{fig:dispatcher} illustrates the execution of the first half
of $\texttt{randUTV\_AB}$ for a matrix $A \in \mathbb{R}^{n \times n}$
with block size $b=n/2$.
Figure~\ref{fig:DAG} shows an actual DAG that illustrate the data dependences
between tasks for the complete execution. From the code perspective, the
main FLAME formulation (see Figure~\ref{fig:FLArandutv}) remains unchanged,
replacing the actual calls to BLAS/LAPACK codes by task definition --including
input/output per-task data-- and addition to the DAG. From that point on,
the scheduling/dispatching stage is transparent for the developer.

\begin{figure}
\footnotesize
\begin{center}
\begin{tabular}{|c|l|c|c|c|}
\hline
	&
\multicolumn{1}{|c|}{Operation} &
\multicolumn{3}{c|}{Operands} \\ \cline{2-5}
	& & \multicolumn{2}{c|}{In} & In/Out \\ \hline
	\cellcolor{color8}	& Generate\_normal\_random & & & $G_0$ \\
\hline
	\cellcolor{color8} & Generate\_normal\_random & & & $G_1$ \\
\hline
	\cellcolor{color9} & Gemm\_tn\_oz: $C = A^* B$ & $A_{00}$ & $G_{0}$ & $Y_0$ \\
\hline
	\cellcolor{color9} & Gemm\_tn\_oz: $C = A^* B$ & $A_{01}$ & $G_{0}$ & $Y_1$ \\
\hline
	\cellcolor{color9} & Gemm\_tn\_oo: $C = C + A^* B$ & $A_{10}$ & $G_{1}$ & $Y_{0}$ \\
\hline
	\cellcolor{color9} & Gemm\_tn\_oo: $C = C + A^* B$ & $A_{11}$ & $G_{1}$ & $Y_{1}$ \\
\hline
	\cellcolor{color1} & Comp\_dense\_QR & & & $Y_{0}$, $S_0$ \\
\hline
	\cellcolor{color7} & Copy & $Y_{0}$ & & $E_0$ \\
\hline
	\cellcolor{color2} & Comp\_td\_QR & & & $Y_0$, $Y_1$, $S_1$ \\
\hline
	\cellcolor{color5} & Apply\_right\_Q\_of\_dense\_QR & $E_0$ & $S_0$ & $A_{00}$ \\
\hline
	\cellcolor{color5} & Apply\_right\_Q\_of\_dense\_QR & $E_0$ & $S_0$ & $A_{10}$ \\
\hline
	\cellcolor{color6} & Apply\_right\_Q\_td\_QR & $Y_1$ & $S_1$ & $A_{00}, A_{01}$ \\
\hline
	\cellcolor{color6} & Apply\_right\_Q\_td\_QR & $Y_1$ & $S_1$ & $A_{10}, A_{11}$ \\
\hline
	\cellcolor{color1} & Comp\_dense\_QR & & & $A_{00}, X_0$ \\
\hline
	\cellcolor{color7} & Copy & $A_{00}$ & & $D_0$ \\
\hline
	\cellcolor{color2} & Comp\_td\_QR & & & $A_{00}, A_{10}, X_1$ \\
\hline
	\cellcolor{color3} & Apply\_left\_Qt\_of\_dense\_QR & $D_0$ & $X_0$ & $A_{01}$  \\
\hline
	\cellcolor{color4} & Apply\_left\_Qt\_of\_td\_QR & $A_{10}$ & $X_1$ & $A_{01}, A_{11}$  \\
\hline
	\cellcolor{color13} & Keep\_upper\_triang & & & $A_{00}$ \\
\hline
	\cellcolor{color14} & Set\_to\_zero & & & $A_{10}$ \\
\hline
	\cellcolor{color15} & Svd\_of\_block & & & $A_{00}, P_0, Q_0$ \\
\hline
	\cellcolor{color9} & Gemm\_abta: $A = B^* A$ & $P_0$ & & $A_{01}$ \\
\hline
	\cellcolor{color15} & Svd\_of\_block & & & $A_{11}, P_0, Q_0$ \\
\hline
	\cellcolor{color9} & Gemm\_aabt: $A = A B^*$ & $Q_0$ & & $A_{01}$ \\
\hline
\end{tabular}
\end{center}
\caption{A list of the operations queued up by the runtime during the analyzer stage in the simplified case that the block size is $b = n/2$. The ``In'' column specifies pieces of required input data. The ``In/Out'' column specifies required pieces of data that will be altered upon completion of the operation. At run time, the operations may be completed in any order that does not violate the data dependencies encoded in the table. \label{fig:analyzer}}
\end{figure}

\begin{figure}
\begin{center}
\footnotesize
\begin{subfigure}{.6\linewidth}
\centering
\begin{tabular}{|l|c|c|c|}
\hline
\multicolumn{1}{|c|}{Operation} &
\multicolumn{3}{c|}{Operands} \\ \cline{2-4}
& \multicolumn{2}{c|}{In} & In/Out \\
\hline
Generate\_normal\_random & & & $G_0$ \checkmark \\
\hline
Generate\_normal\_random & & & $G_1$ \checkmark \\
\hline
Gemm\_tn\_oz: $C = A^* B$ & $A_{00}$ \checkmark & $G_{0}$ & $Y_0$ \checkmark \\
\hline
Gemm\_tn\_oz: $C = A^* B$ & $A_{01}$ \checkmark & $G_{0}$ & $Y_1$ \checkmark \\
\hline
Gemm\_tn\_oo: $C = C + A^* B$ & $A_{10}$ \checkmark & $G_{1}$ & $Y_{0}$ \\
\hline
Gemm\_tn\_oo: $C = C + A^* B$ & $A_{11}$ \checkmark & $G_{1}$ & $Y_{1}$ \\
\hline
Comp\_dense\_QR & & & $Y_{0}$, $S_0$ \checkmark \\
\hline
Copy & $Y_{0}$ & & $E_0$ \checkmark \\
\hline
Comp\_td\_QR & & & $Y_0$, $Y_1$, $S_1$ \checkmark \\
\hline
Apply\_right\_Q\_of\_dense\_QR & $E_0$ & $S_0$ & $A_{00}$ \checkmark \\
\hline
Apply\_right\_Q\_of\_dense\_QR & $E_0$ & $S_0$ & $A_{10}$ \checkmark \\
\hline
Apply\_right\_Q\_td\_QR & $Y_1$ & $S_1$ & $A_{00}, A_{01}$ \checkmark \\
\hline
Apply\_right\_Q\_td\_QR & $Y_1$ & $S_1$ & $A_{10}, A_{11}$ \checkmark \\
\hline
\end{tabular}
\minusvs
\caption{First half of the original table}
\vs

\end{subfigure}
\begin{subfigure}{0.3\linewidth}
\centering
\begin{tabular}{| c | c | c |}
\hline
\multicolumn{3}{| c |}{Operands} \\
\cline{1-3}
\multicolumn{2}{| c |}{In} & In/Out \\
\hline
& & \\
\hline
& & \\
\hline
$A_{00}$ \checkmark & $G_{0}$ \checkmark & $Y_0$ \checkmark \\
\hline
$A_{01}$ \checkmark & $G_{0}$ \checkmark & $Y_1$ \checkmark \\
\hline
$A_{10}$ \checkmark & $G_{1}$ \checkmark & $Y_{0}$ \\
\hline
$A_{11}$ \checkmark & $G_{1}$ \checkmark & $Y_{1}$ \\
\hline
& & $Y_{0}$, $S_0$ \checkmark \\
\hline
$Y_{0}$ & & $E_0$ \checkmark \\
\hline
& & $Y_0$, $Y_1$, $S_1$ \checkmark \\
\hline
$E_0$ & $S_0$ & $A_{00}$ \checkmark \\
\hline
$E_0$ & $S_0$ & $A_{10}$ \checkmark \\
\hline
$Y_1$ & $S_1$ & $A_{00}, A_{01}$ \checkmark \\
\hline
$Y_1$ & $S_1$ & $A_{10}, A_{11}$ \checkmark \\
\hline
\end{tabular}
\minusvs
\caption{After second operation}
\vs

\end{subfigure}
\begin{subfigure}{0.3\linewidth}
\centering
\begin{tabular}{| c | c | c |}
\hline
\multicolumn{3}{| c |}{Operands} \\
\cline{1-3}
\multicolumn{2}{| c |}{In} & In/Out \\
\hline
& & \\
\hline
& & \\
\hline
 & & \\
\hline
 & & \\
\hline
$A_{10}$ \checkmark & $G_{1}$ \checkmark & $Y_{0}$ \checkmark \\
\hline
$A_{11}$ \checkmark & $G_{1}$ \checkmark & $Y_{1}$ \checkmark \\
\hline
& & $Y_{0}$, $S_0$ \checkmark \\
\hline
$Y_{0}$ & & $E_0$ \checkmark \\
\hline
& & $Y_0$, $Y_1$, $S_1$ \checkmark \\
\hline
$E_0$ & $S_0$ & $A_{00}$ \checkmark \\
\hline
$E_0$ & $S_0$ & $A_{10}$ \checkmark \\
\hline
$Y_1$ & $S_1$ & $A_{00}, A_{01}$ \checkmark \\
\hline
$Y_1$ & $S_1$ & $A_{10}, A_{11}$ \checkmark \\
\hline
\end{tabular}
\minusvs
\caption{After fourth operation}
\vs

\end{subfigure}
\begin{subfigure}{0.3\linewidth}
\centering
\begin{tabular}{| c | c | c |}
\hline
\multicolumn{3}{| c |}{Operands} \\
\cline{1-3}
\multicolumn{2}{| c |}{In} & In/Out \\
\hline
& & \\
\hline
& & \\
\hline
 & & \\
\hline
 & & \\
\hline
 & & \\
\hline
 & & \\
\hline
& & $Y_{0}$ \checkmark , $S_0$ \checkmark \\
\hline
$Y_{0}$ & & $E_0$ \checkmark \\
\hline
& & $Y_0$, $Y_1$ \checkmark , $S_1$ \checkmark \\
\hline
$E_0$ & $S_0$ & $A_{00}$ \checkmark \\
\hline
$E_0$ & $S_0$ & $A_{10}$ \checkmark \\
\hline
$Y_1$ & $S_1$ & $A_{00}, A_{01}$ \checkmark \\
\hline
$Y_1$ & $S_1$ & $A_{10}, A_{11}$ \checkmark \\
\hline
\end{tabular}
\minusvs
\caption{After sixth operation}
\vs

\end{subfigure}
\begin{subfigure}{0.3\linewidth}
\centering
\begin{tabular}{| c | c | c |}
\hline
\multicolumn{3}{| c |}{Operands} \\
\cline{1-3}
\multicolumn{2}{| c |}{In} & In/Out \\
\hline
& & \\
\hline
& & \\
\hline
 & & \\
\hline
 & & \\
\hline
 & & \\
\hline
 & & \\
\hline
& & \\
\hline
$Y_{0}$ \checkmark & & $E_0$ \checkmark \\
\hline
& & $Y_0$, $Y_1$ \checkmark , $S_1$ \checkmark \\
\hline
$E_0$ & $S_0$ \checkmark & $A_{00}$ \checkmark \\
\hline
$E_0$ & $S_0$ \checkmark & $A_{10}$ \checkmark \\
\hline
$Y_1$ & $S_1$ & $A_{00}, A_{01}$ \checkmark \\
\hline
$Y_1$ & $S_1$ & $A_{10}, A_{11}$ \checkmark \\
\hline
\end{tabular}
\minusvs
\caption{After seventh operation}
\vs

\end{subfigure}
\begin{subfigure}{0.4\linewidth}
\centering
\begin{tabular}{| c | c | c |}
\hline
\multicolumn{3}{| c |}{Operands} \\
\cline{1-3}
\multicolumn{2}{| c |}{In} & In/Out \\
\hline
& & \\
\hline
& & \\
\hline
 & & \\
\hline
 & & \\
\hline
 & & \\
\hline
 & & \\
\hline
& & \\
\hline
& & \\
\hline
& & $Y_0$ \checkmark, $Y_1$ \checkmark , $S_1$ \checkmark \\
\hline
$E_0$ \checkmark & $S_0$ \checkmark & $A_{00}$ \checkmark \\
\hline
$E_0$ \checkmark & $S_0$ \checkmark & $A_{10}$ \checkmark \\
\hline
$Y_1$ & $S_1$ & $A_{00}, A_{01}$ \checkmark \\
\hline
$Y_1$ & $S_1$ & $A_{10}, A_{11}$ \checkmark \\
\hline
\end{tabular}
\minusvs
\caption{After eighth operation}
\vs

\end{subfigure}
\begin{subfigure}{0.4\textwidth}
\centering
\begin{tabular}{| c | c | c |}
\hline
\multicolumn{3}{| c |}{Operands} \\
\cline{1-3}
\multicolumn{2}{| c |}{In} & In/Out \\
\hline
& & \\
\hline
& & \\
\hline
 & & \\
\hline
 & & \\
\hline
 & & \\
\hline
 & & \\
\hline
& & \\
\hline
& & \\
\hline
& & \\
\hline
 & & \\
\hline
 & & \\
\hline
$Y_1$ \checkmark & $S_1$ \checkmark & $A_{00} \checkmark, A_{01}$ \checkmark \\
\hline
$Y_1$ \checkmark & $S_1$ \checkmark & $A_{10} \checkmark, A_{11}$ \checkmark \\
\hline
\end{tabular}
\minusvs
\caption{After eleventh operation}
\vs
\end{subfigure}
\end{center}

\caption{An illustration of the execution order of the first half of \texttt{randUTV\_AB} for an $n \times n$ matrix using the SuperMatrix runtime system when the block size is $n/2$. A check mark `\checkmark' indicates the value is available.
The execution order may change depending on the number of available cores
in the system.}
\label{fig:dispatcher}
\end{figure}

\section{Efficient distributed memory \texttt{randUTV} implementation.}
\label{sec:distributed}

Distributed memory computing architectures are commonly used for solving large problems as they extend both memory and processing power over single systems. In this section, we discuss an efficient implementation of \texttt{randUTV} for distributed memory. In Section \ref{sec:dist-overview}, we discuss the algorithmic overview and present the software used in the implementation. Section \ref{sec:dist-software} familiarizes the reader with ScaLAPACK's structure and software dependencies. In Section \ref{sec:dist-data} we review ScaLAPACK's data distribution scheme, and in Section \ref{sec:randutv-building-blocks} we describe how the building blocks of \texttt{randUTV} operate in the distributed memory environment.

\subsection{Implementation overview.}
\label{sec:dist-overview}

The distributed memory implementation of \texttt{randUTV} uses the
standard blocked algorithm of \cite{martinsson2017randutv}
rather than the algorithm-by-blocks
(as discussed in Section \ref{sec:abb-randutv})
since this methodology usually does not render high performances on
distributed memory machines.
Like in some other factorizations (QR, SVD, etc.),
when applying \texttt{randUTV} to a matrix with $m \gg n$,
it is best to perform an unpivoted QR factorization first and then
perform the \texttt{randUTV} factorization
on the resulting square triangular factor.
This method is usually applied in other architectures such as
shared memory.

The ScaLAPACK software library \cite{blackford1997scalapack,choi1996scalapack,choi1992scalapack} was used in the presented implementation.
This library provides much of the functionality of LAPACK
for distributed memory environments.
It hides most of the communication details from the developer
with an object-based API, where each matrix's object information is passed
to library routines.
This design choice enhances the programmability of the library,
enabling codes to be written similarly to a standard LAPACK implementation.
However, as it is implemented in Fortran-77,
its object orientation is not perfect and
the programming effort is larger.

\subsection{Software dependencies.}
\label{sec:dist-software}

ScaLAPACK (scalable LAPACK) was designed to be portable
to a variety of computing distributed memory architectures and
relies on only two external libraries
(since PBLAS is considered an internal module).
The first one is the sequential BLAS (Basic Linear Algebra Subroutines)
\cite{lawson1979basic,dongarra1988algorithm,dongarra1990algorithm},
providing specifications for the most common operations
involving vectors and matrices.
The second one is the BLACS (Basic Linear Algebra Communication Subroutines),
which, as the name suggests, is a specification for
common matrix and vector communication tasks \cite{anderson1991basic}.

\begin{figure}
\begin{center}
\includegraphics[width=.7\textwidth]{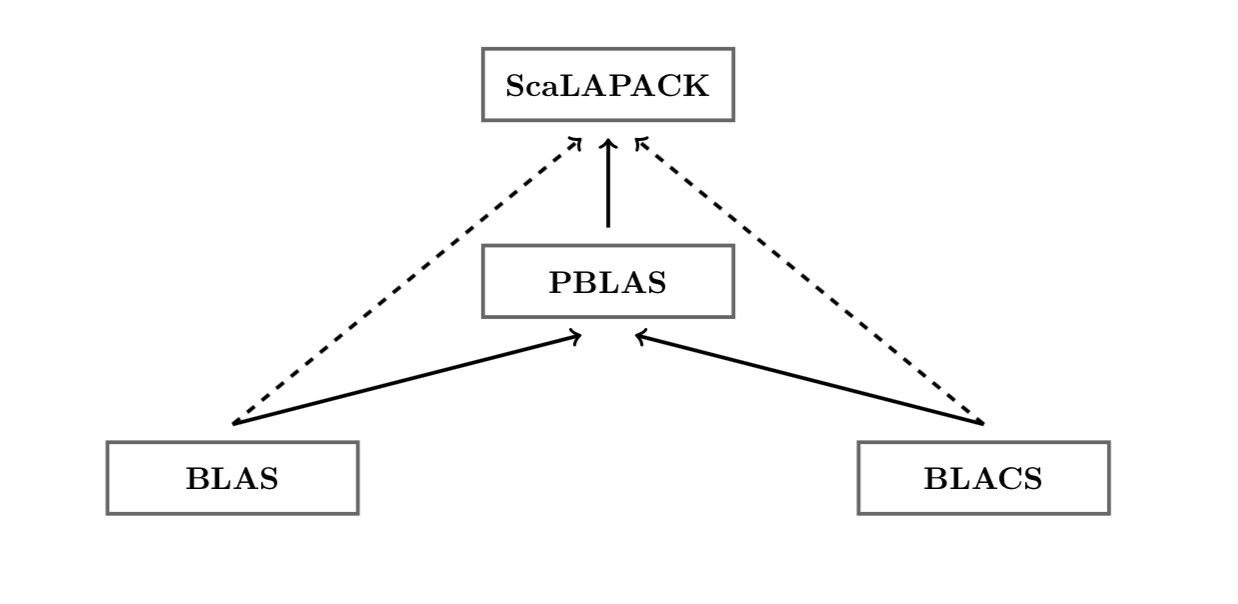}
\end{center}
\caption{The dependencies of the modules of ScaLAPACK. A solid line means the dependence occurs in the main routines (\textit{drivers}),
and a dashed line means the dependence only occurs in auxiliary routines.}
\label{fig:scalapack-deps}
\end{figure}

The PBLAS library is a key module inside ScaLAPACK.
It comprises most of BLAS routines re-written for use in
distributed memory environments.
This library is written using a combination of the sequential BLAS library
and the BLACS library.
Just as the BLAS library contains the primary building blocks
for LAPACK routines,
the PBLAS library contains the foundation for the routines in ScaLAPACK.
The diagram in Figure \ref{fig:scalapack-deps} illustrates the dependencies
of the ScaLAPACK modules.
The PBLAS library serves a dual purpose in the library.
First, because the PBLAS library mirrors the sequential BLAS in function,
the top level of code in main ScaLAPACK routines look largely the same
as the corresponding LAPACK routines.
Second, the PBLAS library adds a layer of flexibility to the code
regarding the mapping of operations.
Traditionally, one process is assigned to each core during execution,
but with a parallel BLAS implementation,
a combination of processes and threads may be used.
This adjustability gives more options
when mapping processes onto cores just before the program execution starts.

\subsection{ScaLAPACK data distribution scheme.}
\label{sec:dist-data}

The strategy for storing data in a distributed memory computation has a significant impact on the communication cost and load balance during computation. All ScaLAPACK routines assume the so-called ``block-cyclic distribution'' scheme \cite{choi1996design}. Since it involves several user-defined parameters, understanding  this method is vital to building an efficient implementation.

The block-cyclic distribution scheme involves four parameters. The first two, $m_b$ and $n_b,$ define the block size, \textit{i.e.}~the dimensions of the submatrices used as the fundamental unit for communication among processes.
Despite this flexibility, nearly all the main routines
usually employ $m_b = n_b$ for the purpose of simplicity.
The last two parameters, typically called $P$ and $Q$,
determine the shape of the logical process grid.

\begin{figure}
\begin{center}
\includegraphics[width=0.7\textwidth]{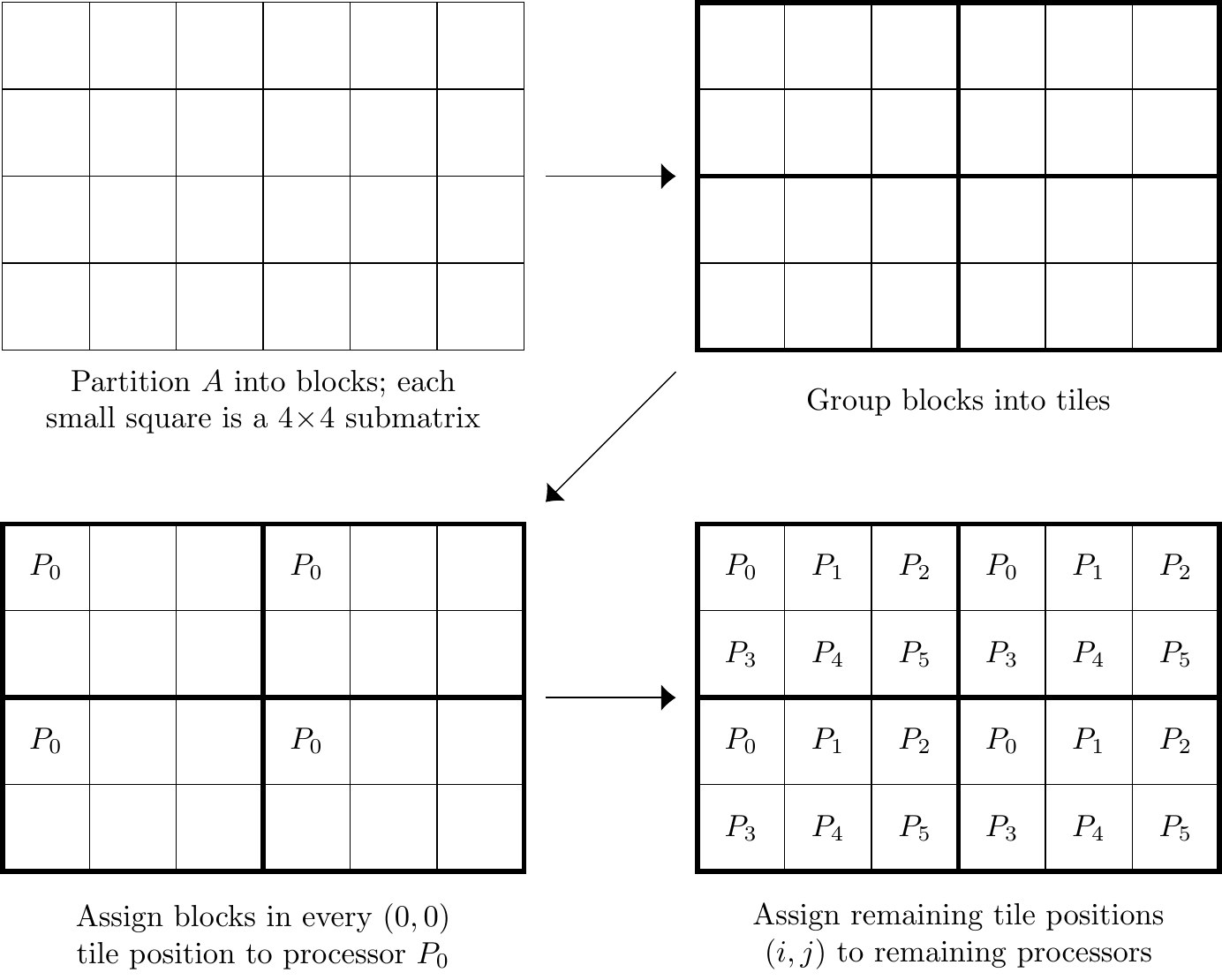}
\end{center}
\caption{A depiction of the block-cyclic data distribution method for a matrix $A$ with $m = 16, n = 24$. The parameters for this distribution are $m_b = 4, n_b = 4, P = 2, Q = 3$. Each tile is a grid of $P \times Q$ blocks. \label{fig:data-distribution}}
\end{figure}

To understand which elements of the input matrix $A$ are stored in which process, we may visualize the matrix as being partitioned into ``tiles.'' In the simple case where $m_B P$ and $n_b Q$ divide $m$ and $n$, respectively, every tile is of uniform size. Each tile is composed of $P \times Q$ blocks, each of size $m_b \times n_b$. Finally, every process is assigned a position on the tile grid. The block in that position on every tile is stored in the corresponding process. For example, the block in the $(0,0)$ spot in each tile belongs with the first process $P_0$, the block in the $(0,1)$ spot in each tile belongs to the second process $P_1$, and so on. An example is given in Figure \ref{fig:data-distribution} to demonstrate this.

\subsection{Building blocks of \texttt{randUTV}}
\label{sec:randutv-building-blocks}

In this section we examine further the \texttt{randUTV} algorithm in order to understand which portions of the computation are most expensive
(when no orthonormal matrices are built)
and how these portions perform in the distributed memory environment.
Judging by numbers of flops required,
the three portions of the computation that take the most time
are the following:

\begin{enumerate}
\item applying $V^{\si},$ stage $\alpha$ to $A$, \label{item:one}
\item applying $U^{\si},$ stage $\alpha$ to $A$, \label{item:two}
\item building $Y$. \label{item:three}
\end{enumerate}

To determine the fundamental operations involved in items \ref{item:one} and \ref{item:two}, first recall that $V^{\si}$ and $U^{\si}$ are both formed from a Householder reduction on matrices with $b$ columns to upper trapezoidal form. As such, we may express them in the so-called compact $WY$ form (see Section \ref{sec:compactwy})  as
\begin{align*}
V^{\si} & = I - W_V^{\si} T_V^{\si} (W_V^{\si})^*, \\
U^{\si} & = I - W_U^{\si} T_U^{\si} (W_U^{\si})^*,
\end{align*}
where $T_V^{\si}, T_U^{\si} \in \mathbb{R}^{b \times b}$ are upper triangular, and $W_V^{\si} \in \mathbb{R}^{n \times b}$ and $W_U^{\si} \in \mathbb{R}^{m \times b}$ are lower trapezoidal. Thus the computations $A V^{\si}$ and $(U^{\si})^*A$ each require three matrix-matrix multiplications where one dimension of the multiplication is small (recall $b \ll n$).
Note that the first computation ($A V^{\si}$)
is more expensive than the second one ($(U^{\si})^*A$)
because the first one processes all the rows of $A$ (the right part of $A$),
whereas the second one only processes some rows of $A$
(the bottom right part of $A$).

It is now evident that items \ref{item:one} and \ref{item:two} use primarily \texttt{xgemm} and \texttt{xtrmm} from the BLAS. Furthermore, item \ref{item:three} is strictly a series of \texttt{xgemm} operations, so we see that matrix-matrix multiplications form the dominant cost within \texttt{randUTV}.

\texttt{xgemm} for distributed memory, which in the PBLAS library is titled \texttt{pxgemm}, is well-suited for efficiency in this environment. In the reference implementation of PBLAS, \texttt{pxgemm} may execute one of three different algorithms for matrix multiplication:
\begin{enumerate}
\item \texttt{pxgemmAB}: The outer-product algorithm is used; matrix $C$ remains in place.
\item \texttt{pxgemmBC}: The inner-product algorithm is used; matrix $A$ remains in place.
\item \texttt{pxgemmAC}: The inner-product algorithm is used; matrix $B$ remains in place.
\end{enumerate}
\texttt{xgemm} chooses among the algorithms by estimating the communication cost for each, depending on matrix dimensions and parameters of the storage scheme. The inherent flexibility of the matrix-matrix multiply enables good \texttt{pxgemm} implementations to overlap the communication with the processing of flops. Thus, \texttt{randUTV} for distributed memory obtains better speedups when more cores are added than competing implementations of SVD and CPQR algorithms for distributed memory.

\section{Performance analysis}
\label{sec:performance_analysis}

In this section, we investigate the speed of our new implementations of
the algorithm for computing the \randUTV{} factorization,
and compare it to the speeds of highly optimized methods
for computing the SVD and the column pivoted QR (CPQR) factorization.
In all the experiments double-precision real matrices were processed.

To fairly compare the different implementations being assessed,
the flop count or the usual flop rate could not be employed
since the computation of the SVD, the CPQR, and the \randUTV{} factorizations
require a very different number of flops
(the dominant $n^3$-term in the asymptotic flop count is very different).
Absolute computational times are not shown either
since they vary greatly because of the large range of matrix dimensions
employed in the experiments.
Therefore, scaled computational times
(absolute computational times divided by $n^3$) are employed.
Hence, the lower the scaled computational times,
the better the performances are.
Since all the implementations being assessed
have asymptotic complexity $O(n^{3})$
when applied to an $n\times n$ matrix,
these graphs better reveal the computational efficiency.
Those scaled times are multiplied by a constant (usually $10^{10}$)
to make the figures in the vertical axis more readable.

Although most of the plots show scaled computational times,
a few plots show speedups.
The speedup is usually computed as the quotient of
the time obtained by the serial implementation (on one core)
and the time obtained by the parallel implementation (on many cores).
Thus, this concept communicates how many times faster
the parallel implementation is compared to the serial one.
Hence, the higher the speedups,
the better the performances of the parallel implementation are.
This measure is usually very useful in checking the scalability of an implementation.
Note that in this type of plots
every implementation compares against itself on one core.

\subsection{Computational speed on shared-memory architectures}
\label{subsec:pasm}

We employed the following two computers in the experiments
with shared-memory architectures:

\begin{itemize}

\item
\texttt{marbore}:
It featured two Intel Xeon\circledR\ CPUs E5-2695 v3 (2.30 GHz),
with 28 cores and 128 GiB of RAM in total.
In this computer the so-called {\em Turbo Boost} mode of the two CPUs
was turned off in our experiments.

Its OS was GNU/Linux (Kernel Version 2.6.32-504.el6.x86\_64).
GCC compiler (version 6.3.0 20170516) was used.
Intel(R) Math Kernel Library (MKL)
Version 2018.0.1 Product Build 20171007 for Intel(R) 64 architecture
was employed
since LAPACK routines from this library
usually deliver much higher performances
than LAPACK routines from the Netlib repository.

Unless explicitly stated otherwise,
experiments have been run in this machine
since it was not so busy.

\item
\texttt{mimir}:
It featured two Intel Xeon\circledR\ CPUs Gold 6254 (3.10 GHz),
with 36 cores and 791 GB of RAM in total.
The {\em Max Turbo Frequency} of the CPUs was 4.00 GHz.

Its OS was GNU/Linux (Kernel Version 5.0.0-32-generic).
Intel C compiler (version 19.0.5.281 20190815) was used.
Intel(R) Math Kernel Library (MKL)
Version 2019.0.5 Product Build 20190808 for Intel(R) 64 architecture
was employed
because of the same reason as above.

\end{itemize}

When using routines of MKL's LAPACK,
optimal block sizes determined by that software were employed.
In a few experiments, in addition to MKL's LAPACK routines,
we also assessed Netlib's LAPACK 3.4.0 routines.
In this case, the {\sc Netlib} term is used.
When using routines of Netlib's LAPACK, several block sizes were employed
and best results were reported.
For the purpose of a fair comparison,
these routines from Netlib were linked to the BLAS library from MKL.

All the matrices used in the experiments were randomly generated.
Similar results for randUTV were obtained on other types of matrices,
since one of the main advantages of the randTUV algorithm is that
its performances do not depend on the matrix being factorized.


Unless explicitely stated otherwise,
all the experiments employed the 28 cores in the computer.

The following implementations were assessed
in the experiments of this subsection:

\begin{itemize}

\item
{\sc MKL SVD}:
The routine called \texttt{dgesvd} from MKL's LAPACK
was used to compute the Singular Value Decomposition.

\item
{\sc Netlib SVD}:
Same as the previous one,
but the code for computing the SVD from Netlib's LAPACK
was employed, instead of MKL's.

\item
{\sc MKL SDD}:
The routine called \texttt{dgesdd} from MKL's LAPACK
was used to compute the Singular Value Decomposition.
Unlike the previous SVD, this one uses the divide-and-conquer approach.
This code is usually faster,
but it requires a much larger auxiliary workspace
when the orthonormal matrices are built
(about four additional matrices of the same dimension as
the matrix being factorized).

\item
{\sc Netlib SDD}:
Same as the previous one,
but the code for computing the SVD with the divide-and-conquer approach
from Netlib's LAPACK was employed, instead of MKL's.

\item
{\sc MKL CPQR}:
The routine called \texttt{dgeqp3} from MKL's LAPACK
was used to compute the column-pivoting QR factorization.

\item
{\sc randUTV PBLAS}
(\randUTV{} with parallel BLAS):
This is the traditional implementation
for computing the \randUTV{} factorization
that relies on the parallel BLAS
to take advantage of all the cores in the system.
The parallel BLAS library from MKL was employed with these codes
for the purpose of a fair comparison.
Our implementations were coded
with {\tt libflame}~\cite{CiSE09,libflame_ref} (Release 11104).

\item
{\sc randUTV AB}
(\randUTV{} with Algorithm-by-Blocks):
This is the new implementation for computing the \randUTV{} factorization
by scheduling all the tasks to be computed in parallel,
and then executing them with serial BLAS.
The serial BLAS library from MKL was employed with these new codes
for the purpose of a fair comparison.
Our implementations were coded
with {\tt libflame}~\cite{CiSE09,libflame_ref} (Release 11104).

\item
{\sc MKL QR}:
The routine called \texttt{dgeqrf} from MKL's LAPACK
was used to compute the QR factorization.
Although this routine does not reveal the rank,
it was included in some experiments as a performance reference for the others.

\end{itemize}

For every experiment, two plots are shown.
The left plot shows the performances
when no orthonormal matrices are computed.
In this case, just the singular values are computed for the SVD,
just the upper triangular factor $R$ is computed for the CPQR and the QR,
and just the upper triangular factor $T$ is computed for the \randUTV{}.
In contrast, the right plot shows the performances
when all orthonormal matrices are explicitly formed
in addition to the singular values (SVD),
the upper triangular matrix $R$ (CPQR),
or the upper triangular matrix $T$ (\randUTV{}).
In this case,
matrices $U$ and $V$ are computed for the SVD and the \randUTV{},
and matrix $Q$ is computed for the CPQR and the QR.
The right plot slightly favors CPQR and QR
since only one orthonormal matrix is formed.

\begin{figure}[ht!]
\tfvspace
\begin{center}
\begin{tabular}{cc}
\includegraphics[width=0.45\textwidth]{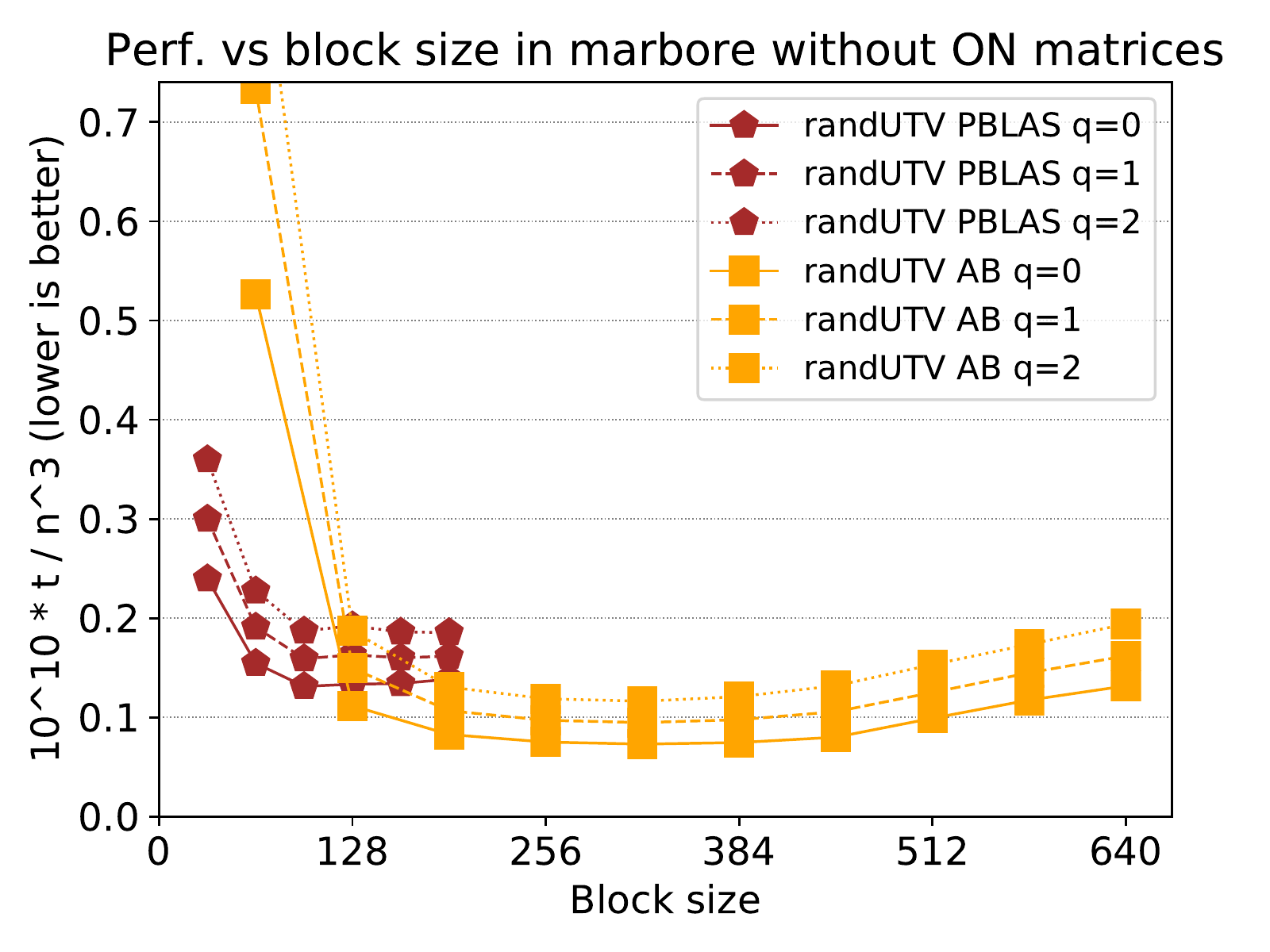} &
\includegraphics[width=0.45\textwidth]{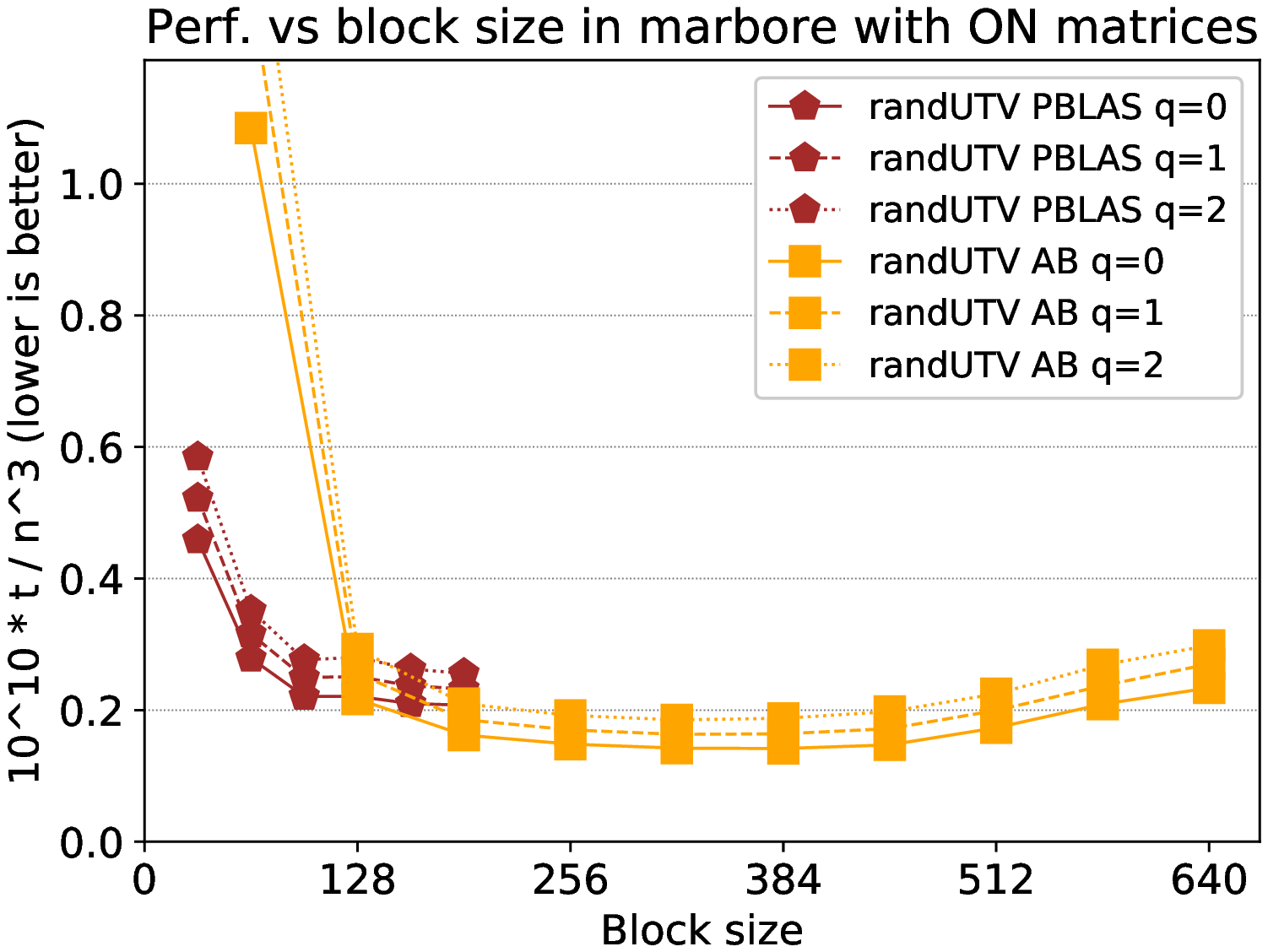} \\
\end{tabular}
\end{center}
\bfvspace
\caption{Performances of \randUTV{} implementations versus block size
on matrices of dimension $14000 \times 14000$.}
\label{fig:sm_block_size}
\end{figure}

Figure~\ref{fig:sm_block_size}
shows the scaled computational times
obtained by both implementations for computing the \randUTV{} factorization
({\sc randUTV PBLAS} and {\sc randUTV AB})
on several block sizes
when processing matrices of dimension $14000 \times 14000$.
The aim of these two plots is to determine the optimal block sizes.
The other factorizations (SVD and CPQR) are not shown
since in those cases we used the optimal block sizes
determined by Intel's software.
Optimal block sizes were around 128 for {\sc randUTV PBLAS};
on the other hand,
optimal block sizes were around 384 for {\sc randUTV AB}.

\begin{figure}[ht!]
\tfvspace
\begin{center}
\begin{tabular}{cc}
\includegraphics[width=0.45\textwidth]{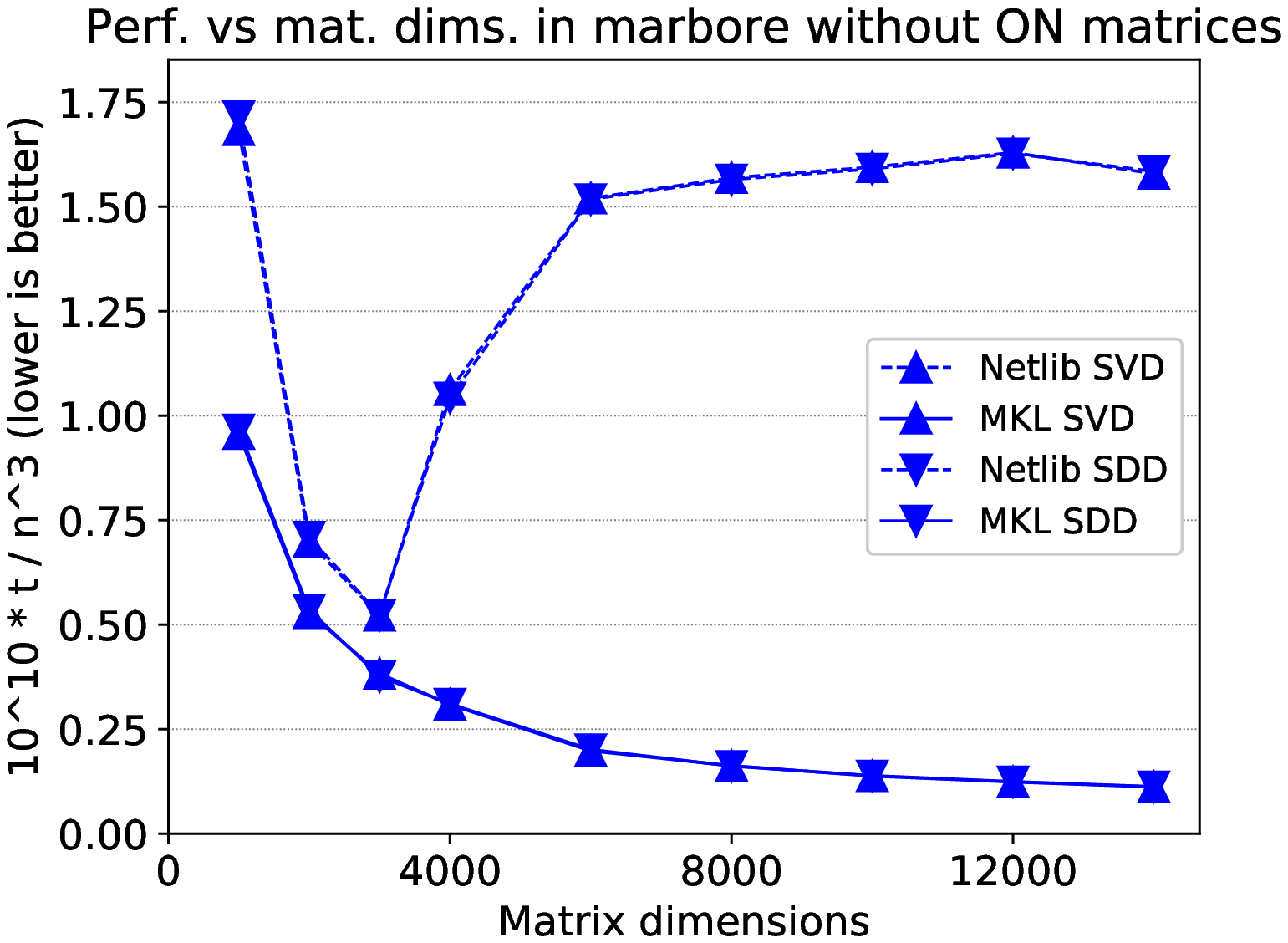} &
\includegraphics[width=0.45\textwidth]{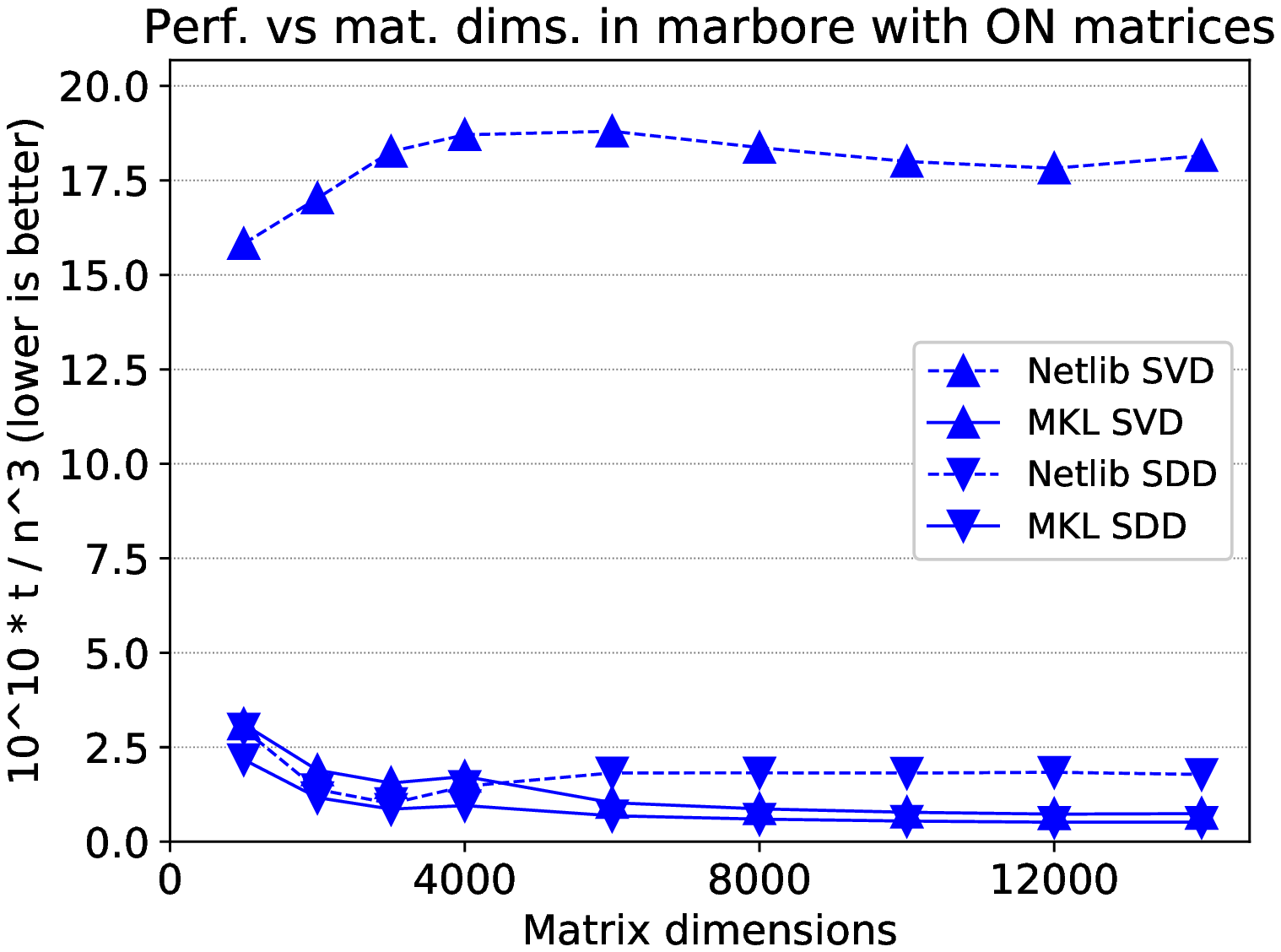} \\
\end{tabular}
\end{center}
\bfvspace
\caption{Performances versus matrix dimensions
for SVD implementations for both Netlib and MKL libraries.}
\label{fig:sm_svd}
\end{figure}

Figure~\ref{fig:sm_svd}
compares the performances of four implementations
for computing the SVD factorization:
{\sc MKL SVD} (usual SVD from the MKL library),
{\sc MKL SDD} (divide-and-conquer SVD from the MKL library),
{\sc Netlib SVD} (usual SVD from the Netlib library), and
{\sc Netlib SDD} (divide-and-conquer SVD from the Netlib library).
Performances are shown with respect to matrix dimensions.
Block sizes similar to those in the previous figure were used
for Netlib's routines and the best results were reported.
When no orthonormal matrices are computed,
both the traditional SVD and the divide-and-conquer SVD
render similar performances for this matrix type.
In this case, MKL routines are up to 14.1 times as fast as Netlib's routines.
When orthonormal matrices are computed,
the traditional SVD is much slower than the divide-and-conquer SVD.
In this case, the {\sc MKL SVD} routine is
up to 24.4 times as fast as the {\sc Netlib SVD},
and the {\sc MKL SDD} routine is
up to 3.4 times as fast as the {\sc Netlib SDD}.
As can be seen,
MKL's codes for computing the SVD are
up to more than one order of magnitude faster than Netlib's codes,
thus showing the great performances achieved by Intel.
This is a remarkable achievement for so complex codes.
Outperforming these highly optimized codes can be really a difficult task.

\begin{figure}[ht!]
\tfvspace
\begin{center}
\begin{tabular}{cc}
\includegraphics[width=0.45\textwidth]{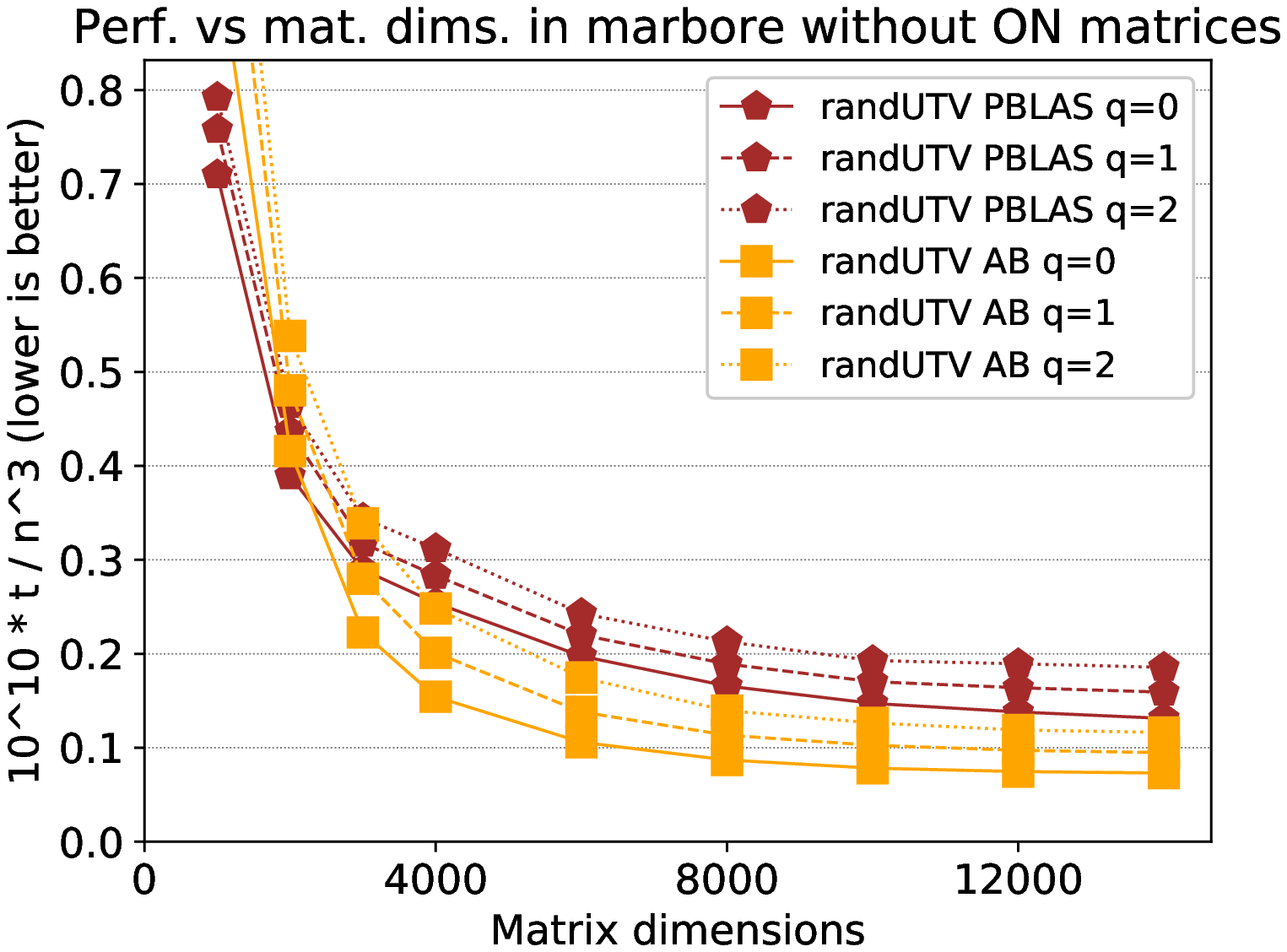} &
\includegraphics[width=0.45\textwidth]{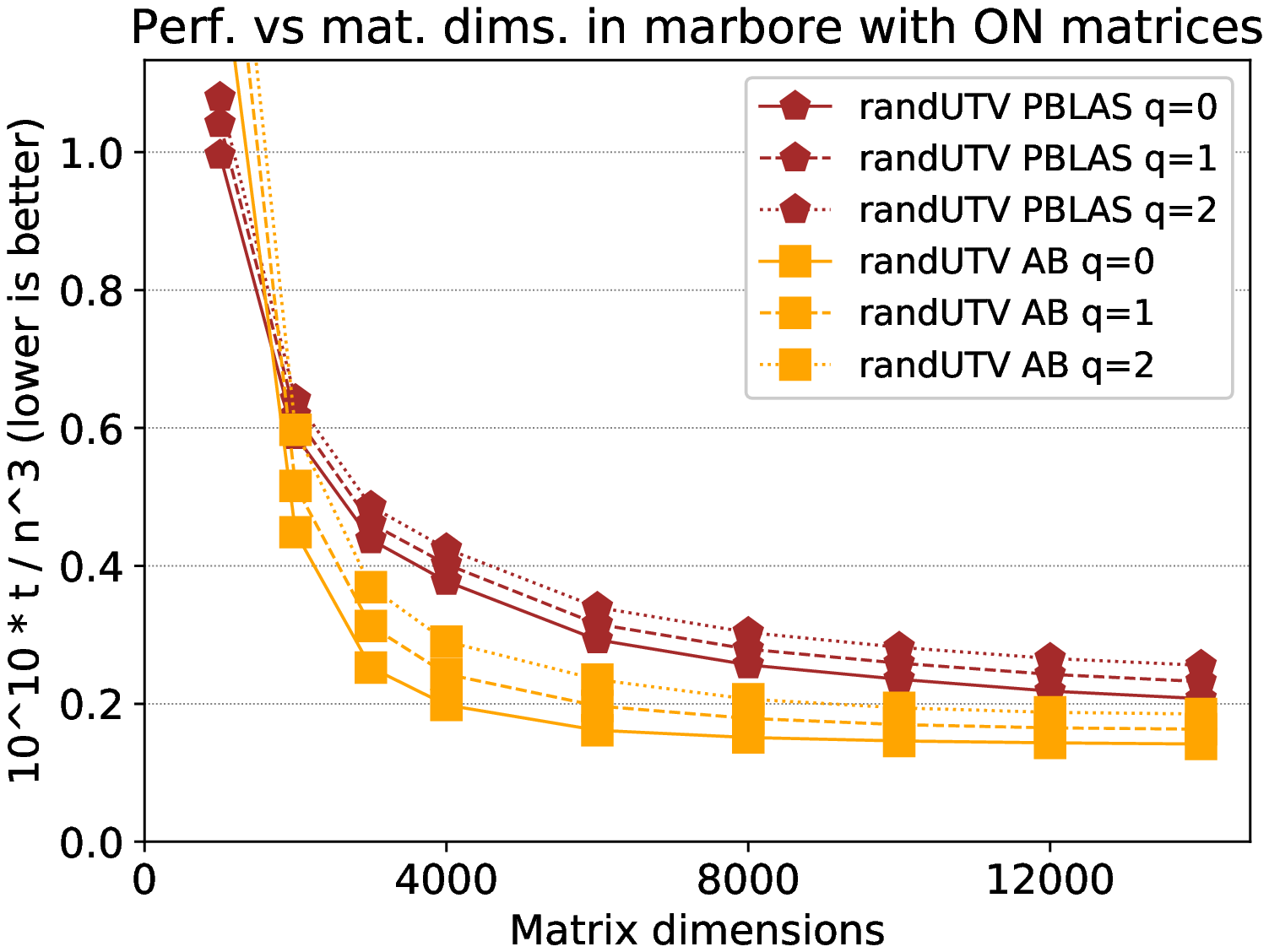} \\
\end{tabular}
\end{center}
\bfvspace
\caption{Performances versus matrix dimensions
for \randUTV{} implementations.}
\label{fig:sm_randutv}
\end{figure}

Figure~\ref{fig:sm_randutv}
compares the performances of both implementations of \randUTV{}
({\sc randUTV PBLAS} and {\sc randUTV AB})
as a function of matrix dimensions.
In both implementations, several block sizes were tested (see above),
and best results were reported.
When no orthonormal matrices are built,
{\sc randUTV AB} is between 1.80 ($q=0$) and 2.54 ($q=2$) times
as fast as {\sc randUTV PBLAS} for the largest matrix size.
When orthonormal matrices are built,
{\sc randUTV AB} is between 1.73 ($q=0$) and 1.80 ($q=2$) times
as fast as {\sc randUTV PBLAS} for the largest matrix size.

\begin{figure}[ht!]
\tfvspace
\begin{center}
\begin{tabular}{cc}
\includegraphics[width=0.45\textwidth]{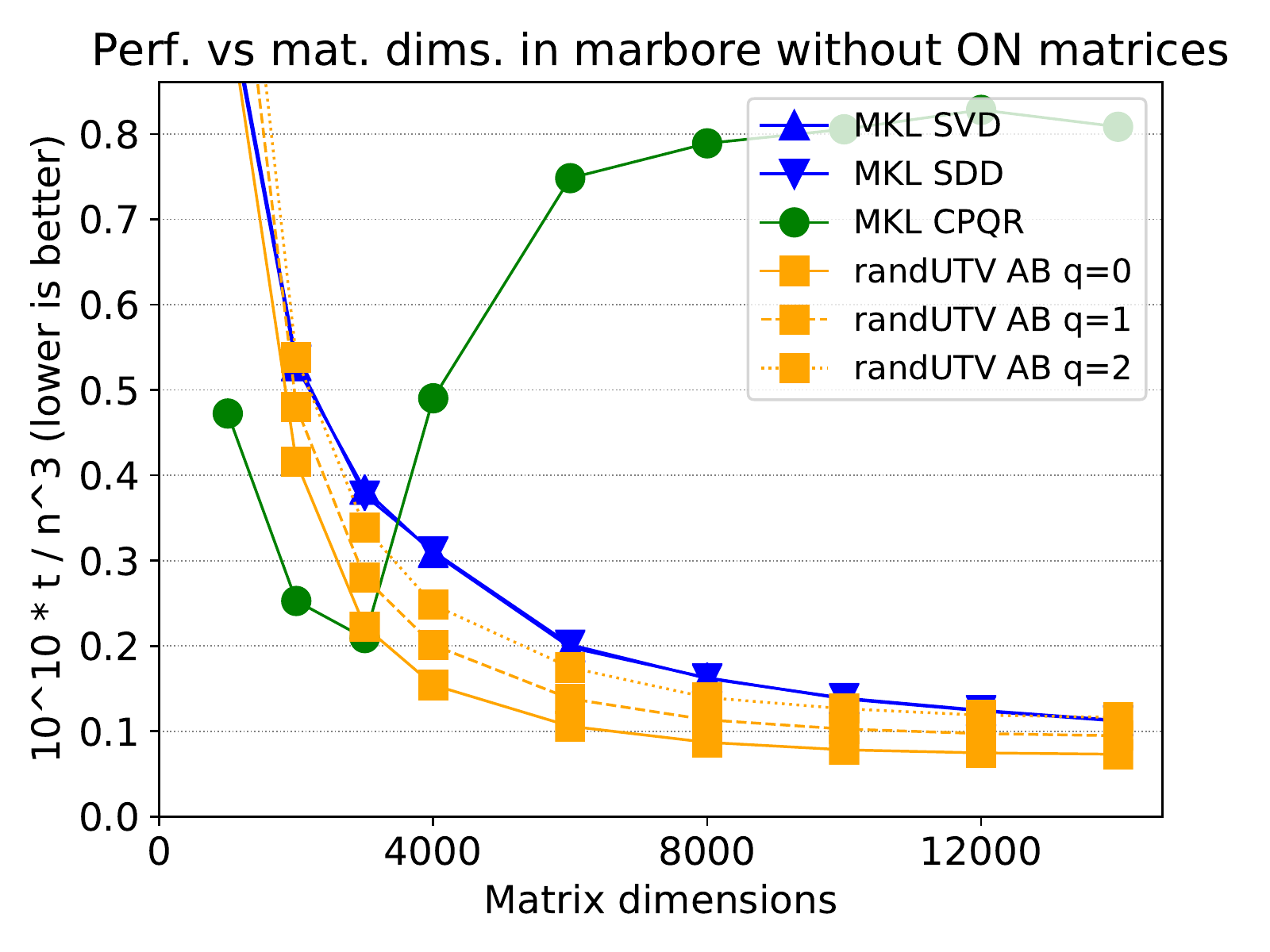} &
\includegraphics[width=0.45\textwidth]{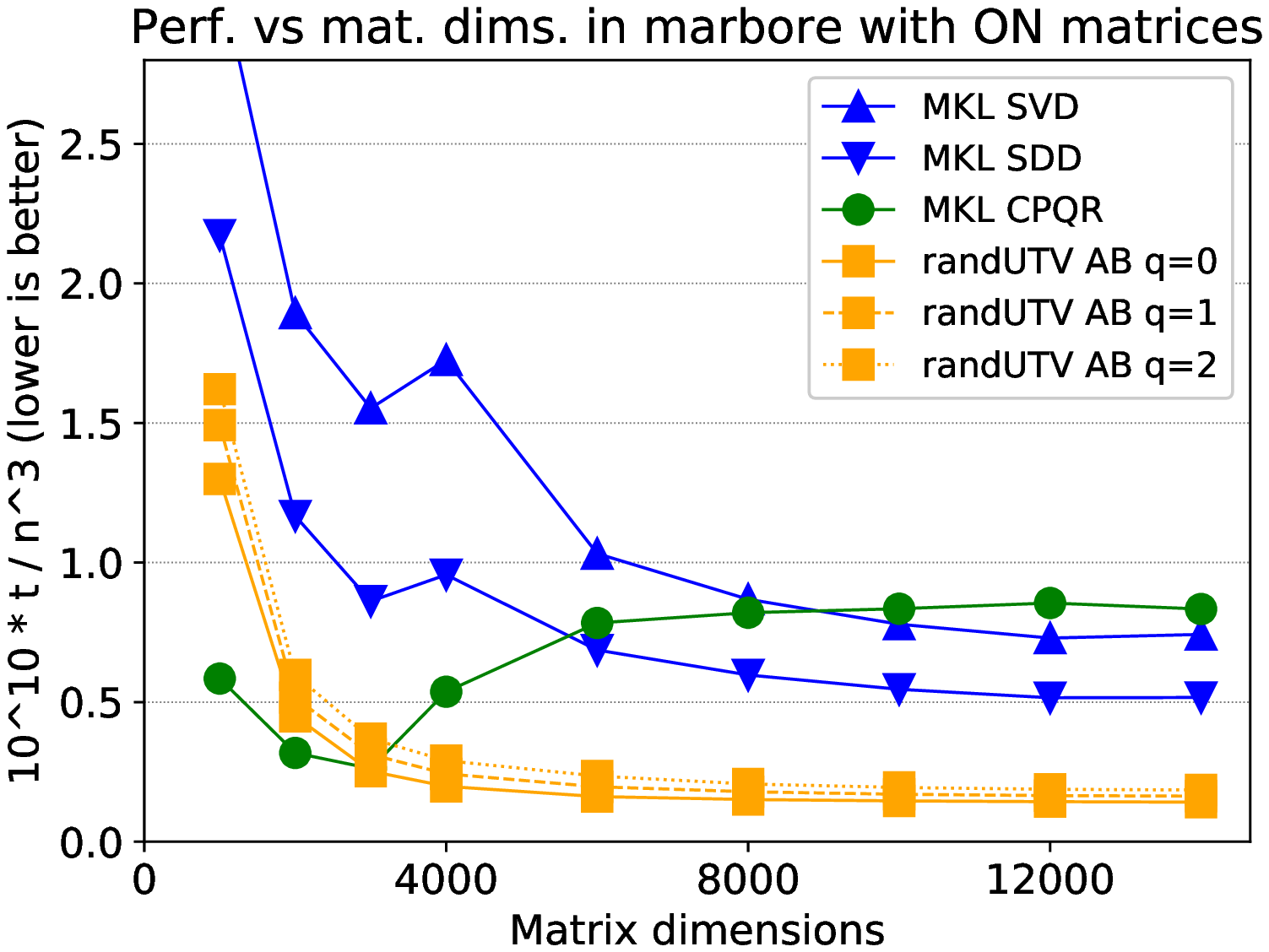} \\
\includegraphics[width=0.45\textwidth]{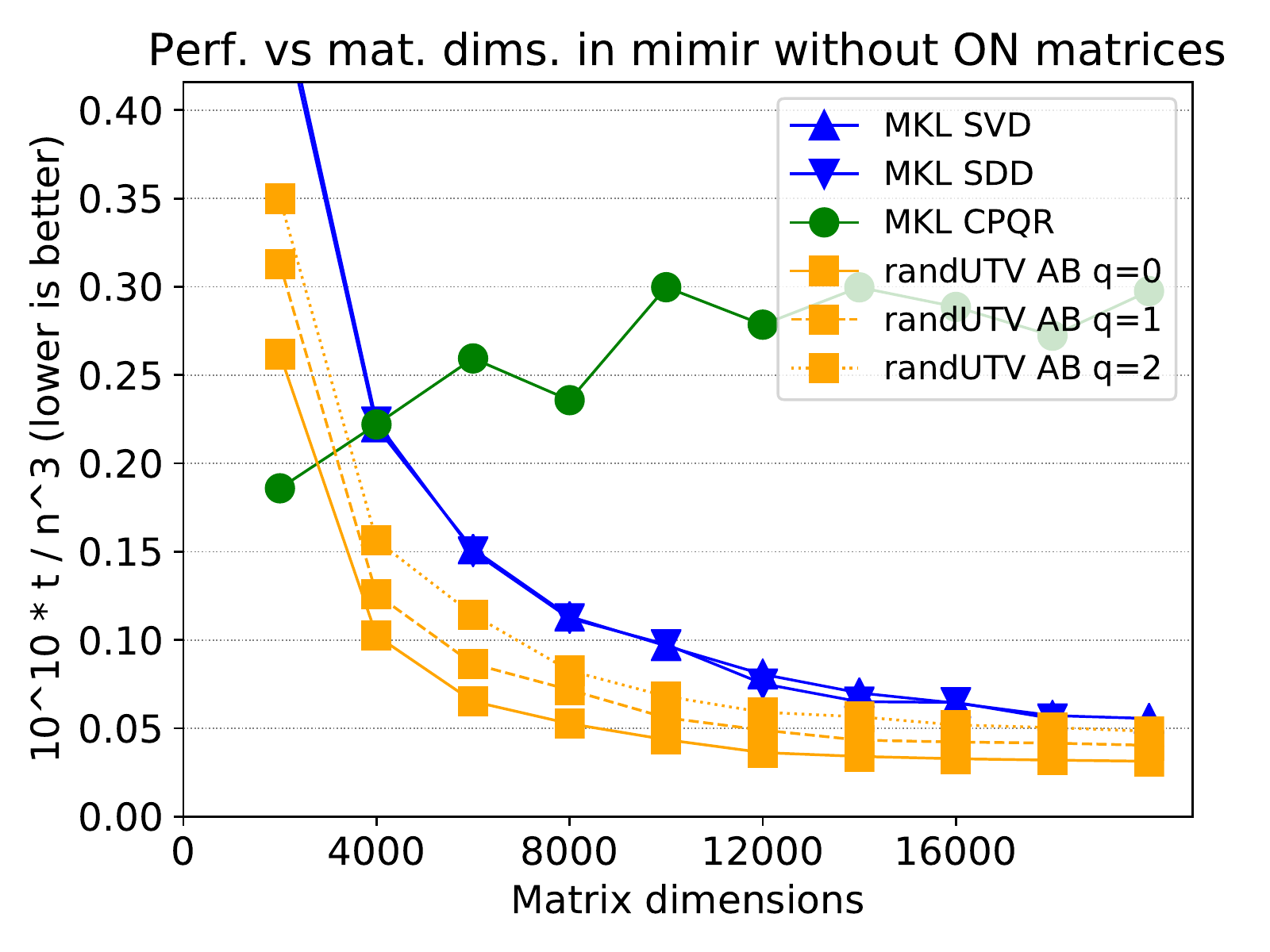} &
\includegraphics[width=0.45\textwidth]{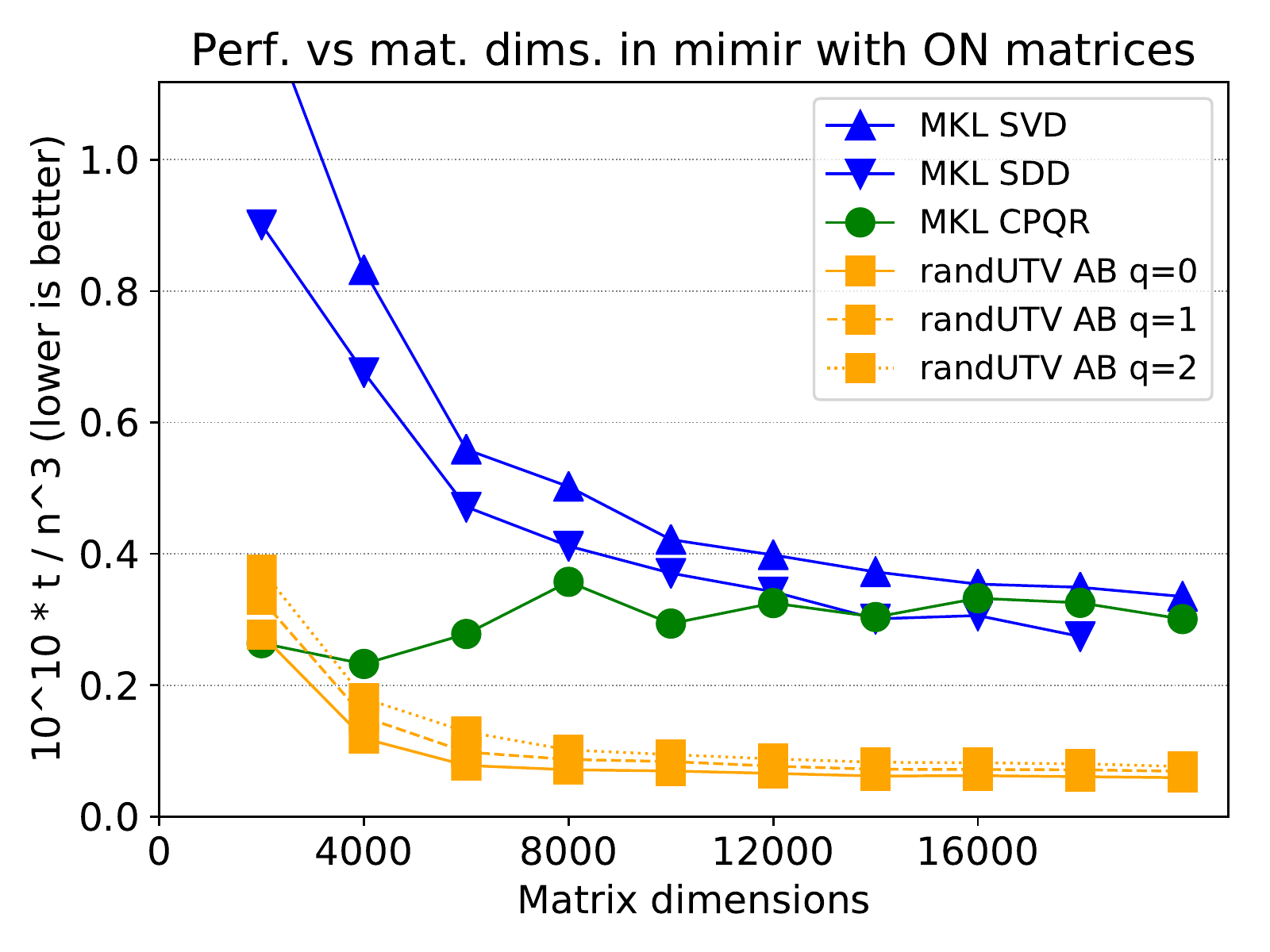} \\
\end{tabular}
\end{center}
\bfvspace
\caption{Performances versus matrix dimensions for the best implementations.
The top row shows results for \texttt{marbore} with 28 cores;
the bottom row shows results for \texttt{mimir} with 36 cores.}
\label{fig:sm_matrix_dimensions}
\end{figure}

Figure~\ref{fig:sm_matrix_dimensions}
shows the performances of the best implementations
as a function of the matrix dimensions.
The top row shows results for \texttt{marbore} with 28 cores,
whereas the bottom row shows results for \texttt{mimir} with 36 cores.
When no orthonormal matrices are built on \texttt{marbore},
{\sc randUTV AB} is between 1.54 ($q=0$) and 0.97 ($q=2$) times
as fast as {\sc MKL SVD} for the largest matrix size.
When no orthonormal matrices are built on \texttt{mimir},
{\sc randUTV AB} is between 1.77 ($q=0$) and 1.15 ($q=2$) times
as fast as {\sc MKL SVD} for the largest matrix size.
When orthonormal matrices are built on \texttt{marbore},
{\sc randUTV AB} is between 3.65 ($q=0$) and 2.79 ($q=2$) times
as fast as {\sc MKL SVD} for the largest matrix size.
When orthonormal matrices are built on \texttt{mimir},
{\sc randUTV AB} is between 5.65 ($q=0$) and 4.39 ($q=2$) times
as fast as {\sc MKL SVD} for the largest matrix size.
The previous comparisons have been done against {\sc MKL SVD}
since {\sc MKL SDD} could not be executed on $20,000 \times 20,000$
because of its larger memory requirements for workspace.
Recall that this driver requires a much larger auxiliary workspace
than {\sc randUTV AB} (about four times as large as the original matrix).
The speeds of {\sc randUTV AB}, {\sc MKL SVD} and {\sc MKL SDD}
are so remarkable that they are similar or even much faster
than {\sc MKL CPQR}, a factorization that requires much fewer flops.

\begin{figure}[ht!]
\tfvspace
\begin{center}
\begin{tabular}{cc}
\includegraphics[width=0.45\textwidth]{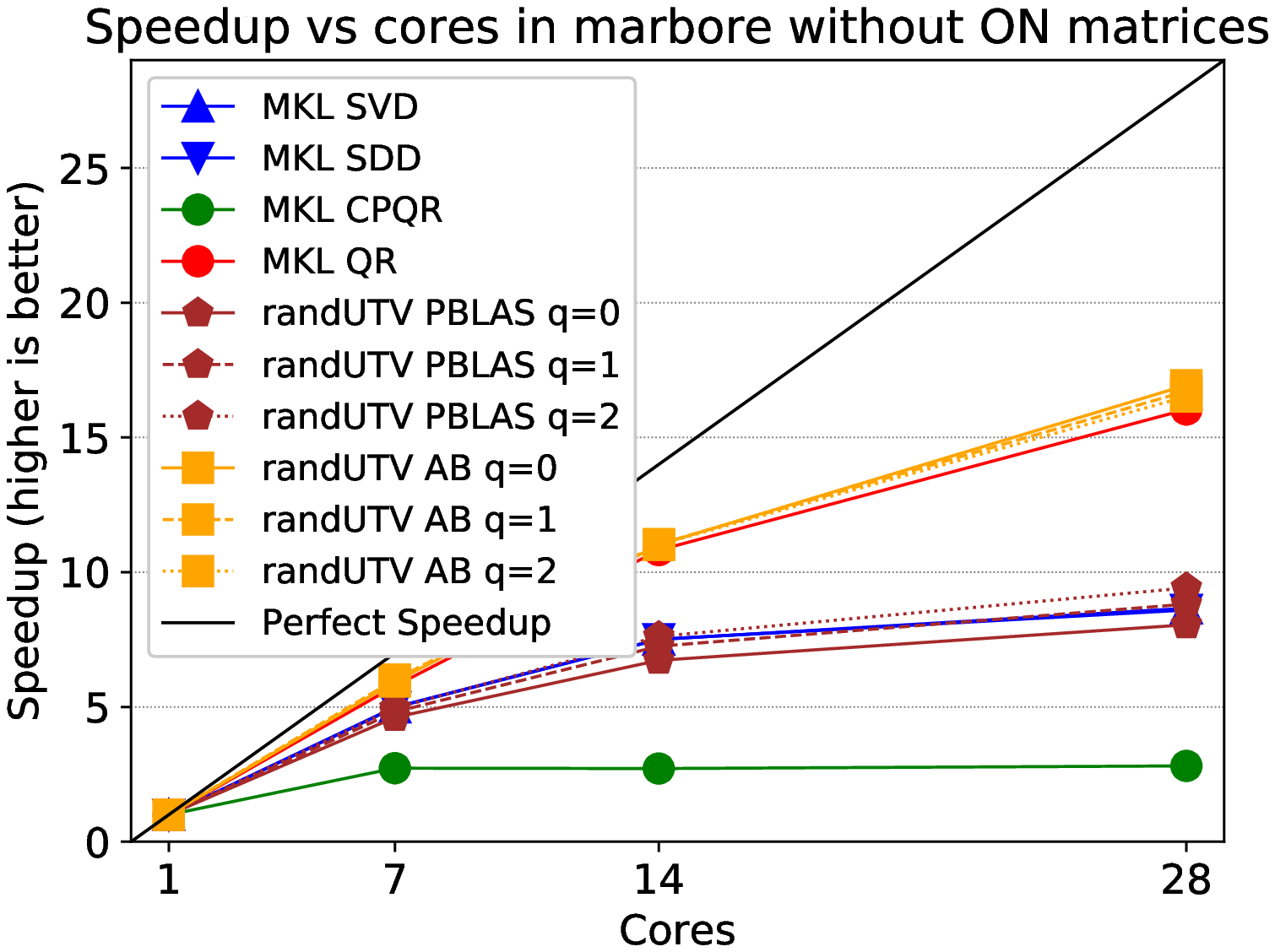} &
\includegraphics[width=0.45\textwidth]{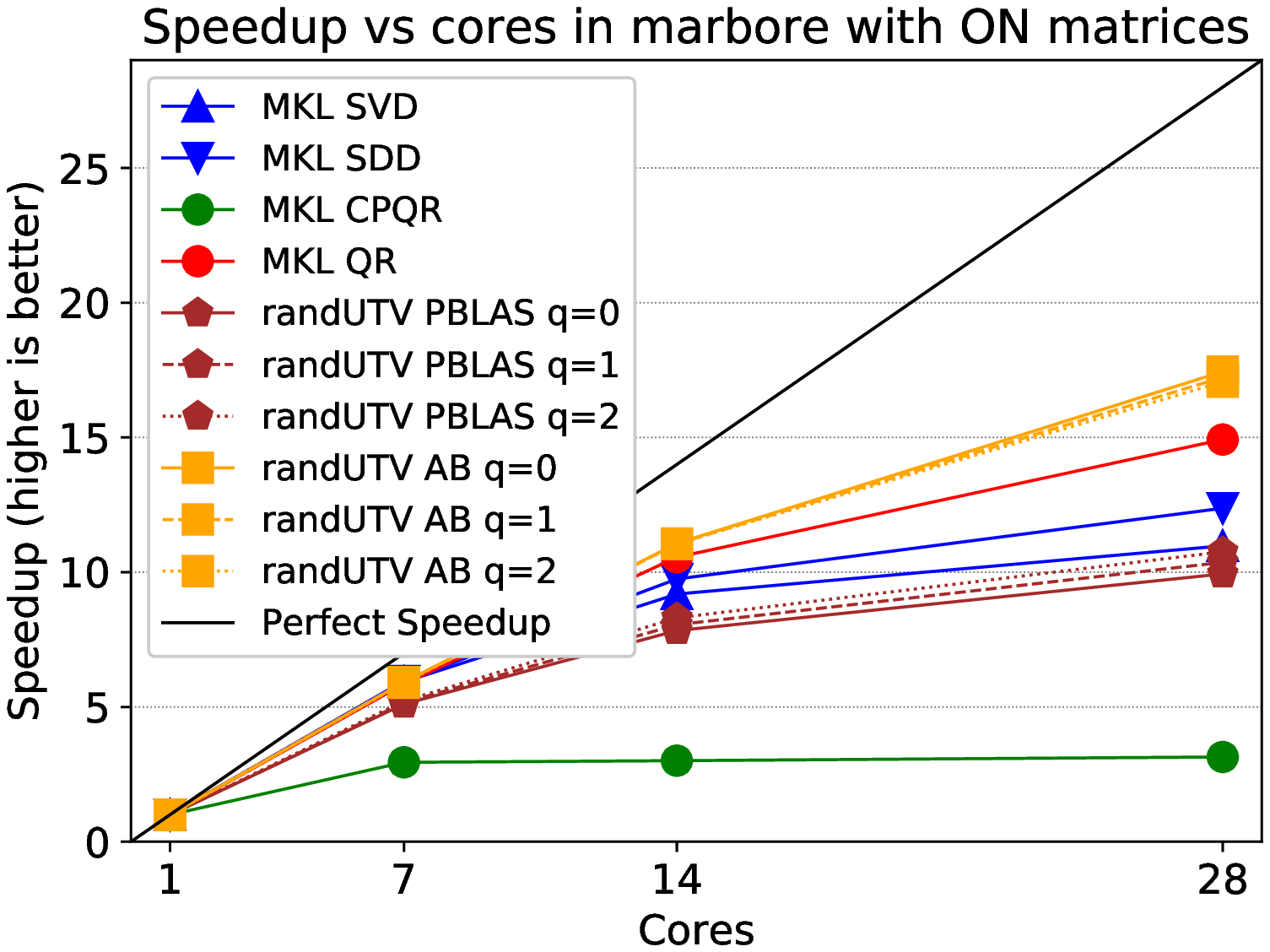} \\
\includegraphics[width=0.45\textwidth]{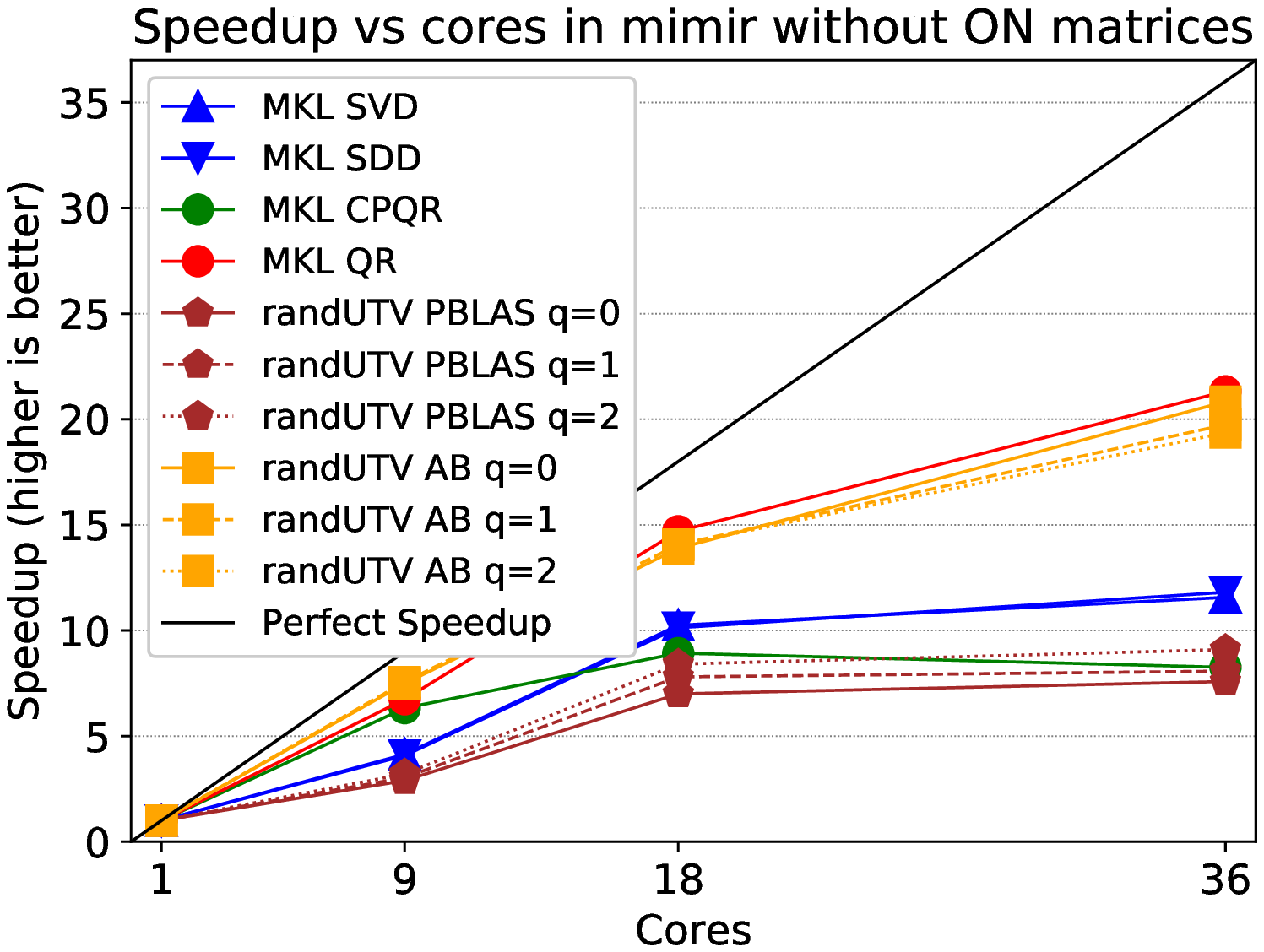} &
\includegraphics[width=0.45\textwidth]{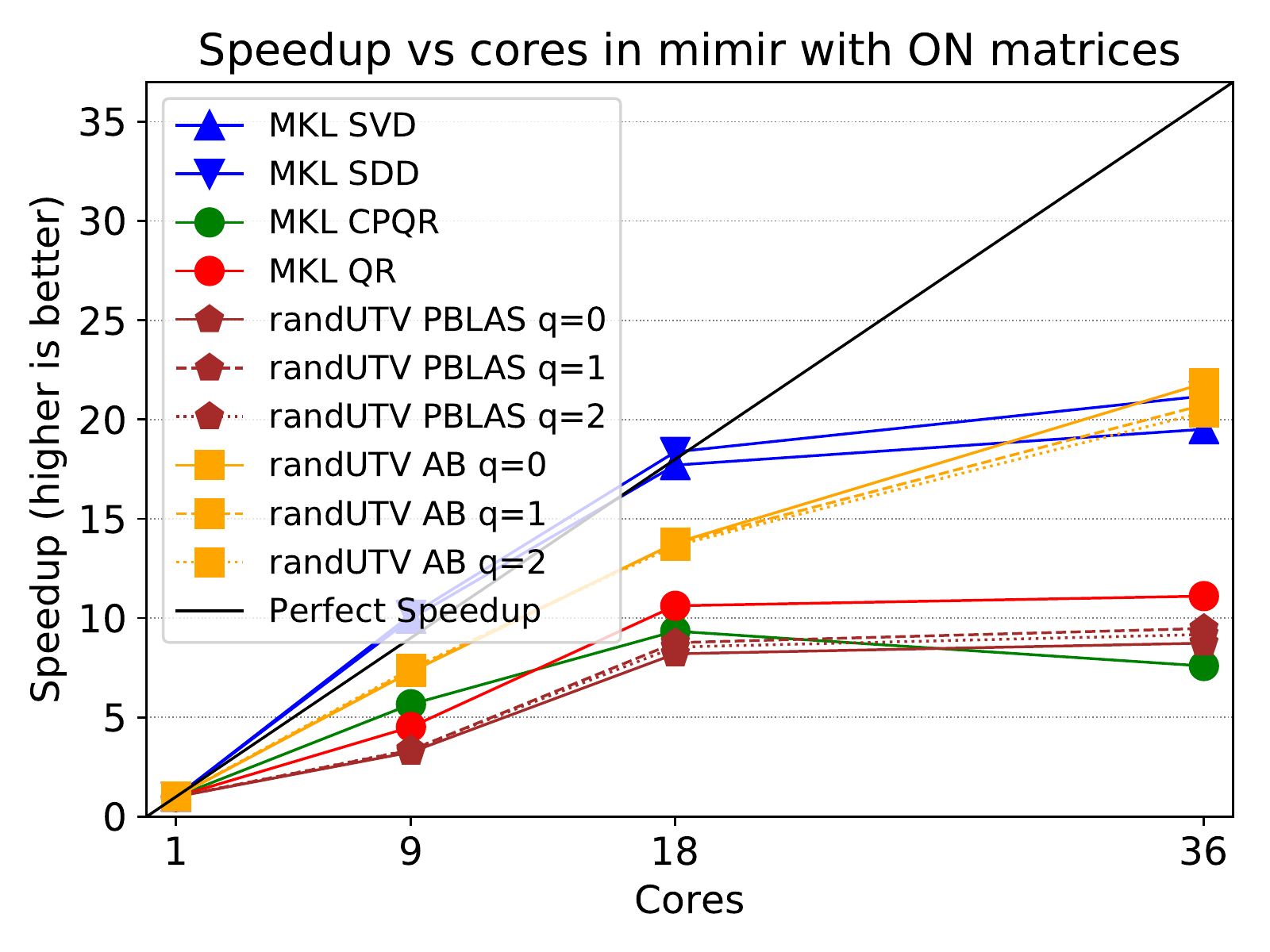} \\
\end{tabular}
\end{center}
\bfvspace
\caption{Speedups versus number of cores for the best implementations.
The top row shows results for \texttt{marbore} with 28 cores and $m=n=14000$;
the bottom row shows results for \texttt{mimir} with 36 cores and $m=n=18000$.}
\label{fig:sm_speedups}
\end{figure}

Figure~\ref{fig:sm_speedups}
shows the speedups obtained by the best implementations on both machines.
The top row shows results of \texttt{marbore}
on matrices of dimension $14000 \times 14000$,
whereas
the bottom row shows results of \texttt{mimir}
on matrices of dimension $18000 \times 18000$
(the largest dimension in which all the best implementations could be run).
Recall that in this plot
every implementation compares against itself on one core.

We see that
the scalability of {\sc randUTV AB} is always similar or even better
than the scalability of the highly efficient unpivoted QR factorization,
and it does not depend on whether the orthonormal matrices are built.
Note that it always grows whenever more cores are employed.
In contrast,
the SVD factorizations perform very well in one case
(even with a slight superspeedup):
when orthonormal matrices are built
using 18 cores or fewer in \texttt{mimir}.
In all the other cases, the speedups are not so good,
and the scalability even drops (the speedups do not grow much)
when going from half the number of cores to the full number of cores.

As can be seen,
{\sc randUTV AB} is the only factorization that achieves
an efficiency similar or higher than 50 \%
(the speedups are higher than half the number of cores)
when employing the maximum number of cores in both architectures,
whereas the efficiency of all the other factorizations are usually always lower.
To conclude this analysis,
the scalability of {\sc randUTV AB}
is similar (or even better) to that of the QR factorization,
and much higher than the rest of the implementations.

\begin{figure}
\begin{center}
\includegraphics[width=\textwidth]{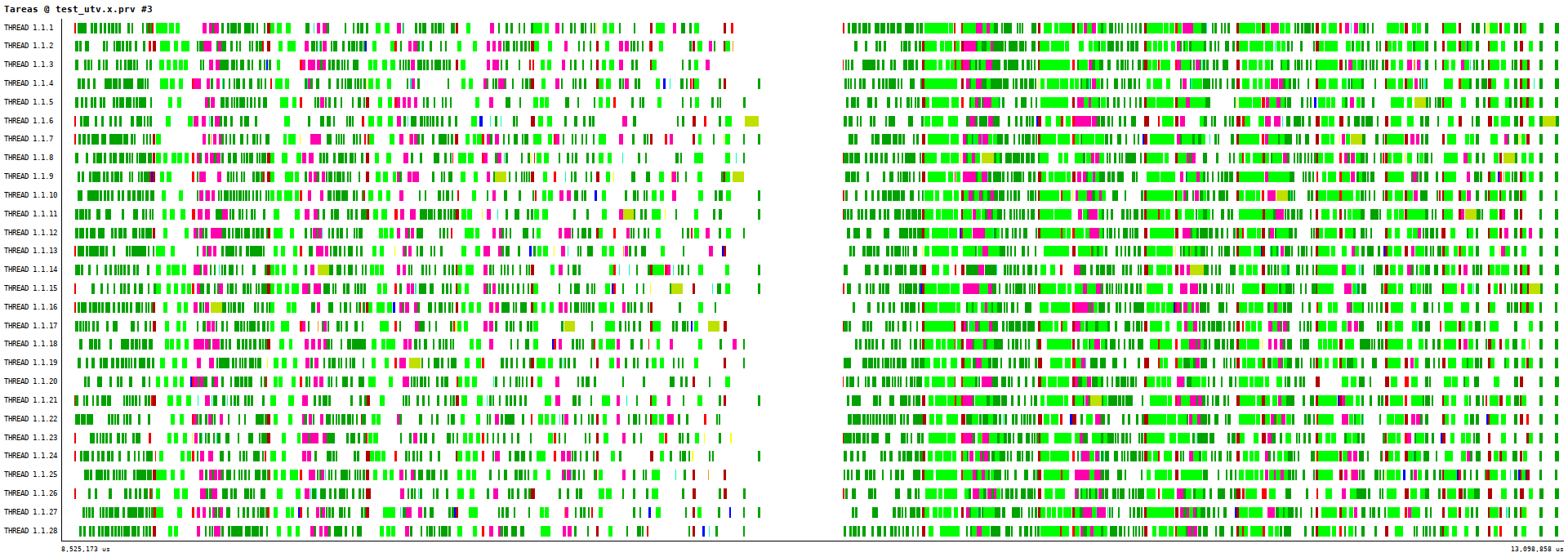}
\end{center}
\caption{Execution traces for the algorithm-by-block of \texttt{randUTV}
with $n = 1920$ and $b = 384$ in \texttt{marbore} (28 threads), without and with
computation of orthonormal matrices (left and right parts of the trace, respectively).\label{fig:exec_trace}}
\end{figure}

Figure~\ref{fig:exec_trace} reports an actual execution trace for an experiment
on \texttt{marbore} for $n = 1920$ and $b = 384$ running on 28 cores,
for a factorization that does not compute orthonormal matrices (left part of the
trace) and computing orthonormal matrices (right part of the trace). Each row
in the trace correspond to a worker thread; colors match those depicted in Table~\ref{fig:analyzer}.

In conclusion, {\sc randUTV AB} is the clear winner over competing factorization methods in terms of raw speed when orthonormal matrices are required and the matrix is not too small ($n \gtrsim 4000$). In terms of scalability, {\sc randUTV AB} outperforms the competition as well. Also, the algorithm-by-blocks implementation gives noticeable speedup over the blocked PBLAS version. That {\textsc randUTV AB} can compete with \textsc{MKL SVD} at all in terms of speed is remarkable, given the large effort usually invested by Intel on its sofware. This is evidenced by the fact that the \textsc{MKL CPQR} is left in the dust by both \textsc{MKL SVD} and \textsc{randUTV AB}, each of which costs far more flops than \textsc{MKL CPQR}. The scalability results of \textsc{randUTV AB} and its excellent timings evince its potential as a high performance tool in shared memory computing.

\subsection{Computational speed on distributed-memory architectures}
\label{subsec:padm}

The experiments on distributed-memory architectures reported in this subsection
were performed on a cluster of HP computers.
Each node of the cluster contained
two Intel Xeon\circledR\ CPU X5560 processors at 2.8 GHz,
with 12 cores and 48 GiB of RAM in total.
The nodes were connected with an Infiniband 4X QDR network.
This network is capable of supporting 40 Gb/s signaling rate,
with a peak data rate of 32 Gb/s in each direction.

Its OS was GNU/Linux (Version 3.10.0-514.21.1.el7.x86\_64).
Intel's \texttt{ifort} compiler (Version 12.0.0 20101006) was employed.
LAPACK and ScaLAPACK routines were taken from
the Intel(R) Math Kernel Library (MKL) Version 10.3.0 Product Build 20100927
for Intel(R) 64 architecture,
since this library usually delivers much higher performances
than LAPACK and ScaLAPACK codes from the Netlib repository.

All the matrices used in these experiments were randomly generated
since they are much faster to be generated, and
the cluster was being heavily loaded by other users.

The following implementations were assessed
in the experiments of this subsection:

\begin{itemize}

\item
{\sc ScaLAPACK SVD}:
The routine called \texttt{pdgesvd} from MKL's ScaLAPACK
is used to compute the Singular Value Decomposition (SVD).

\item
{\sc ScaLAPACK CPQR}:
The routine called \texttt{pdgeqpf} from MKL's ScaLAPACK
is used to compute the column-pivoted QR factorization.

\item
{\sc PLiC CPQR}:
The routine called \texttt{pdgeqp3} from the PLiC library
(Parallel Library for Control)~\cite{benner-2008}
is used to compute the column-pivoted QR factorization
by using BLAS-3.
This source code was linked to the ScaLAPACK library from MKL
for the purpose of a fair comparison.

\item
{\sc randUTV}:
A new implementation for computing the \randUTV{} factorization
based on the ScaLAPACK infrastructure and library.
This source code was linked to the ScaLAPACK library from MKL
for the purpose of a fair comparison.

\item
{\sc ScaLAPACK QR}:
The routine called \texttt{dgeqrf} from MKL's ScaLAPACK
is used to compute the QR factorization.
Although this routine does not reveal the rank,
it was included in some experiments as a reference for the others.

\end{itemize}

Like in the previous subsection on shared-memory architectures,
for every experiment two plots are shown.
The left plot shows the performances
when no orthonormal matrices are computed
(the codes compute just the singular values for the SVD,
the upper triangular matrix $R$ for the CPQR and the QR factorizations,
and the upper triangular matrix $T$ for the \randUTV{} factorization).
The right plot shows the performances
when, in addition to those, all orthonormal matrices are explicitly formed
(matrices $U$ and $V$ for SVD and \randUTV{},
and matrix $Q$ for QR and CPQR).
Recall that the right plot slightly favors CPQR and QR
since only one orthonormal matrix is built.

\begin{figure}[ht!]
\tfvspace
\begin{center}
\begin{tabular}{cc}
\includegraphics[width=0.45\textwidth]{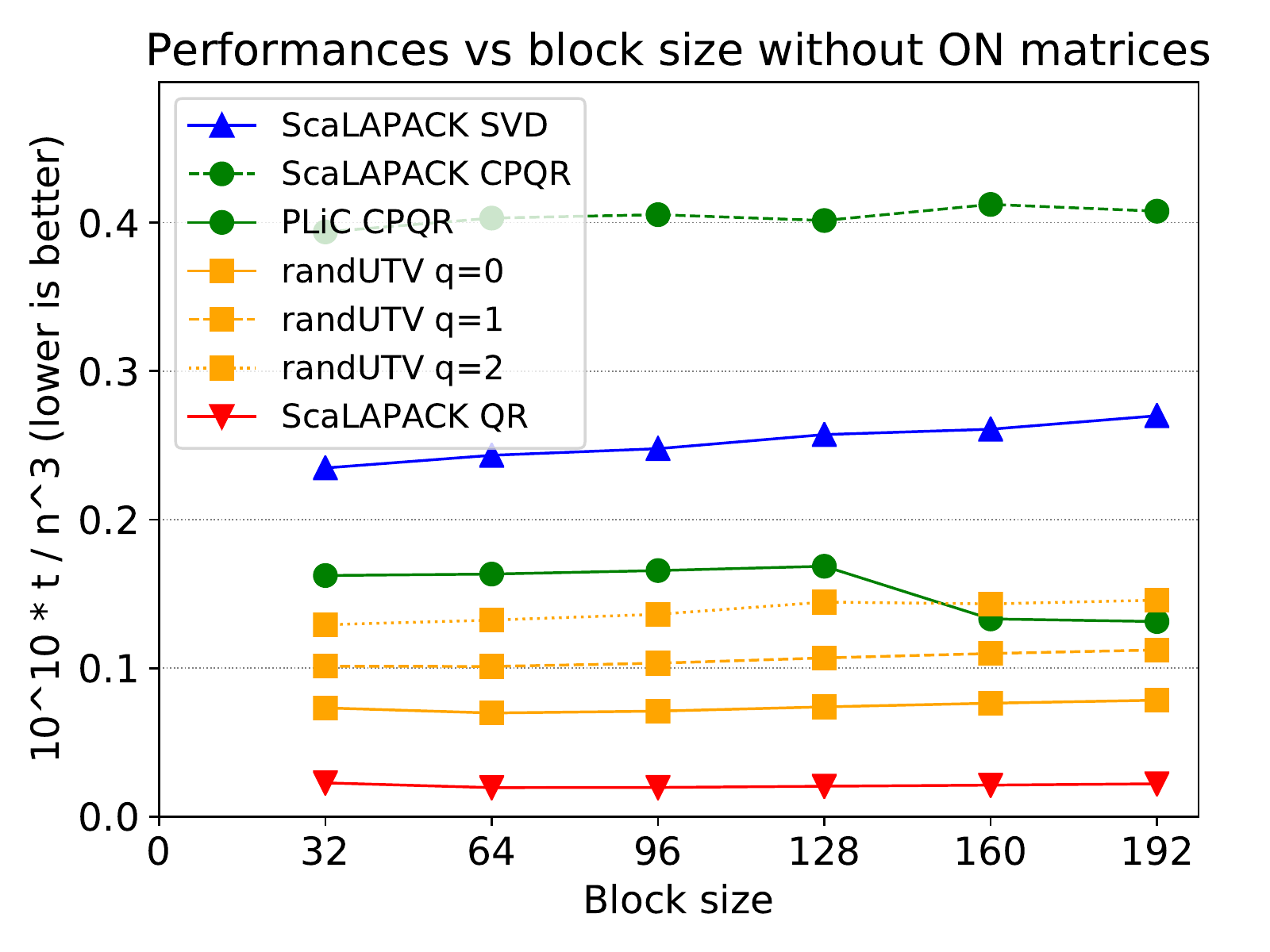} &
\includegraphics[width=0.45\textwidth]{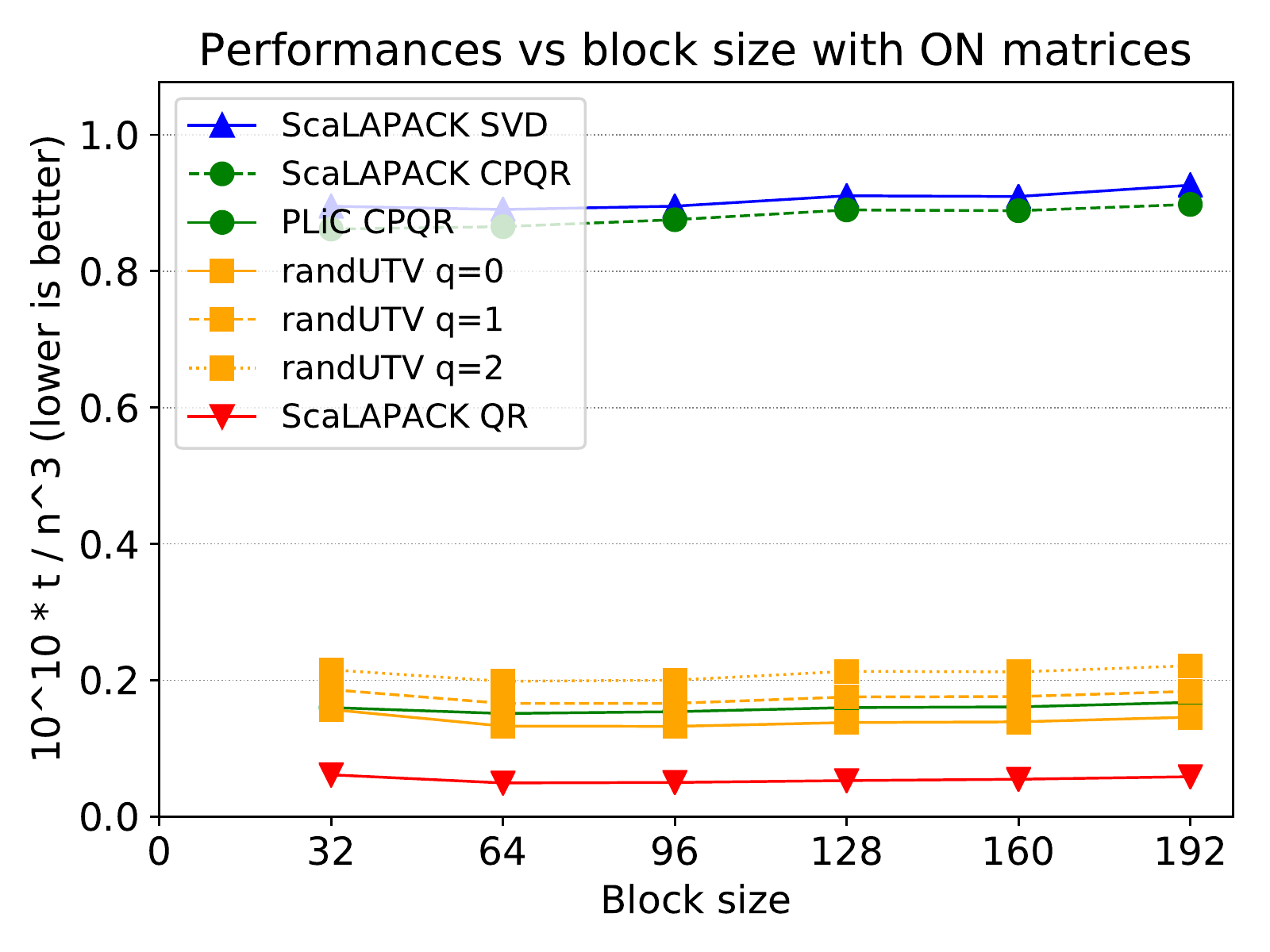} \\
\end{tabular}
\end{center}
\bfvspace
\caption{Performances versus block size on 96 cores
arranged as a $6 \times 16$ mesh.}
\label{fig:dm_block_sizes}
\end{figure}

Figure~\ref{fig:dm_block_sizes}
shows the performances of all the implementations described above
on several block sizes
when using 96 cores arranged as a $6 \times 16$ mesh
on matrices of dimension $25600 \times 25600$.
As can be seen, most implementations perform slightly better
on small block sizes, such as 32 and 64,
the only exception being PLiC CPQR, which performs a bit better
on large block sizes when no orthonormal matrices are built.

\begin{figure}[ht!]
\tfvspace
\begin{center}
\begin{tabular}{cc}
\includegraphics[width=0.45\textwidth]{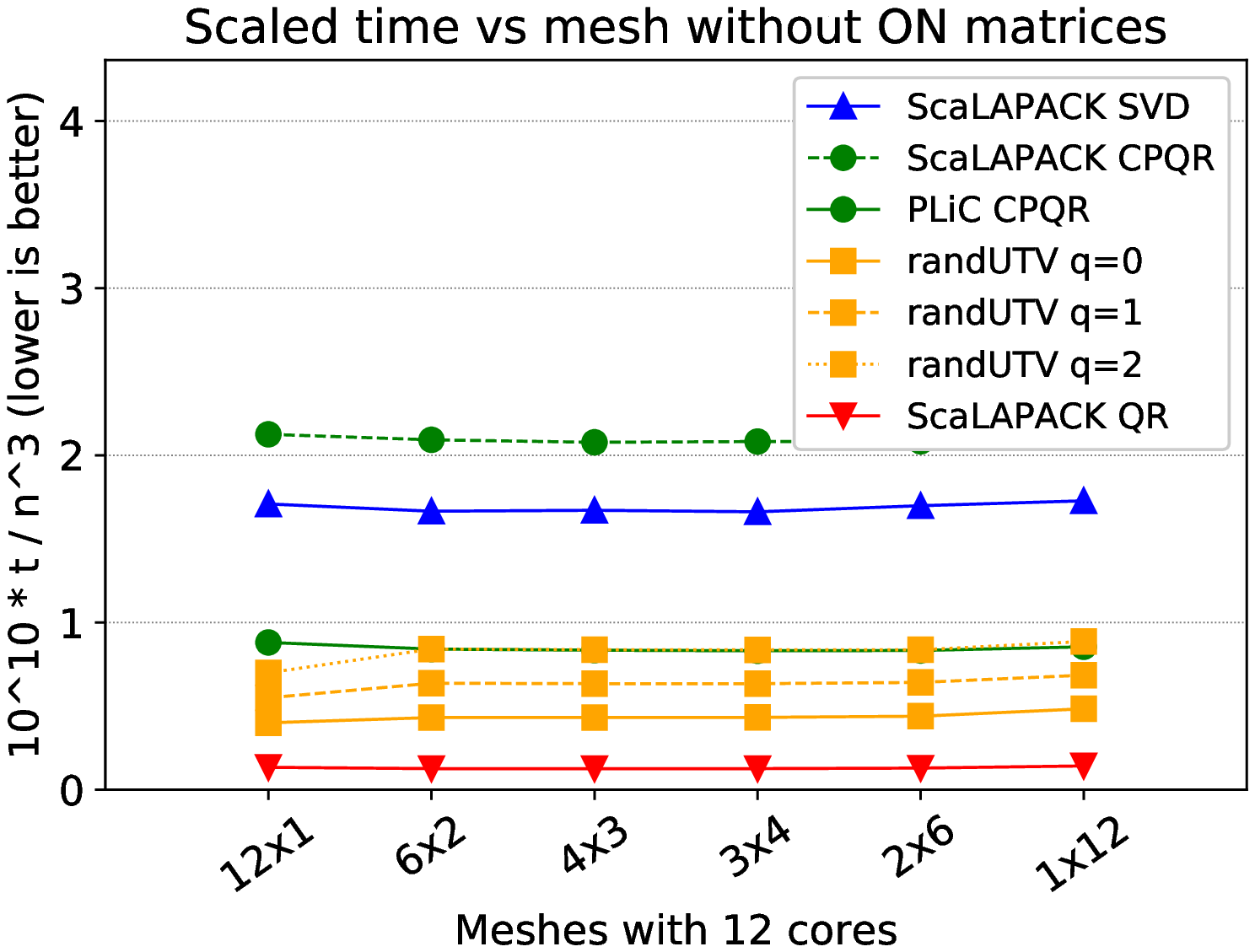} &
\includegraphics[width=0.45\textwidth]{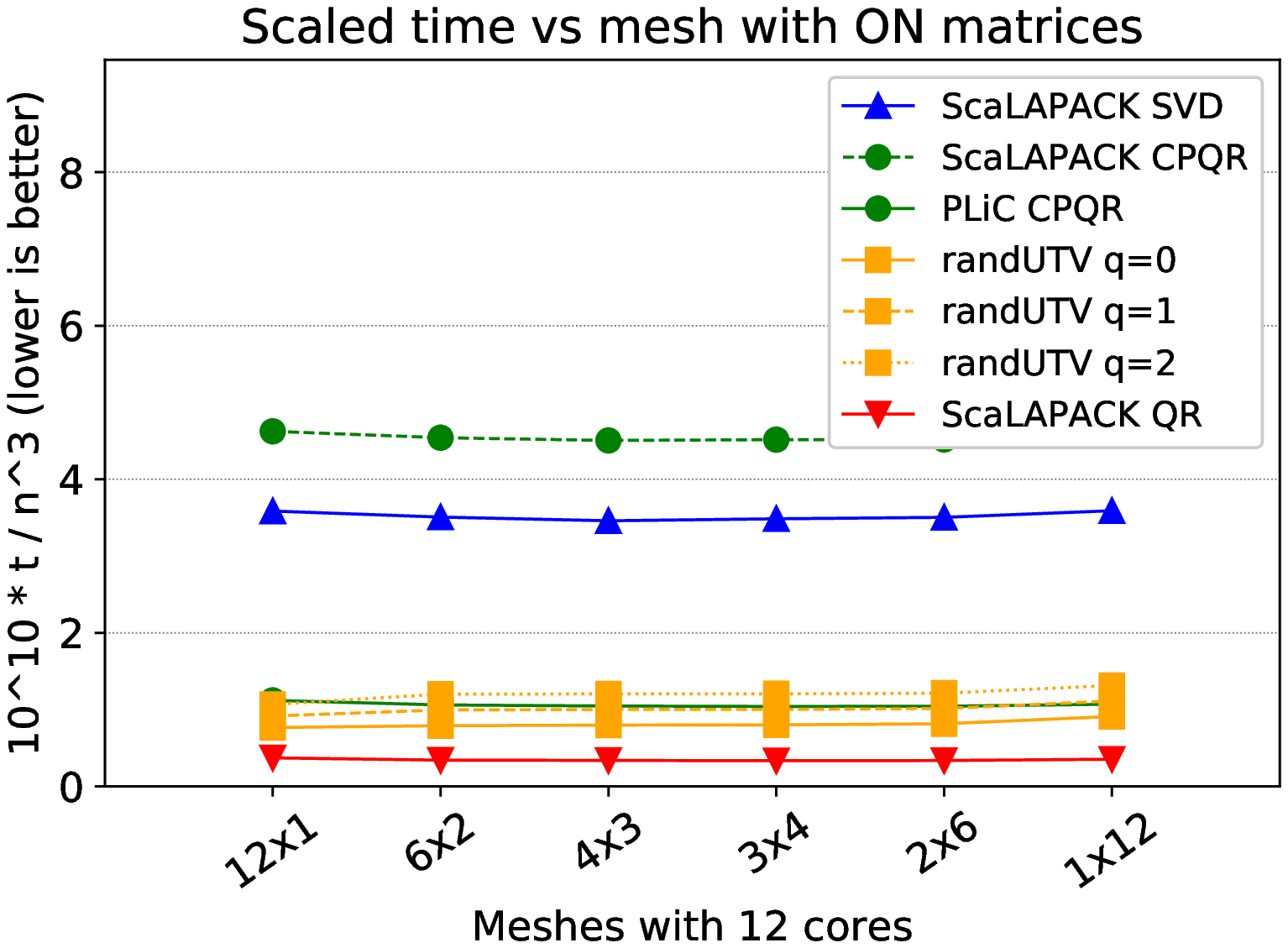} \\
\includegraphics[width=0.45\textwidth]{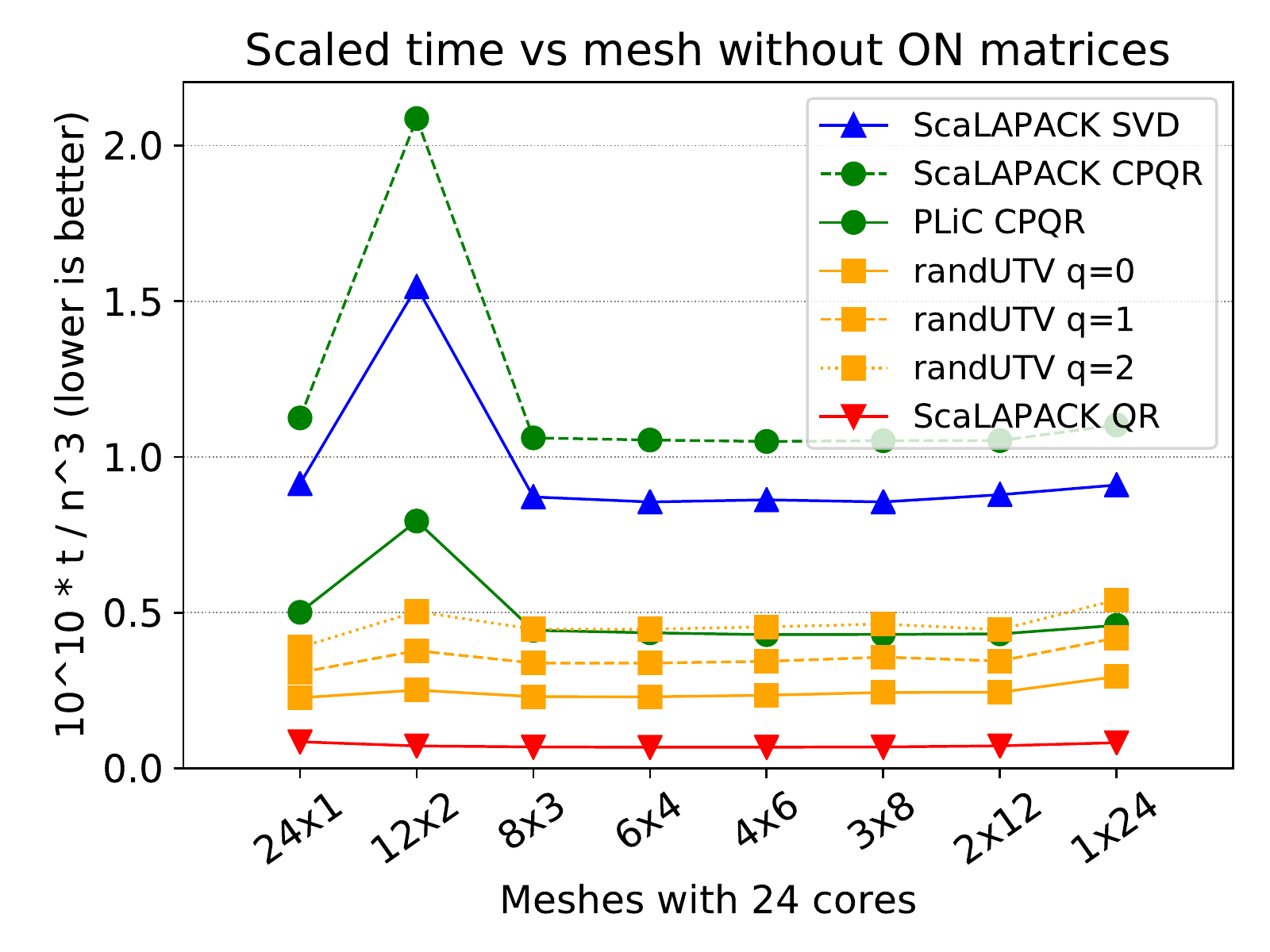} &
\includegraphics[width=0.45\textwidth]{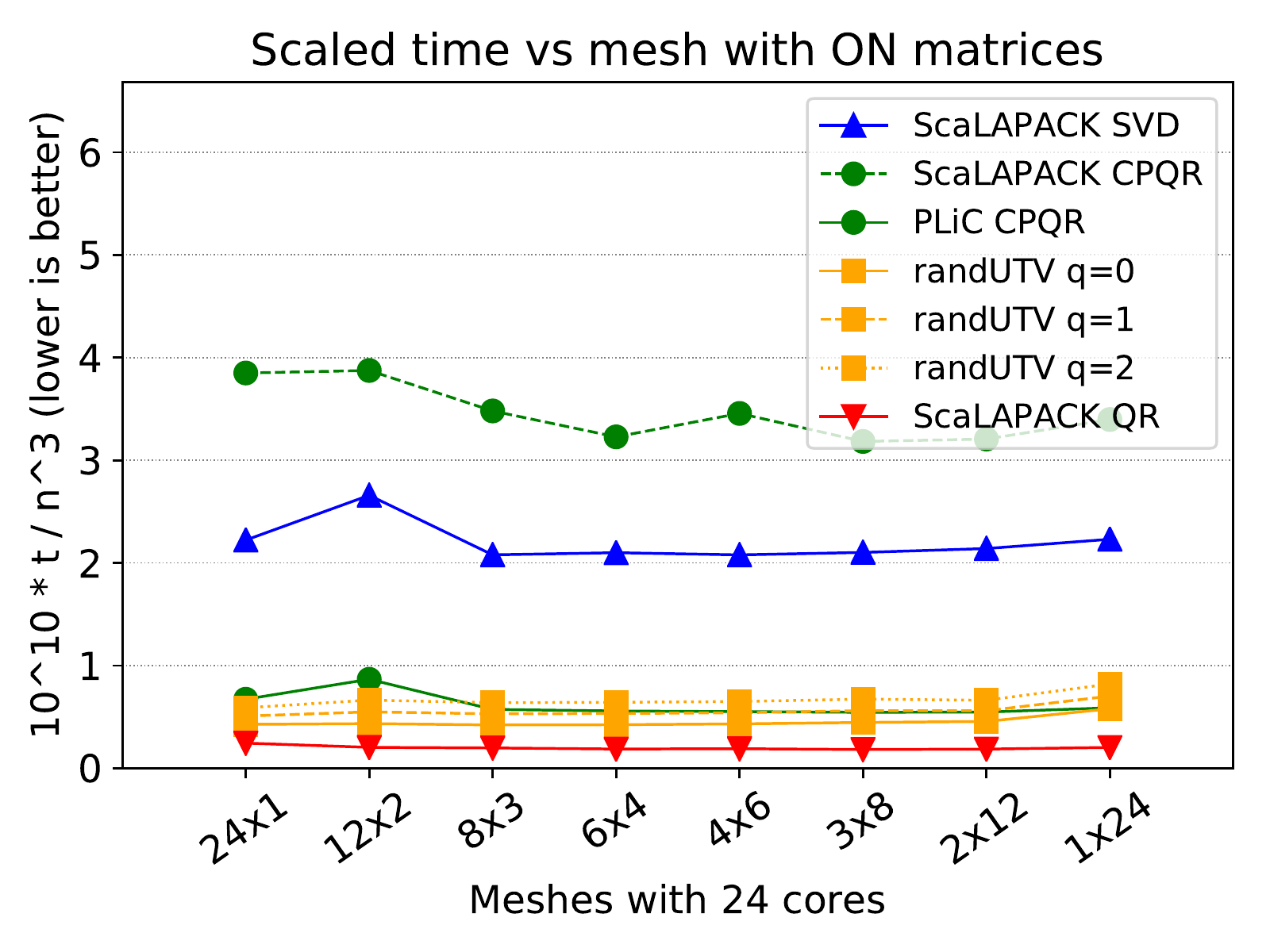} \\
\includegraphics[width=0.45\textwidth]{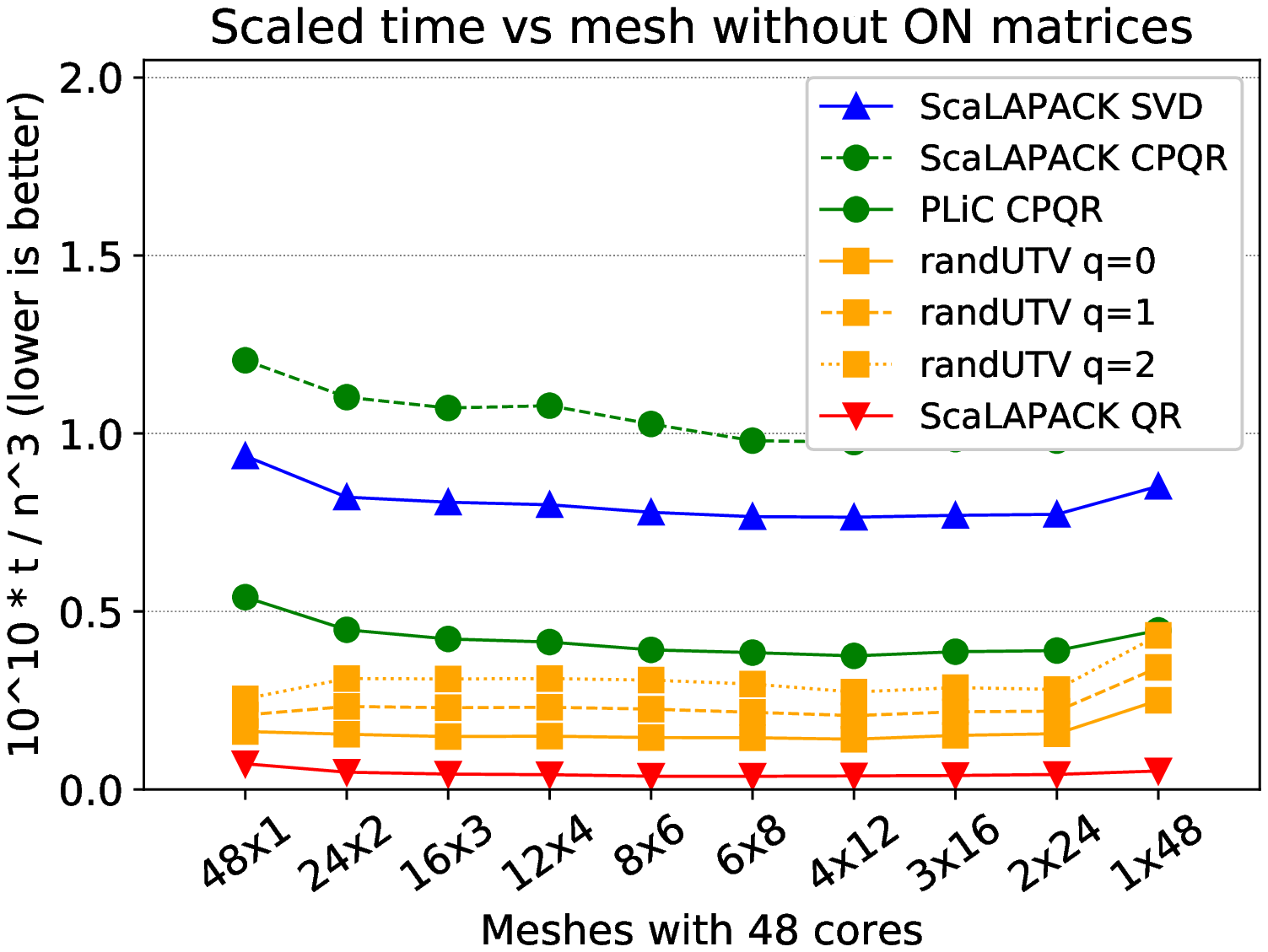} &
\includegraphics[width=0.45\textwidth]{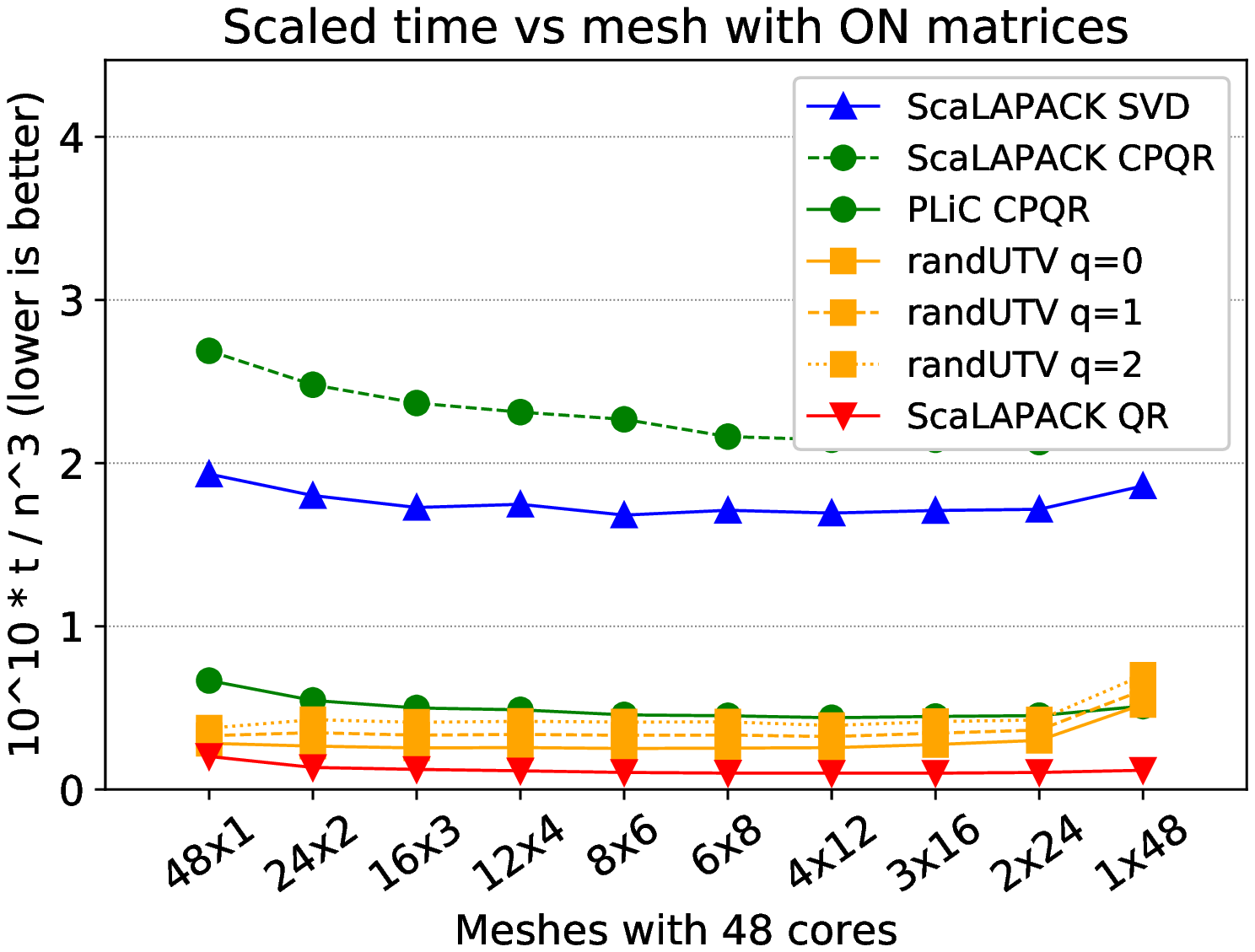} \\
\includegraphics[width=0.45\textwidth]{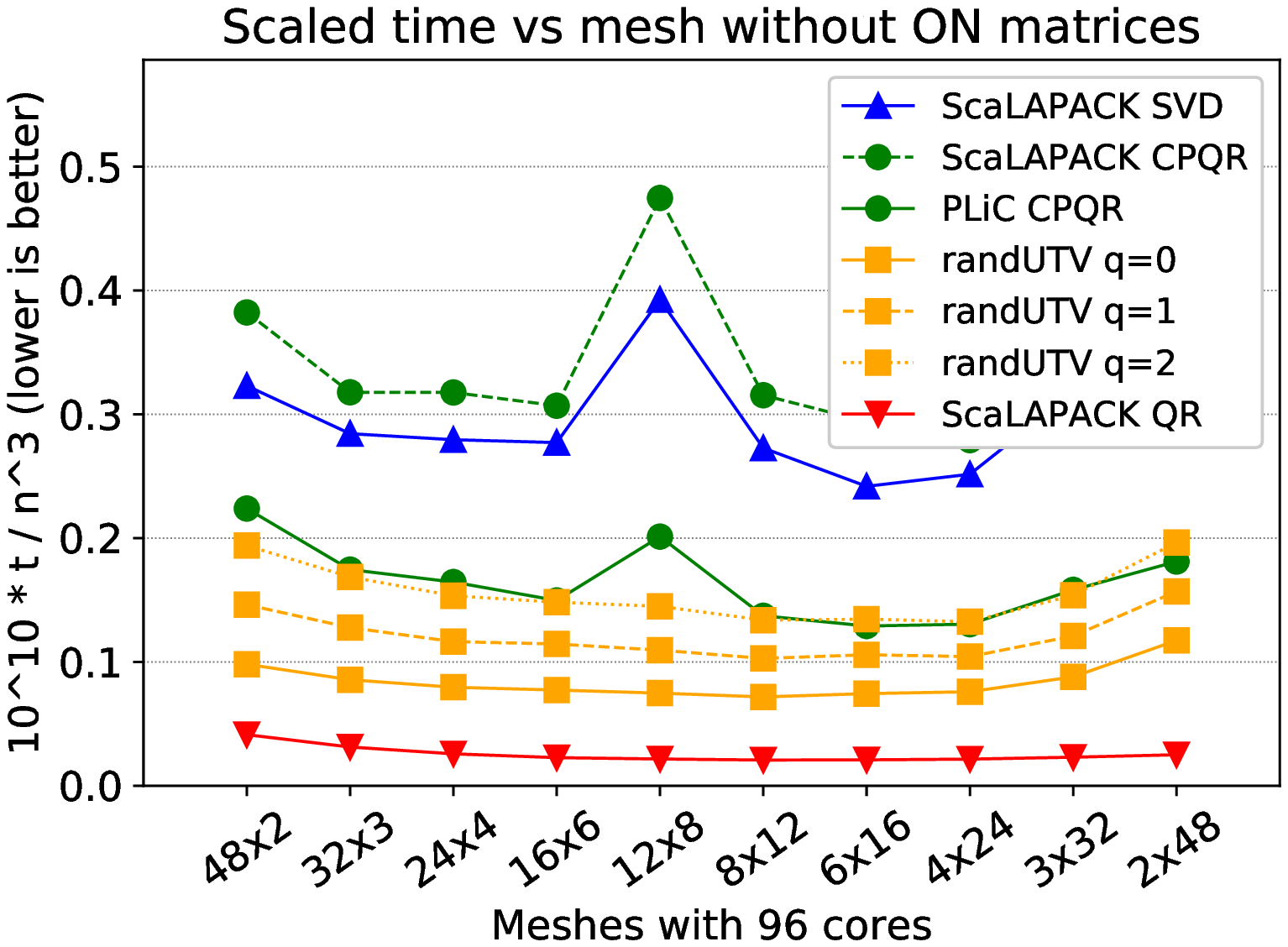} &
\includegraphics[width=0.45\textwidth]{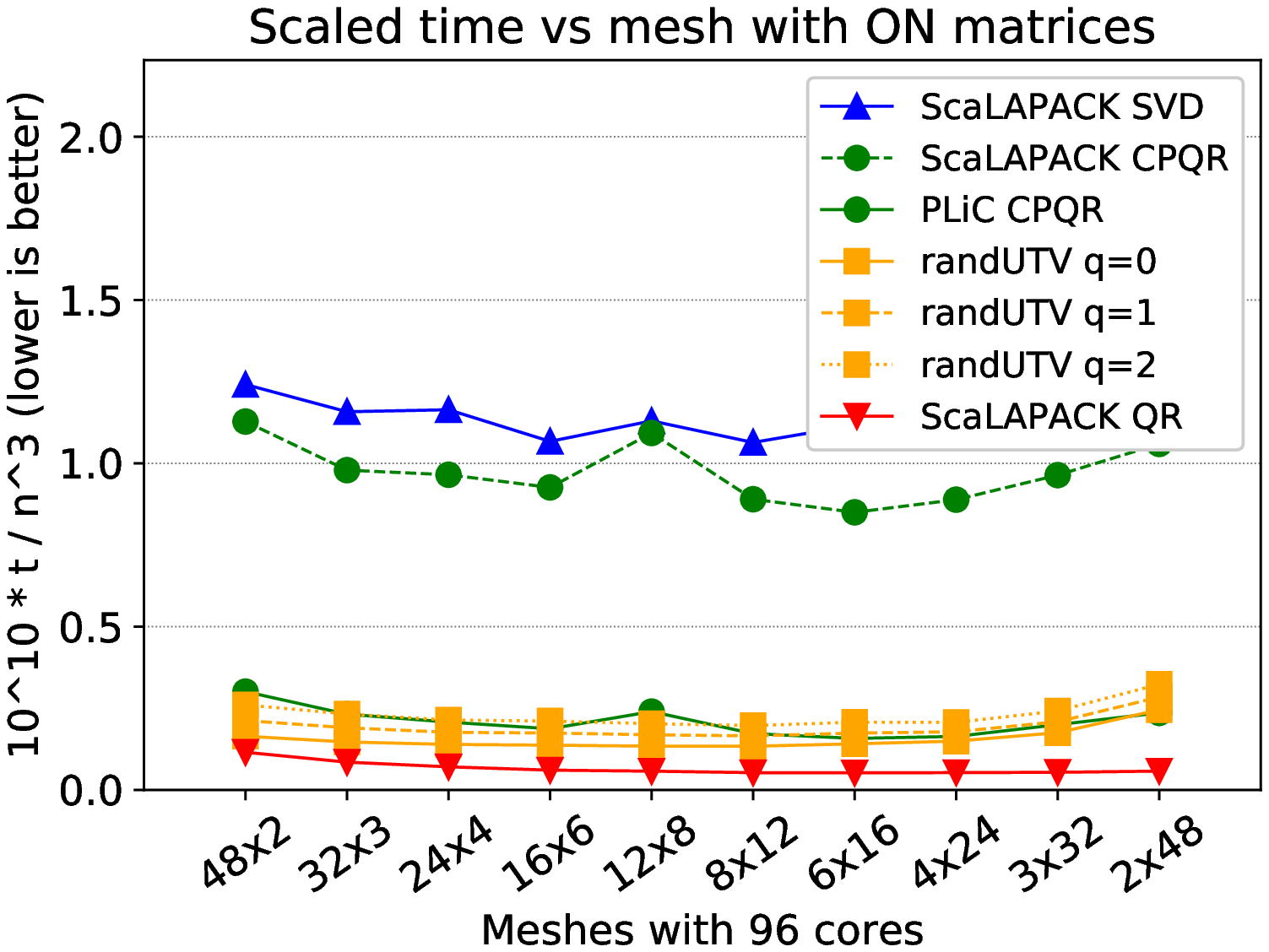} \\
\end{tabular}
\end{center}
\bfvspace
\caption{Performances on several topologies
on matrices of dimension $20480 \times 20480$.}
\label{fig:dm_topologies}
\end{figure}

Figure~\ref{fig:dm_topologies}
shows the performances of all the implementations for many topologies
on matrices of dimension $20480 \times 20480$.
The top row shows the results on one node (12 cores),
the second row shows the results on two nodes (24 cores),
the third row shows the results on four nodes (48 cores), and
the fourth row shows the results on eight nodes (96 cores).
As can be seen, best topologies are usually
$p \times q$ with $p$ slightly smaller than $q$.

\begin{figure}[ht!]
\tfvspace
\begin{center}
\begin{tabular}{cc}
\includegraphics[width=0.45\textwidth]{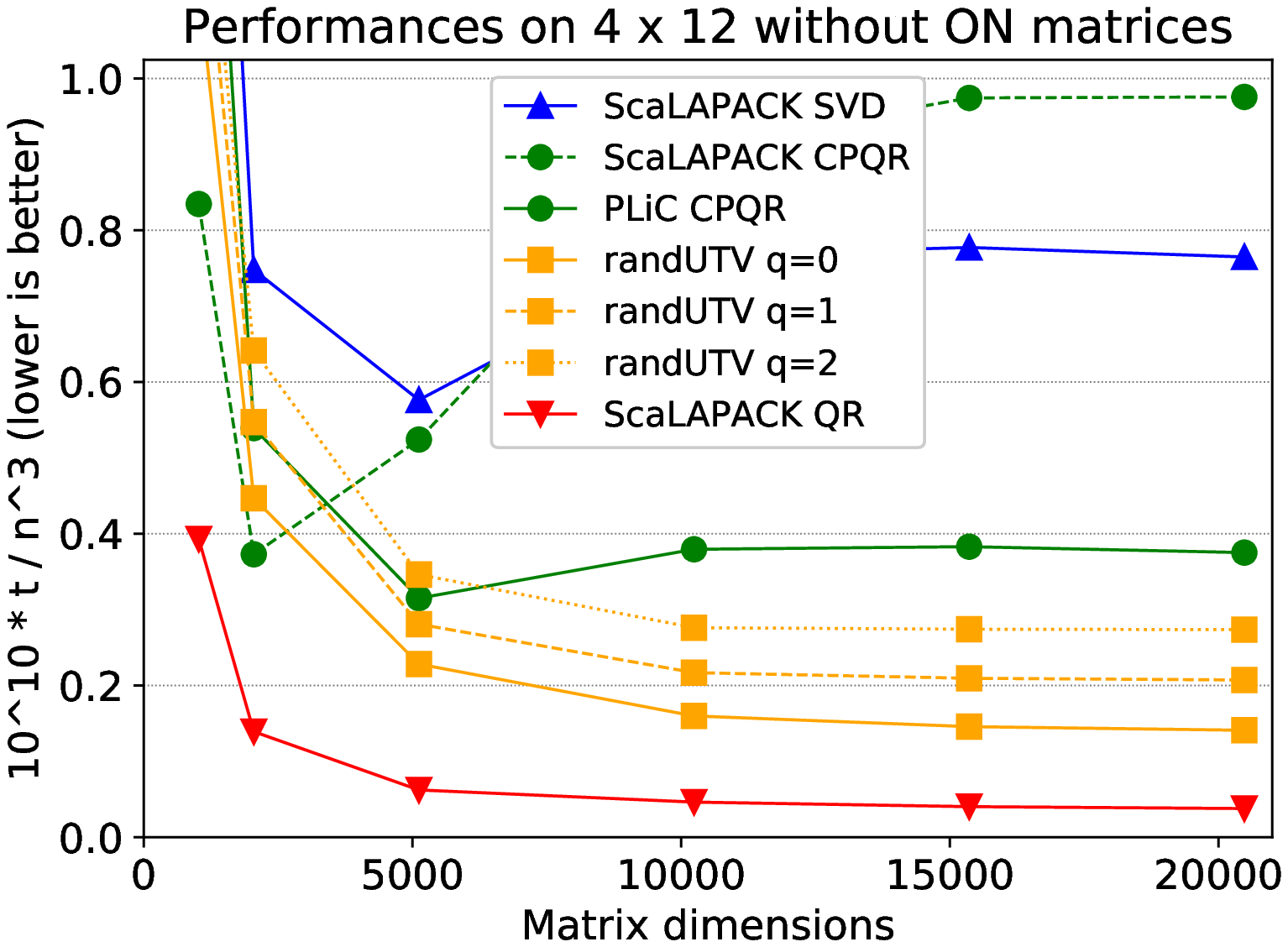} &
\includegraphics[width=0.45\textwidth]{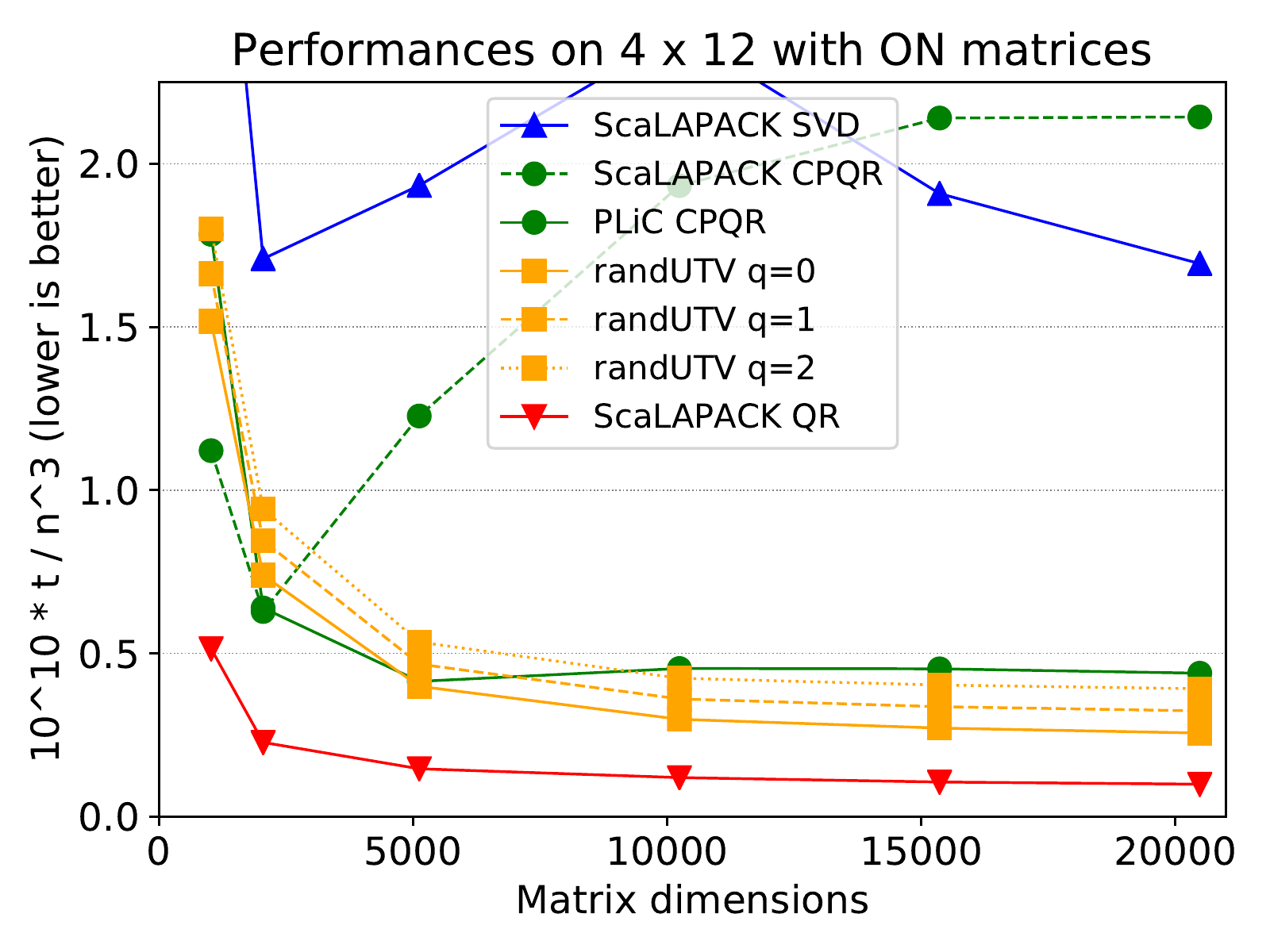} \\
\includegraphics[width=0.45\textwidth]{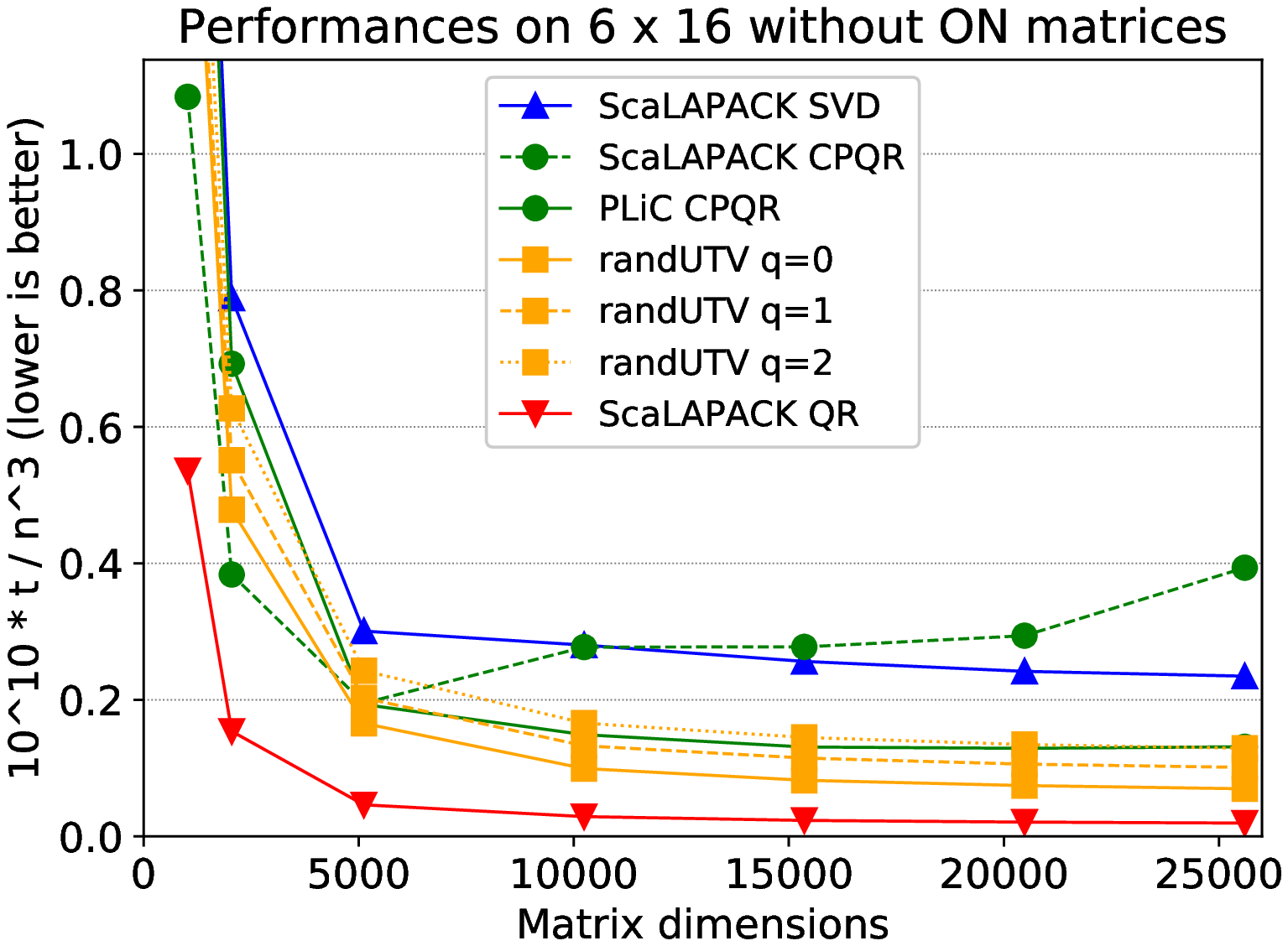} &
\includegraphics[width=0.45\textwidth]{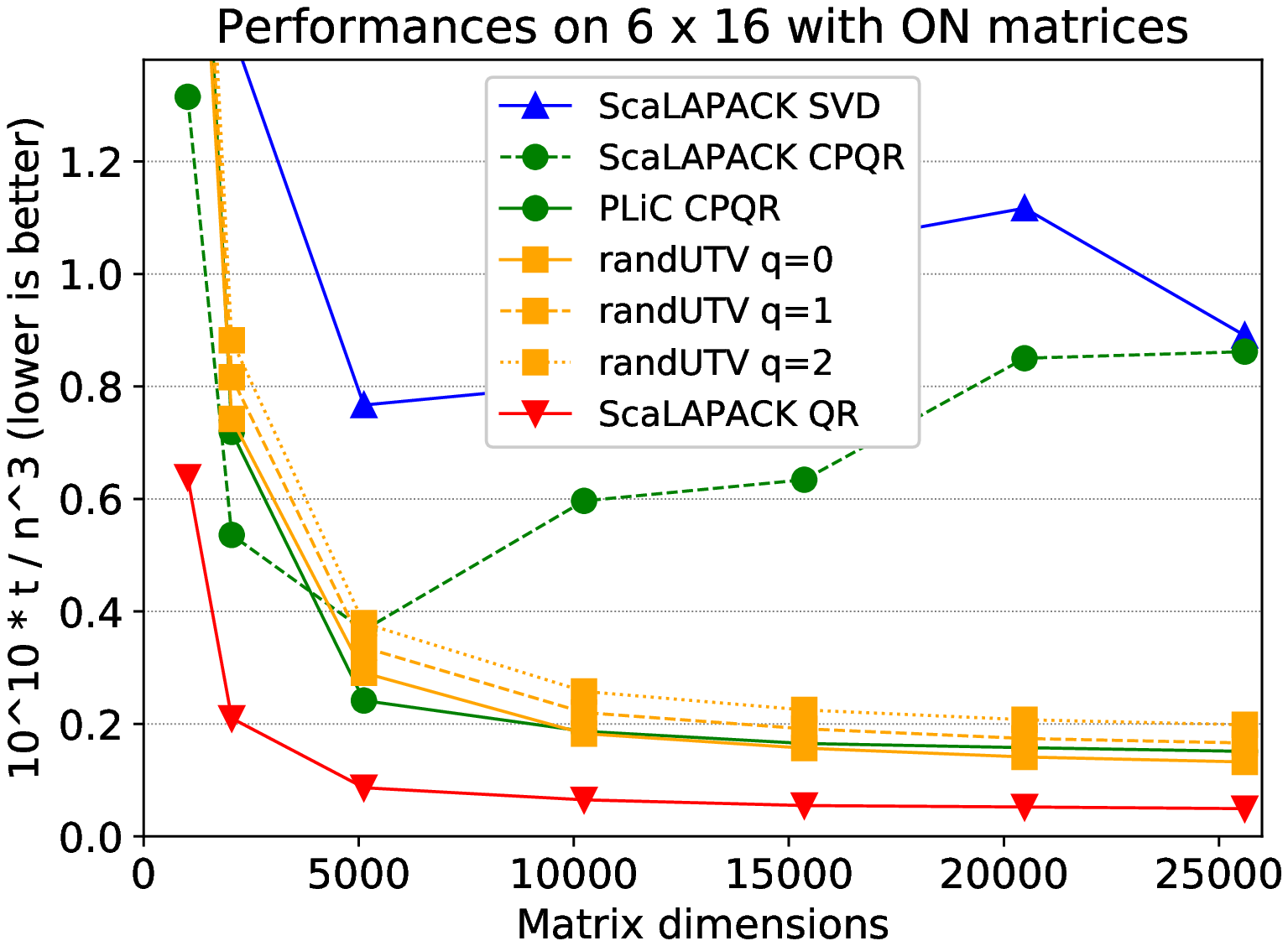} \\
\end{tabular}
\end{center}
\bfvspace
\caption{Performances versus matrix dimensions
on two different number of cores.
The top row shows results on 48 cores arranged as $4 \times 12$;
the bottom row shows results on 96 cores arranged as $6 \times 16$.}
\label{fig:dm_matrix_dimensions}
\end{figure}

Figure~\ref{fig:dm_matrix_dimensions}
shows the performances versus matrix dimensions
on two different number of cores:
48 cores arranged as $4 \times 12$ (top row) and
96 cores arranged as $6 \times 16$ (bottom row).
On the largest matrix dimension on 48 cores,
when no orthonormal matrices are built,
\randUTV{} is between 5.4 ($q=0$) and 2.8 ($q=2$) times
as fast as the SVD,
whereas
when orthonormal matrices are built,
\randUTV{} is between 6.6 ($q=0$) and 4.3 ($q=2$) times
as fast as the SVD.
On the largest matrix dimension on 96 cores,
when no orthonormal matrices are built,
\randUTV{} is between 3.4 ($q=0$) and 1.8 ($q=2$) times
as fast as the SVD,
whereas
when orthonormal matrices are built,
\randUTV{} is between 6.7 ($q=0$) and 4.5 ($q=2$) times
as fast as the SVD.
On medium and large matrices,
performances of ScaLAPACK CPQR are much lower
than those of \randUTV{},
whereas performances of PLiC CPQR
are more similar to those of \randUTV{}.
Nevertheless, recall that the precision of CPQR
is usually much smaller than that of \randUTV{}.


In distributed-memory applications
the traditional approach creates one process per core.
However, creating fewer processes
and then a corresponding number of threads per process
can improve performances in some cases.
Obviously, the product of the number of processes and
the number of threads per process
must be equal to the total number of cores.
The advantage of this approach is that the creation of fewer processes
reduces the communication cost,
which is usually the main bottleneck in distributed-memory applications.
In the case of linear algebra applications,
creating and using several threads per process can be easily achieved
by employing shared-memory parallel LAPACK and BLAS libraries.
Nevertheless, great care must be taken to ensure a proper pinning of processes
to cores, since otherwise performances drop markedly.
This was achieved by using the \texttt{-genv I\_MPI\_PIN\_DOMAIN socket} flag
when executing the \texttt{mpirun}/\texttt{mpiexec} command
in the machine used in the experiments.

\begin{figure}[ht!]
\tfvspace
\begin{center}
\begin{tabular}{cc}
\includegraphics[width=0.45\textwidth]{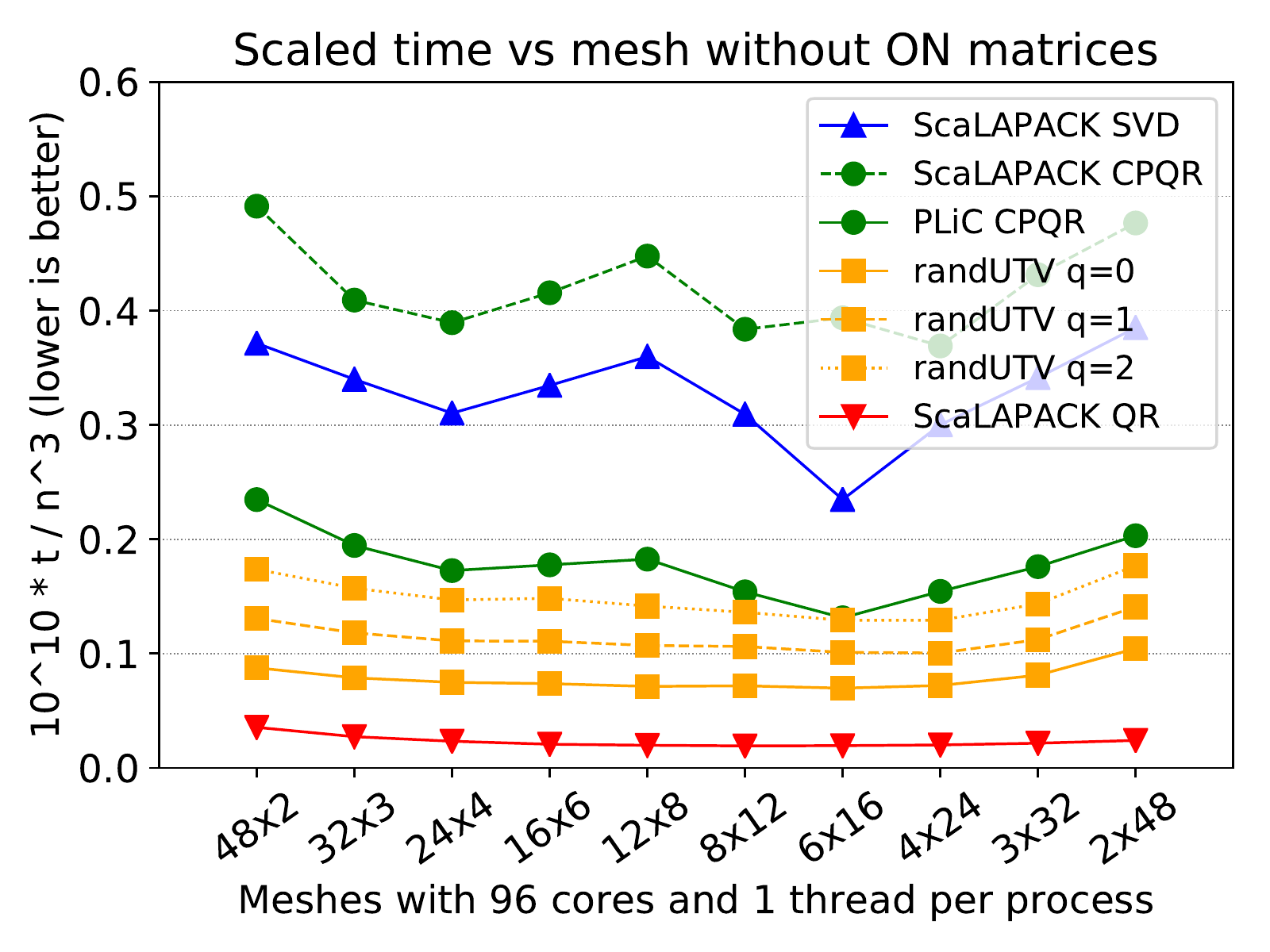} &
\includegraphics[width=0.45\textwidth]{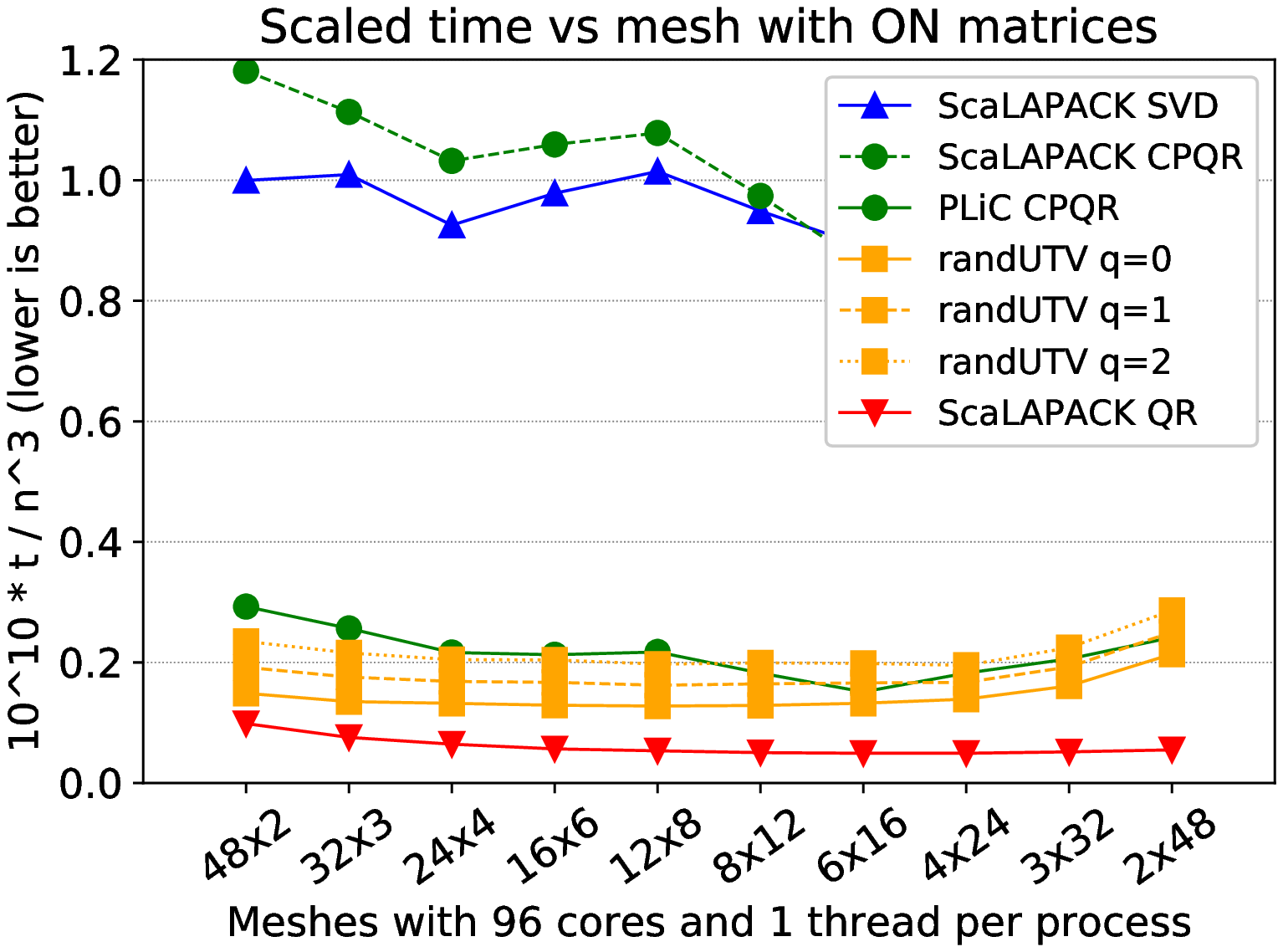} \\
\includegraphics[width=0.45\textwidth]{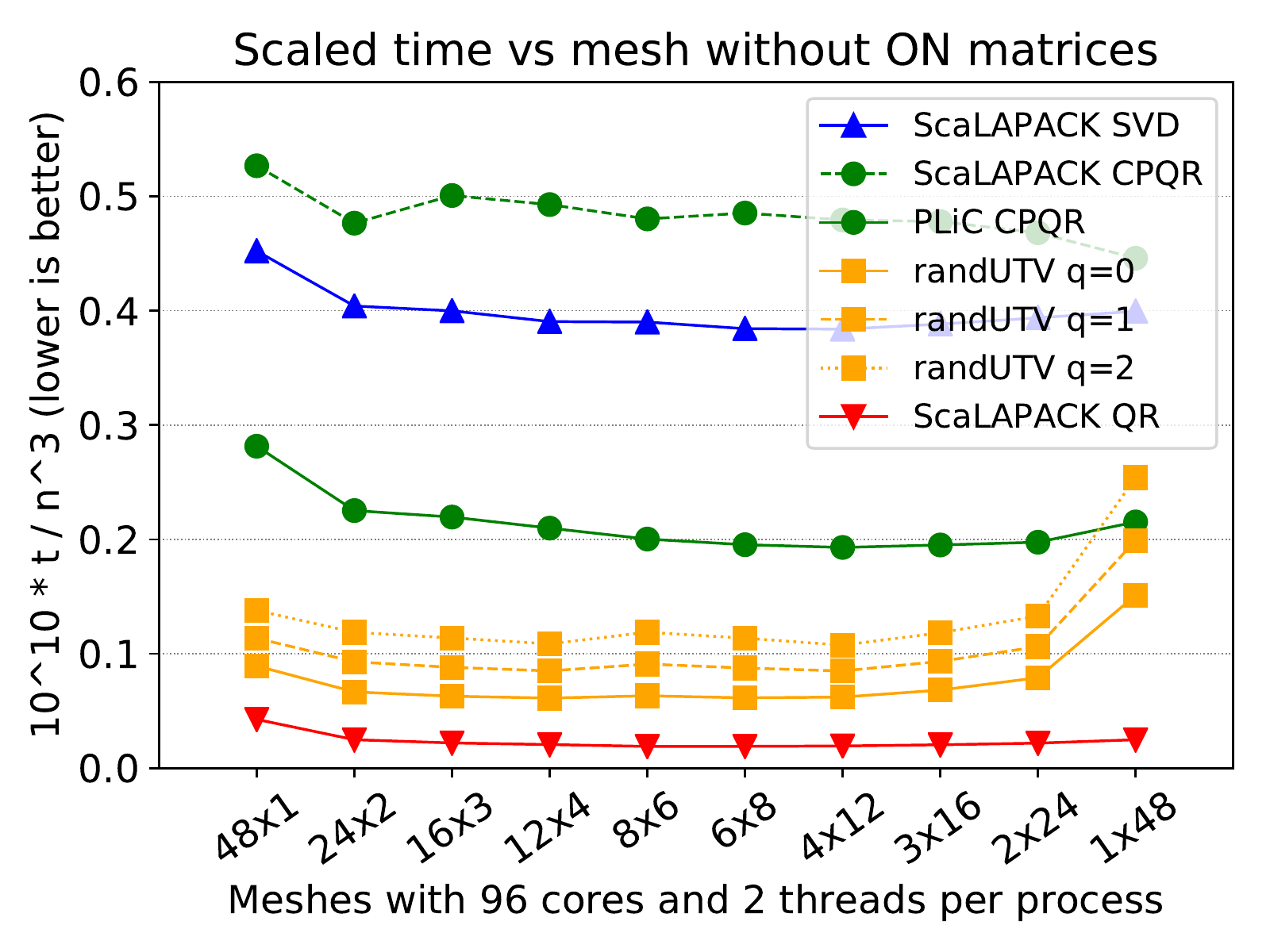} &
\includegraphics[width=0.45\textwidth]{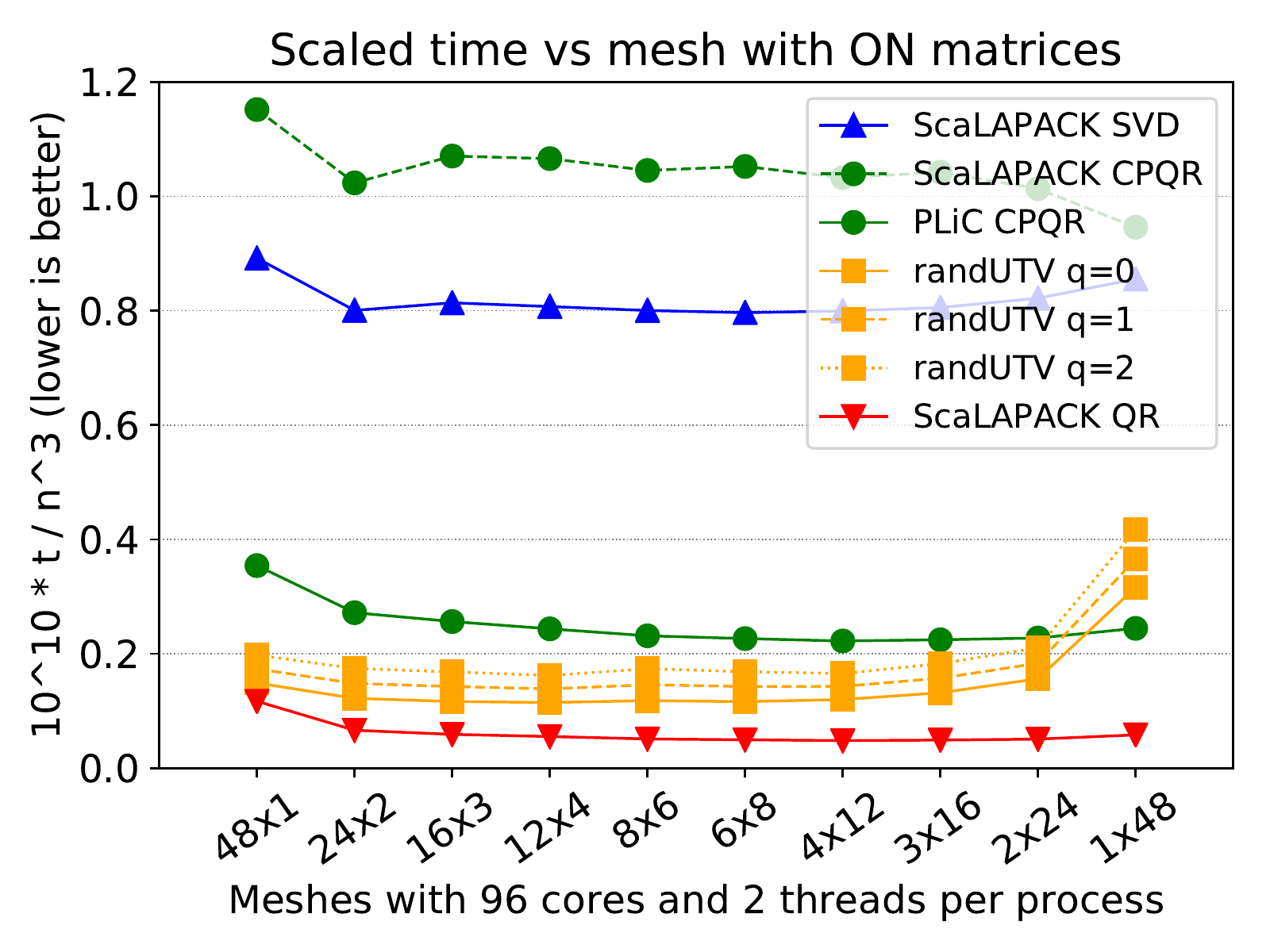} \\
\includegraphics[width=0.45\textwidth]{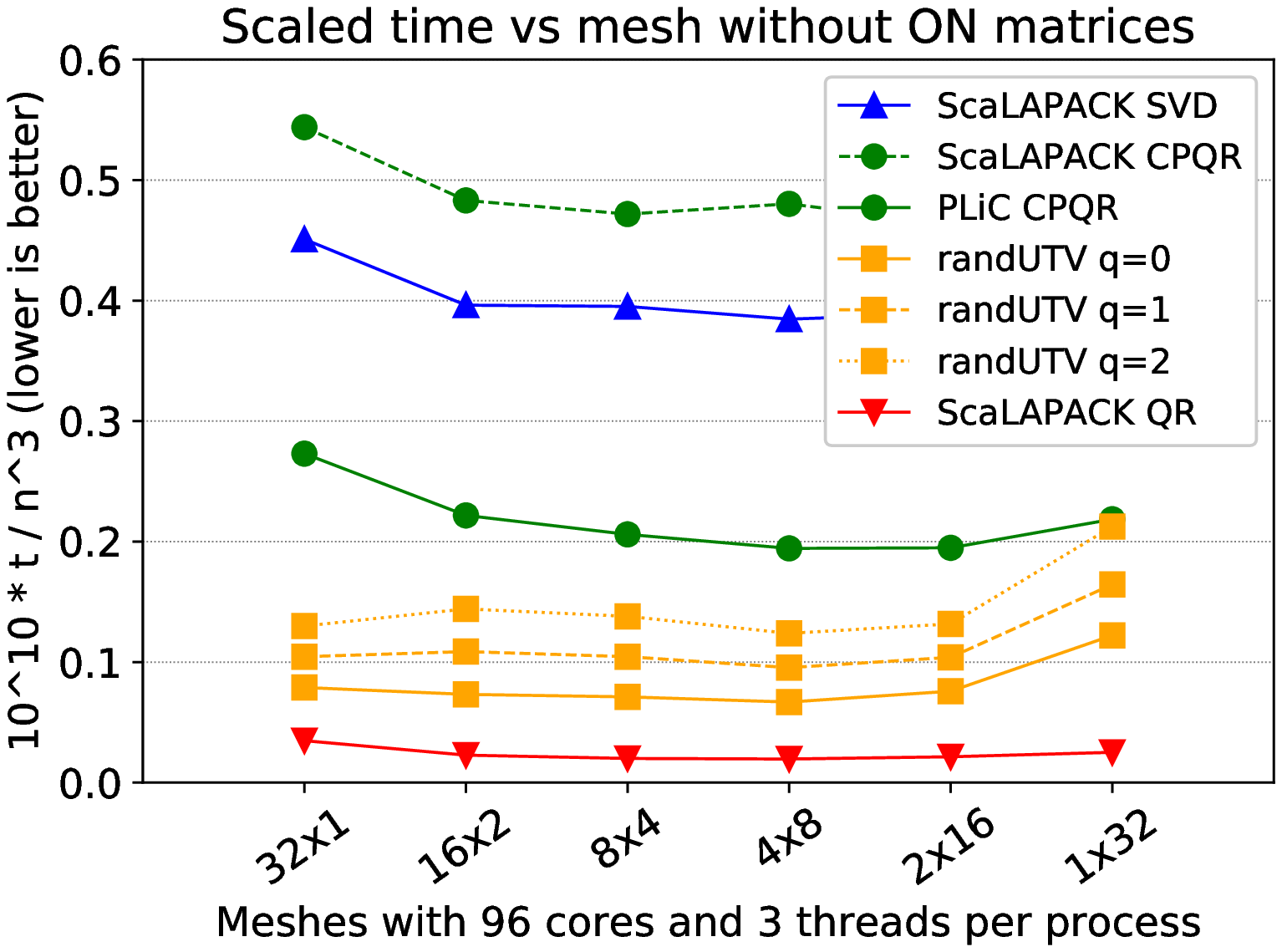} &
\includegraphics[width=0.45\textwidth]{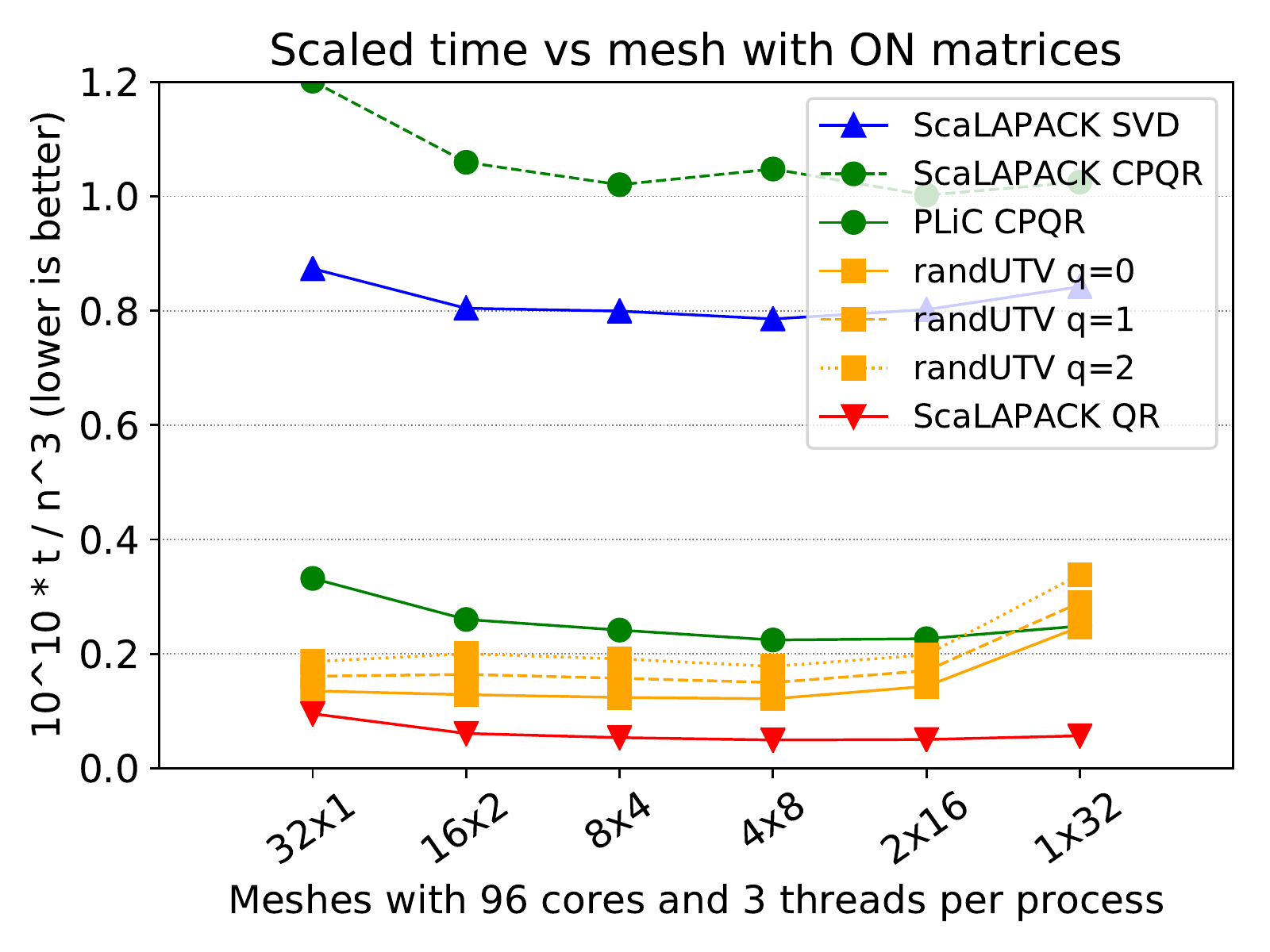} \\
\end{tabular}
\end{center}
\bfvspace
\caption{Performances on several topologies
on matrices of dimension $25600 \times 25600$.}
\label{fig:dm_threads_per_process}
\end{figure}

Figure~\ref{fig:dm_threads_per_process}
shows the scaled timings of the factorizations
of matrices of dimension $25600 \times 25600$ on 96 cores
when using several configurations with
different numbers of threads per process.
These plots include the results on a complete set of topologies
to isolate the effect of the increased number of threads.
As usual, the left three plots show performances
when no orthonormal matrices are built,
whereas
the right three plots show performances when orthonormal matrices are built.
The top row shows performances
when one process per core (96 processes) and
then one thread per process are created
($96 \times 1 = 96$).
The second row shows performances
when one process per two cores (48 processes) and
then two threads per process are created
($48 \times 2 = 96$).
The third row shows performances
when one process per three cores (32 processes) and
then three threads per process are created
($32 \times 3 = 96$).
As can be seen,
the SVD only increases performances when orthonormal matrices are created,
whereas \randUTV{} increases performances in both cases
(both with and without orthonormal matrices).


\begin{table}[ht!]
\begin{center}
\begin{tabular}{|l|rrr|rrr|}
  \multicolumn{1}{c}{} &
  \multicolumn{3}{|c}{No ON matrices} &
  \multicolumn{3}{|c|}{ON matrices} \\
  \multicolumn{1}{c}{} &
  \multicolumn{3}{|c}{Threads per process} &
  \multicolumn{3}{|c|}{Threads per process} \\ \cline{2-7}
  Factorization &
  \multicolumn{1}{c}{1} &
  \multicolumn{1}{c}{2} &
  \multicolumn{1}{c|}{3} &
  \multicolumn{1}{c}{1} &
  \multicolumn{1}{c}{2} &
  \multicolumn{1}{c|}{3} \\ \hline
  SVD              & 393.9 & 644.0 & 645.3 & 1494.1 & 1336.4 & 1318.0 \\
  \randUTV{} $q=0$ & 117.0 & 102.7 & 112.2 &  214.2 &  192.3 &  203.7 \\
  \randUTV{} $q=1$ & 168.6 & 142.5 & 160.1 &  272.1 &  232.3 &  251.4 \\
  \randUTV{} $q=2$ & 216.7 & 180.4 & 207.8 &  327.5 &  271.7 &  298.7 \\ \hline
\end{tabular}
\end{center}
\caption{Best timings in seconds of several topologies with 96 cores
on matrices of dimension $25600 \times 25600$
considering several number of threads per process.}
\label{tab:dm_threads_per_process}
\end{table}

Table~\ref{tab:dm_threads_per_process}
shows the best timings (in seconds) for several topologies with 96 cores
so a finer detail comparison can be achieved.
Matrices being factorized are $25600 \times 25600$.
As can be seen,
SVD increases performances 13 \% when orthonormal matrices are built,
whereas \randUTV{} with $q=2$ improves performances 20 \% in both cases.
Performances usually increase when using two threads per process,
but they remain similar or drop when using more than two threads per process.


\begin{figure}[ht!]
\tfvspace
\begin{center}
\begin{tabular}{cc}
\includegraphics[width=0.45\textwidth]{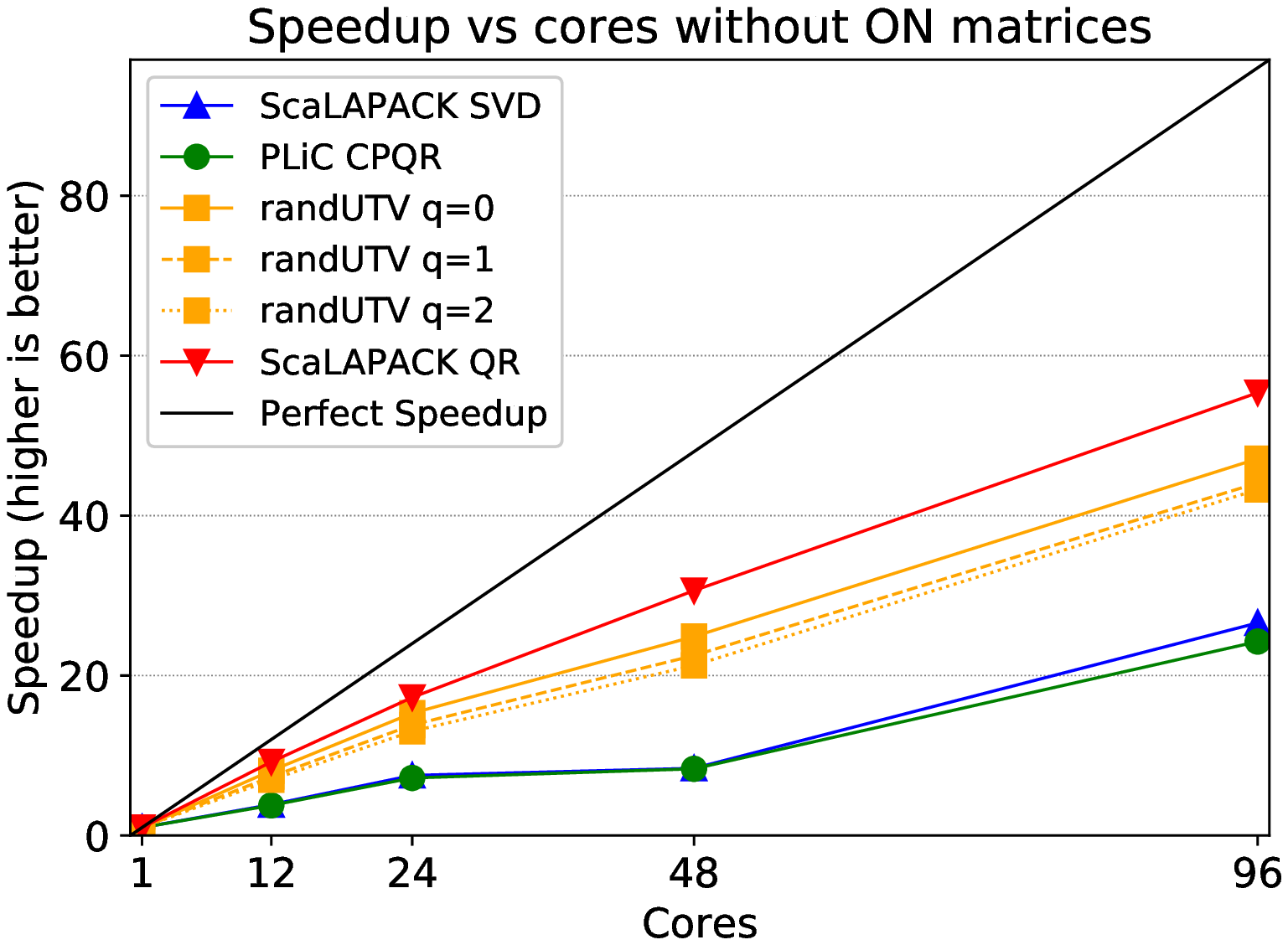} &
\includegraphics[width=0.45\textwidth]{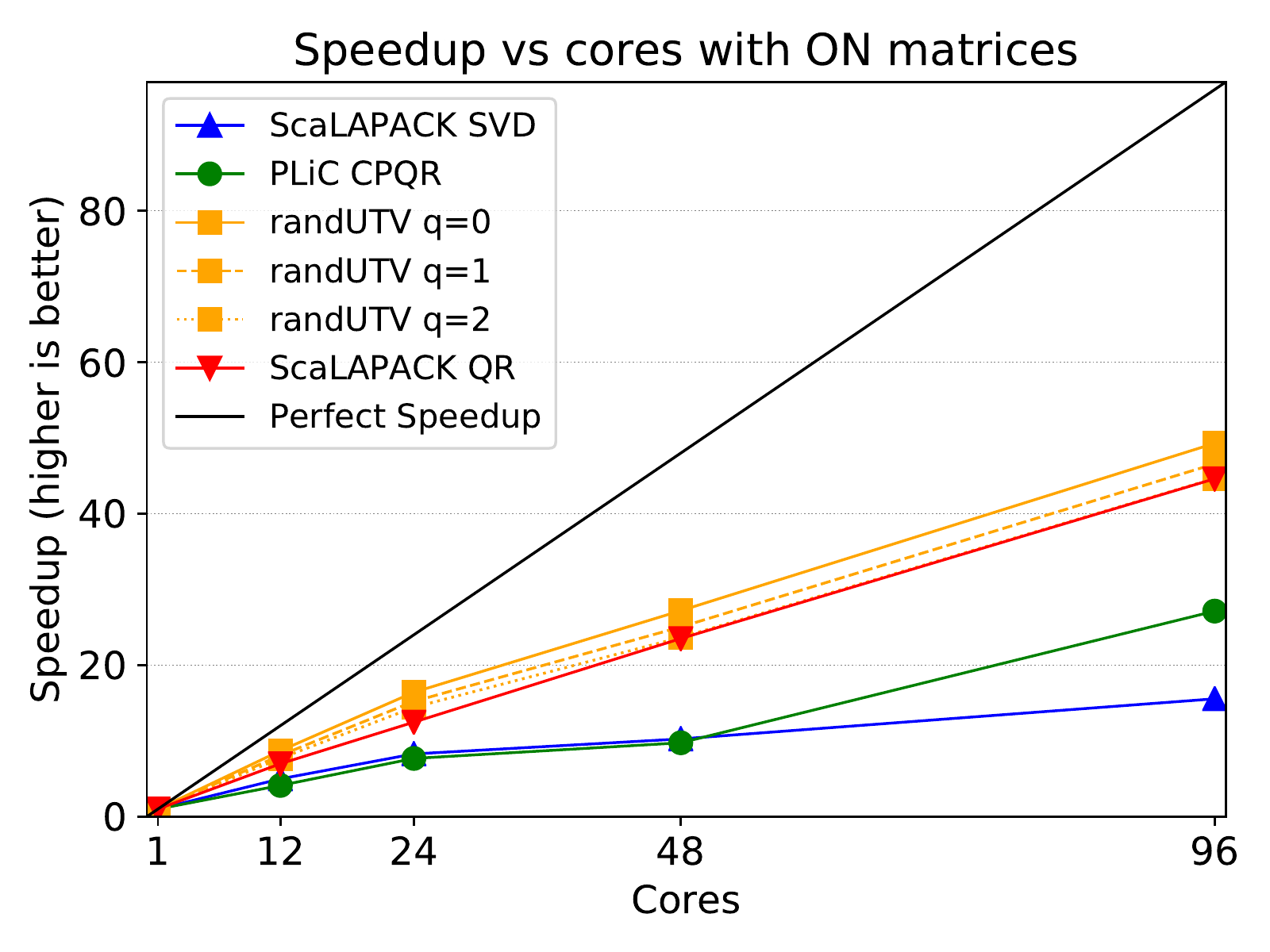} \\
\end{tabular}
\end{center}
\bfvspace
\caption{Speedups versus number of cores for all the implementations
on matrices of dimension $20480 \times 20480$.}
\label{fig:dm_speedups}
\end{figure}

Figure~\ref{fig:dm_speedups}
shows the speedups obtained by all the implementations
on matrices of dimension $20480 \times 20480$.
Recall that in this plot
every implementation compares against itself on one core.
The best topologies have been selected for the following number of cores:
$3 \times 4$ for 12 cores,
$6 \times 4$ for 24 cores,
$4 \times 12$ for 48 cores, and
$6 \times 16$ for 96 cores.
When no orthonormal matrices are built,
speedups of \randUTV{} on the largest number of cores (96)
are between 47.1 ($q=0$) and 43.3 ($q=2$).
When orthonormal matrices are built,
speedups of \randUTV{} on the largest number of cores (96)
are between 49.3 ($q=0$) and 44.7 ($q=2$).
In both cases, the efficiency is close to 50 \%.
When no orthonormal matrices are built,
speedups of \randUTV{} are a bit lower than those of QR factorization;
when orthonormal matrices are built,
speedups of \randUTV{} are a bit higher than those of QR factorization.
In both cases, the speedups of \randUTV{} are much higher than those
obtained by the SVD and the CPQR factorization,
thus showing the great scalability potential of this factorization.

In conclusion, \randUTV{} is significantly faster than the available distributed memory implementations of SVD. It also matches the best CPQR implementation tested. \randUTV{} is known to reveal rank far better than CPQR \cite{martinsson2017randutv}, {and it also furnishes orthonormal bases for the row-space, and for the (numerical) null-space of the matrix.} This means that just matching the speed of CPQR represents a major gain in information at no additional computational cost. Furthermore, \randUTV{} is faster than even CPQR in the case that orthonormal matrices are required. We finally observe that the potential for scalability of \randUTV{} is a clear step above competing implementations for rank-revealing factorizations in distributed memory.

\section{Conclusions.}
\label{sec:conclusions}

We have described two new implementations of the {\tt randUTV} algorithm for computing
the SVD of a matrix, targeting shared-memory and distributed-memory architectures, respectively.

Regarding shared memory, the new implementation proposes an {\em algorithm-by-blocks} that,
built on top of a runtime task scheduler ({\tt libflame}'s SuperMatrix) implements
a dataflow execution model. Based on a DAG, this model reduces the amount of synchronization
points and hence increases performance on massively parallel architectures. Actually, performance
results on a up to 36 cores reveal excellent performance and scalability results compared
with state-of-the-art proprietary libraries.

We have also proposed a distributed-memory algorithm for {\tt randUTV}. This proposal leverages
the classic blocked algorithm rather than the algorihm-by-blocks, and makes heavy use of
ScaLAPACK. Performance results show competitive performance and excellent scalability
compared with alternative state-of-the-art implementations.

In this article, we focused exclusively on the case of multicore CPUs with shared memory and
homogeneous distributed-memory architectures. We expect that the
relative advantages of randUTV will be even more pronounced in more severely communication-
constrained environments, such as GPU-based architectures (composed by one or many nodes).
Work on variations of the method modied for such environments is proposed as future work.

\section*{Acknowledgements}

F.~D.~Igual was supported by the EU (FEDER) and
Spanish MINECO (GA No. RTI2018-093684-B-I00),
and by Spanish CM (GA No. S2018/TCS-4423).

G.~Quintana-Ort\'{\i} was supported by
the Spanish Ministry of Science, Innovation and Universities
under Grant RTI2018-098156-B-C54 co-financed with FEDER funds.

P.~G.~Martinsson was supported by the Office of Naval Research
(grant N00014-18-1-2354) and
by the National Science Foundation (grant DMS-1620472).

The authors would also like to thank Javier Navarrete (Universitat d'Alacant)
for granting access to the distributed-memory server.


\nocite{golub}
\bibliography{main_bib}
\bibliographystyle{amsplain}

\end{document}